\documentclass[oneside,11pt]{article}

\usepackage[utf8]{inputenc}
\usepackage[T1]{fontenc}

\usepackage{amsthm}
\usepackage{amsmath}
\usepackage{amsfonts}
\usepackage{amssymb}

\usepackage{graphicx} 
\usepackage{float}
\usepackage{booktabs}
\usepackage{multirow}
\usepackage{rotating}
\usepackage{array}

\usepackage{breakcites}

\usepackage{enumitem}

\usepackage[top=1.5cm,bottom=2cm,right=2.5cm,left=2.5cm]{geometry}
\usepackage{titling}
\usepackage{natbib}

\usepackage{ifplatform}

\ifwindows
\fi

\ifwindows
	\usepackage{bbm}
\else
	\iflinux
		\usepackage{bbm}
	\else
		\usepackage{bbm}
	\fi
\fi

\newcommand{\lp}{\left(}
\newcommand{\rp}{\right)}
\newcommand{\lc}{\left[}
\newcommand{\rc}{\right]}
\newcommand{\lb}{\left\{}
\newcommand{\rb}{\right\}}

\newcommand{\ri}{\right.}
\newcommand{\R}{\mathbb{R}}

\newcommand{\bx}{\mathbf{x}}

\newcommand{\by}{\mathbf{y}}
\newcommand{\bX}{\mathbf{X}}
\newcommand{\bY}{\mathbf{Y}}
\newcommand{\bmu}{\boldsymbol\mu}
\newcommand{\bu}{\mathbf{u}}

\newcommand{\bb}{\mathbf{b}}

\newcommand{\bz}{\mathbf{z}}

\newcommand{\bxi}{\boldsymbol\xi}

\newcommand{\ba}{\boldsymbol\alpha}

\newcommand{\bbeta}{\boldsymbol\eta}

\newcommand{\btheta}{\boldsymbol\theta}

\newcommand{\Hcal}{\mathcal{H}}
\newcommand{\bHcal}{\boldsymbol{\mathcal{H}}}

\newcommand{\bnab}{\boldsymbol\nabla}
\newcommand{\bB}{\mathbf{B}}
\newcommand{\bA}{\mathbf{A}}

\newcommand{\bI}{\mathbf{I}}

\newcommand{\lrp}[1]{\left(#1\right)}
\newcommand{\lrc}[1]{\left[#1\right]}
\newcommand{\lrb}[1]{\left\{#1\right\}}
\newcommand{\Prob}[1]{\mathbb{P}\lb #1\rb}
\newcommand{\E}[1]{\mathbb{E}\lc #1\rc}
\newcommand{\V}[1]{\mathbb{V}\mathrm{ar}\lc #1\rc}

\newcommand{\Cov}[2]{\mathbb{C}\mathrm{ov}\lc #1,#2\rc}

\newcommand{\pf}[2]{\frac{\partial #1}{\partial #2}}
\newcommand{\pftwo}[2]{\frac{\partial^2 #1}{\partial #2^2}}
\newcommand{\pfmix}[3]{\frac{\partial^2 #1}{\partial #2\partial #3}}
\newcommand{\norm}[1]{\left|\left| #1\right|\right|}
\newcommand{\abs}[1]{\left| #1\right|}
\newcommand{\tr}[1]{\mathrm{tr}\left[#1\right]}

\newcommand{\vlinel}[1]{\multicolumn{1}{|c}{#1}}

\newcommand{\Om}[1]{\Omega_{#1}}
\newcommand{\om}[1]{\omega_{#1}}
\newcommand{\Iq}[2]{\int_{\Omega_{q}} #1\,\omega_{q}(d #2)}
\newcommand{\Iqr}[3]{\int_{\Omega_{q}\times\R} #1\,d #3\,\omega_{q}(d #2)}

\newcommand{\Ir}[2]{\int_{\R} #1\,d #2}
\newcommand{\Iqq}[3]{\int_{\Omega_{q_1}\times\Omega_{q_2}} #1\,\omega_{q_2}(d #3)\,\omega_{q_1}(d #2)}

\DeclareFontFamily{OT1}{pzc}{}
\DeclareFontShape{OT1}{pzc}{m}{it}{<-> s * [1.10] pzcmi7t}{}
\DeclareMathAlphabet{\mathpzc}{OT1}{pzc}{m}{it}
\newcommand{\order}[1]{\mathpzc{o}\lp#1\rp}
\newcommand{\Order}[1]{\mathcal{O}\lp#1\rp}

\newcommand{\Orderp}[1]{\mathcal{O}_\mathbb{P}\lp#1\rp}

\usepackage[pdftex,final,bookmarksnumbered,bookmarksopen=false,breaklinks,colorlinks]{hyperref}
\hypersetup{
	final,
    bookmarksnumbered=true,
    bookmarksopen=true,
    bookmarksopenlevel=0,
    unicode=false,          % non-Latin characters in Acrobat?s bookmarks
    pdftoolbar=true,        % show Acrobat?s toolbar?
    pdfmenubar=true,        % show Acrobat?s menu?
    pdffitwindow=false,     % window fit to page when opened
    pdftitle={Central limit theorems for directional and linear random variables with applications},    % title
	pdfdisplaydoctitle=true, %display document title instead of file name in title bar
	pdftoolbar=true,
	pdfmenubar=true,
	pdflang={English},
    pdfauthor={Eduardo Garcia-Portugues, Rosa M. Crujeiras, Wenceslao Gonzalez-Manteiga},     % author
    pdfsubject={arXiv paper},   % subject of the document
    pdfcreator={Eduardo Garcia-Portugues},   % creator of the document
    pdfproducer={Eduardo Garcia-Portugues}, % producer of the document
    pdfkeywords={Directional data} {Goodness-of-fit} {Independence test} {Kernel density estimation} {Limit distribution}, % list of keywords
    pdfnewwindow=true,      % links in new window
    breaklinks=true,
    hidelinks,       
    linkcolor=black,          % color of internal links (change box color with linkbordercolor)
    citecolor=black,        % color of links to bibliography
    filecolor=magenta,      % color of file links
    urlcolor=cyan           % color of external links
}

\newtheorem{theo}{Theorem}
\newtheorem{coro}{Corollary}

\newtheorem{lem}{Lemma}
\newtheorem{algo}{Algorithm}

\allowdisplaybreaks

\newif\ifmain
\maintrue
\ifmain

\else

\fi
\newif\ifsupplement
\supplementtrue

\begin{document}

\ifmain

%---------- TITLE ---------------%
\title{Central limit theorems for directional and linear random variables with applications}
\setlength{\droptitle}{-1cm}
\predate{}%
\postdate{}%
%---------- AUTHORS ---------------%
\date{}

\author{Eduardo Garc\'ia-Portugu\'es$^{1,2}$, Rosa M. Crujeiras$^{1}$, and Wenceslao Gonz\'alez-Manteiga$^{1}$}

\footnotetext[1]{
Department of Statistics and Operations Research, University of Santiago de Compostela (Spain).}
\footnotetext[2]{Corresponding author. e-mail: \href{mailto:eduardo.garcia@usc.es}{eduardo.garcia@usc.es}.}

\maketitle

%-------------------------------------%

\begin{abstract}
A central limit theorem for the integrated squared error of the directional-linear kernel density estimator is established. The result enables the construction and analysis of two testing procedures based on squared loss: a nonparametric independence test for directional and linear random variables and a goodness-of-fit test for parametric families of directional-linear densities. Limit distributions for both test statistics, and a consistent bootstrap strategy for the goodness-of-fit test, are developed for the directional-linear case and adapted to the directional-directional setting. Finite sample performance for the goodness-of-fit test is illustrated in a simulation study. This test is also applied to datasets from biology and environmental sciences.
\end{abstract}
\begin{flushleft}
\end{flushleft}

%-------------------------------------------------%
\section{Introduction}
\label{gofdens:sec:introduction}
%-------------------------------------------------%

Statistical inference on random variables comprises estimation and testing procedures that allow one to characterize the underlying distribution, regardless the variables nature and/or dimension. Specifically, density estimation stands out as a basic problem in statistical inference for which parametric and nonparametric approaches have been explored. In nonparametrics, kernel density estimation (see \cite{Silverman1986}, \cite{Scott1992}, or \cite{Wand1995}, as comprehensive references for scalar random variables) provides a simple and intuitive way to explore and do inference on random variables. Among other contexts, kernel density estimation has been also adapted to directional data (see \cite{Mardia2000}). Data on the $q$-dimensional sphere arises, for example, in meteorology when measuring wind direction; in proteomics, when studying the angles in protein structure (circular data, $q=1$, see \cite{Fern'andez-Dur'an2007}); in astronomy, with the stars positions in the celestial sphere ($q=2$, see \cite{Garcia-Portugues:exact}); in text mining, when codifying documents in the vector space model (large $q$, see Chapter 6 in \cite{Srivastava2010}). Some early works on kernel density estimation with directional data are the papers by \cite{Hall1987} and \cite{Bai1988}, who introduced kernel density estimators and their properties (bias, variance and uniformly strong consistency, among others). The estimation of the density derivatives was studied by \cite{Klemela2000}, and \cite{Zhao2001} stated a Central Limit Theorem (CLT) for the Integrated Squared Error (ISE) of the directional kernel density estimator. Some recent works deal with the bandwidth selection problem, such as \cite{Taylor2008} and \cite{Oliveira2012}, devoted to circular data and \cite{Garcia-Portugues:exact}, for a general dimension. In some contexts, joint density models for directional and linear random variables are useful (e.g. for describing wind direction and SO$_2$ concentration \citep{Garcia-Portugues:so2}). In this setting, a kernel density estimator for directional-linear data was proposed and analysed by \cite{Garcia-Portugues:dirlin}.\\

Regardless of estimation purposes, kernel density estimators have been extensively used for the development of goodness-of-fit tests (see \cite{Gonzalez-Manteiga2013} for a review) and independence tests. For example, \cite{Bickel1973} and \cite{Fan1994} provided goodness-of-fit tests for parametric densities for real random variables. Similarly, in the directional setting, \cite{Boente2013} presented a goodness-of-fit test for parametric directional densities. For assessing independence between two linear random variables, Rosenblatt (1975) proposed a test statistic based on the squared difference between the joint kernel density estimator and the product of the marginal ones (see also \cite{Rosenblatt1992}). This idea was adapted to the directional-linear setting by \cite{Garcia-Portugues:testindep}, who derived a permutation independence test and compared its performance with the testing proposals given by \cite{Mardia1976}, \cite{Johnson1978}, and \cite{Fisher1981} in this\nopagebreak[4] context.\\

The main device for the goodness-of-fit and independence tests is the CLT for the ISE of the kernel density estimator, and the aim of this work is to provide such a result for the directional-linear kernel estimator and use it to derive a goodness-of-fit test for parametric families of directional-linear densities and an independence test for directional and linear variables. The CLT is obtained by proving an extended version of Theorem 1 in \cite{Hall1984}. The goodness-of-fit test follows by taking the ISE between the joint kernel estimator and a smoothed parametric estimate of the unknown density as a test statistic. For the independence test, the test statistic introduced in \cite{Garcia-Portugues:testindep} is considered and its asymptotic properties are studied. Jointly with the asymptotic distribution, a bootstrap resampling strategy to calibrate the goodness-of-fit test is investigated. Finite sample performance of the goodness-of-fit test is checked through an extensive simulation study, and this methodology is applied to analyse datasets from forestry and proteomics. In addition, the results obtained for the directional-linear case are adapted to the directional-directional context.\\

The rest of this paper is organized as follows. Section \ref{gofdens:sec:background} presents some background on kernel density estimation for directional and linear random variables. Section \ref{gofdens:sec:clt} includes the CLT for the ISE of the directional-linear estimator and its extension to the directional-directional setting. The independence test for directional and linear variables is presented in Section \ref{gofdens:sec:indep}. The goodness-of-fit test for simple and composite null hypotheses, its bootstrap calibration and extensions are given in Section \ref{gofdens:sec:gof}. The empirical performance of the goodness-of-fit test is illustrated with a simulation study in Section \ref{gofdens:sec:sim} and with applications to datasets in Section \ref{gofdens:sec:data}. Appendix \ref{gofdens:ap:main} collects the outline of the main proofs. Technical lemmas and further details on simulations and data analysis are provided as supplementary material, as well as the extensions of the independence test.

%-------------------------------------------------%
\section{Background}
\label{gofdens:sec:background}
%-------------------------------------------------%

For simplicity, $f$ denotes the target density along the paper, which may be linear, directional, directional-linear, or directional-directional, depending on the context.\\

Let $Z$ denote a linear random variable with support $\mathrm{supp}(Z)\subseteq\mathbb R$ and density $f$, and let $Z_1,\ldots,Z_n$ be a random sample of $Z$. The linear kernel density estimator is defined as
\[
\hat f_g(z)=\frac{1}{ng}\sum_{i=1}^n K\lrp{\frac{z-Z_i}{g}},\quad z\in\R,
\]
where $K$ denotes the kernel function and $g>0$ is the bandwidth parameter, which controls the smoothness of the estimator (see \cite{Silverman1986}, among others).\\

Let $\bX$ denote a directional random variable with density $f$ and support the $q$-dimensional sphere, denoted by $\Om{q}=\big\{\bx\in\R^{q+1}:x_1^2+\cdots+x^2_{q+1}=1\big\}$. Lebesgue measure in $\Om{q}$ is denoted by $\om{q}$ and, therefore, a directional density satisfies $\Iq{f(\bx)}{\bx}=1$. When there is no possible confusion, $\om{q}$ will also denote the surface area of $\Om{q}$: $\om{q}=\om{q}\lrp{\Om{q}}=2\pi^\frac{q+1}{2}/\Gamma\big(\frac{q+1}{2}\big)$. The directional kernel density estimator introduced by \cite{Hall1987} and \cite{Bai1988} for a directional density $f$, based on a random sample $\bX_1,\ldots,\bX_n$ in the $q$-sphere, is
\[
\hat f_{h}(\bx)=\frac{c_{h,q}(L)}{n}\sum_{i=1}^n L\lrp{\frac{1-\bx^T\bX_i}{h^2}},\quad\bx\in \Omega_q,
\]
where $L$ is the directional kernel, $h>0$ is the bandwidth parameter and the scalar product of two vectors, $\bx$ and $\by$, is denoted by $\bx^T\by$, where $\bx^T$ is the transpose of the column vector $\bx$. $c_{h,q}(L)$ is a normalizing constant depending on the kernel $L$, the bandwidth $h$ and the dimension $q$. Specifically, \cite{Bai1988} has the inverse of the normalizing constant as
\begin{align}
c_{h,q}(L)^{-1}=\lambda_{h,q}(L) h^{q}\sim\lambda_q(L) h^{q},\label{gofdens:normalizing}
\end{align}
where $\lambda_{h,q}(L)=\om{q-1}\int_0^{2h^{-2}} L(r) r^{\frac{q}{2}-1}(2-rh^2)^{\frac{q}{2}-1}\,dr$ and $\lambda_q(L)=2^{\frac{q}{2}-1}\om{q-1}\allowbreak\int_0^{\infty} L(r) r^{\frac{q}{2}-1}\,dr$. The notation $a_n\sim b_n$ means that $a_n=b_n(1+\order{1})$.\\

A usual choice for the directional kernel is $L(r)=e^{-r}$, also known as the von Mises kernel due to its relation with the von Mises--Fisher density \citep{Watson1983}, $\mathrm{vM}(\bmu,\kappa)$, given by
\[
f_{\mathrm{vM}}(\bx;\bmu,\kappa)=C_q(\kappa) \exp{\lrb{\kappa\bx^T\bmu}},\quad C_q(\kappa)=\frac{\kappa^{\frac{q-1}{2}}}{(2\pi)^{\frac{q+1}{2}}\mathcal{I}_{\frac{q-1}{2}}(\kappa)},%
\]
where $\bmu\in\Omega_q$ is the directional mean, $\kappa>0$ is the concentration parameter around the mean, and $\mathcal{I}_\nu$ is the modified Bessel function of order $\nu$.\\

The kernel estimator for a directional-linear density $f$ based on a random sample $\lrp{\bX_1,Z_1},\ldots,$ $\lrp{\bX_n,Z_n}$, with $\lrp{\bX_i,Z_i}\in\Omega_q\times\R,\,i=1,\ldots,n$, was proposed by \cite{Garcia-Portugues:dirlin}:
\begin{align}
\label{gofdens:kernel_dirlin}
\hat f_{h,g}(\bx,z)=\frac{c_{h,q}(L)}{ng}\sum_{i=1}^nLK\lp\frac{1-\bx^T\bX_i}{h^2},\frac{z-Z_i}{g}\rp,\quad (\bx,z)\in\Omega_q\times\R,
\end{align}
where $LK$ is a directional-linear kernel, $h$ and $g$ are the bandwidths for the directional and the linear components, respectively, and $c_{h,q}(L)$ is the normalizing constant. For simplicity, the product kernel $LK(\cdot,\cdot)=L(\cdot)\times K(\cdot)$ is considered. To quantify the error of the density estimator, the ISE,
\[
\mathrm{ISE}\lrc{\hat f_{h,g}}=\Iqr{\lrp{\hat f_{h,g}(\bx,z)-f(\bx,z)}^2}{\bx}{z},
\]
can be used. In this expression, the integral is taken with respect to the product measure $\om{q}\times m_\R$, with $m_\R$ denoting the usual Lebesgue measure in $\R$.\\

It is possible to define a directional-directional kernel density estimator at $(\bx,\by)\in\Omega_{q_1}\times\Omega_{q_2}$ from a random sample $\lrp{\bX_1,\bY_1},\ldots,\lrp{\bX_n,\bY_n}$, with $\lrp{\bX_i,\bY_i}\in\Omega_{q_1}\times\Omega_{q_2},\,i=1,\ldots,n$, that comes from a directional-directional density $f$:
\[
\hat f_{h_1,h_2}(\bx,\by)=\frac{c_{h_1,q_1}(L_1)c_{h_2,q_2}(L_2)}{n}\sum_{i=1}^nL_1\lrp{\frac{1-\bx^T\bX_i}{h_1^2}}\times L_2\lrp{\frac{1-\by^T\bY_i}{h_2^2}}.
\]

To fix notation, $R(\mathcal\varphi)$ denotes the integral of the squared function $\varphi^2$ along its domain. The following integrals are needed: 
\[
\mu_2(K)=\int_\R z^2K(z)\,dz,\quad b_q(L)=\frac{\int_0^\infty L(r) r^{\frac{q}{2}}\,dr}{\int_0^\infty L(r) r^{\frac{q}{2}-1}\,dr}.
\]
Density derivatives of different orders are denoted as follows:
\begin{align*}
\renewcommand{\arraystretch}{1.4} %
\bnab f(\bx,z)=&\,\lrp{\pf{f(\bx,z)}{x_1},\ldots,\pf{f(\bx,z)}{x_{q+1}},\pf{f(\bx,z)}{z}}^T=\lrp{\bnab_{\bx}f(\bx,z),\nabla_z f(\bx,z)}^T,\\
\bHcal f(\bx,z)=&\,
\lrp{\begin{array}{cc}
	\lrp{\pfmix{f(\bx,z)}{x_i}{x_j}} & \vlinel{\pfmix{f(\bx,z)}{\bx}{z}}\\[0.1cm]\cline{1-2}
	& \vlinel{}\\[-0.35cm]
	\pfmix{f(\bx,z)}{z}{\bx^T} & \vlinel{\pftwo{f(\bx,z)}{z}}
\end{array}}=
\lrp{\begin{array}{cc}
		\bHcal_\bx f(\bx,z) & \vlinel{\bHcal_{\bx,z} f(\bx,z)} \\[0.1cm]\cline{1-2}
		& \vlinel{}\\[-0.35cm]
		\bHcal_{\bx,z} f(\bx,z)^T & \vlinel{\Hcal_z f(\bx,z)}
	\end{array}}.
\end{align*}

%-------------------------------------------------%
\section{Central limit theorem for the integrated squared error}
\label{gofdens:sec:clt}
%-------------------------------------------------%

Our main result is the CLT for the ISE of the kernel density estimator (\ref{gofdens:kernel_dirlin}). 

%-------------------------------------------------%
\subsection{Main result}
%-------------------------------------------------%

We need the following conditions.
\begin{enumerate}[label=\textbf{A\arabic{*}}, ref=\textbf{A\arabic{*}}]
	\item If $f$ is extended from $\Omega_q\times\R$ to $\R^{q+2}\backslash\lrb{(\mathbf{0},z):z\in\R}$ as $f(\bx,z)\equiv f\lrp{\bx/\norm{\bx},z}$ for all $\bx\neq\mathbf{0}$ and $z\in\R$, $f$ and its first three derivatives are bounded and uniformly continuous with respect to the product Euclidean norm in $\Om{q}\times\R$, $\norm{(\bx,z)}=\sqrt{\norm{\bx}^2+\abs{z}^2}$. \label{gofdens:assump:a1}
	\item $L:[0,\infty)\rightarrow[0,\infty)$ and $K:\R\rightarrow[0,\infty)$ are continuous and bounded; $L$ is nonincreasing such that $0<\lambda_q(L),\,\lambda_q(L^2)<\infty$, $\forall q\geq1$ and $K$ is a linear density, symmetric around zero and with $\mu_2(K)<\infty$. \label{gofdens:assump:a2}
	\item $h=h_n$ and $g=g_n$ are sequences of positive numbers such that $h_n\rightarrow0$, $g_n\rightarrow0$, and $n h_n^q g_n\rightarrow\infty$ as $n\rightarrow\infty$. \label{gofdens:assump:a3}
\end{enumerate}

The uniform continuity and boundedness up to the second derivatives of $f$ is a common assumption that appears, among others, in \cite{Hall1984} and \cite{Rosenblatt1992}, while the assumption on the third derivatives is needed for uniform convergence. The assumption of compact support for the directional kernel $L$, stated in \cite{Zhao2001}, is replaced by the nonincreasing requirement and the finiteness of $\lambda_q(L)$ and $\lambda_q(L^2)$. These two conditions are less restrictive and allow for consideration of the von Mises kernel. We provide the limit distribution of the ISE for (\ref{gofdens:kernel_dirlin}). The proof is based on a generalization of Theorem 1 in \cite{Hall1984}, stated as Lemma \ref{gofdens:lemma:extendHn} in Appendix \ref{gofdens:ap:main}.

\begin{theo}[CLT for the directional-linear ISE]
	\label{gofdens:theo:clt}
	Denote the ISE of  $\hat f_{h,g}$ by $I_n$. If \ref{gofdens:assump:a1}--\ref{gofdens:assump:a3} hold, then
	\begin{enumerate}[label=\roman{*}., ref=\textit{\roman{*}}]
		\item $n^{\frac{1}{2}}\phi(h,g)^{-\frac{1}{2}}\lrp{I_n-\E{I_n}}\stackrel{d}{\longrightarrow}\mathcal{N}(0,1)$, if $n\phi(h,g)h^qg\to\infty$,\label{gofdens:theo:clt:i}
		\item $n(h^qg)^{\frac{1}{2}}\lrp{I_n-\E{I_n}}\stackrel{d}{\longrightarrow} \mathcal{N}\lrp{0,2\sigma^2}$, if $n\phi(h,g)h^qg\to0$,\label{gofdens:theo:clt:ii}
		\item $n(h^qg)^{\frac{1}{2}}\lrp{I_n-\E{I_n}}\stackrel{d}{\longrightarrow} \mathcal{N}\lrp{0,\delta+2\sigma^2}$, if  $n\phi(h,g)h^qg\to\delta$,\label{gofdens:theo:clt:iii}
	\end{enumerate}
	where $0<\delta<\infty$ and
	\begin{align*}
		\phi(h,g)=&\,\frac{4b_{q}(L)^2}{q^2}\sigma_\bX^2h^4+\mu_2(K)^2\sigma_Z^2g^4+\frac{4b_q(L)\mu_2(K)}{q}\sigma_{\bX,Z}h^2g^2,
	\end{align*}
	with $\sigma_{\bX,Z}=\Cov{\tr{\bHcal_\bx(f,\bX,Z)}}{\Hcal_z f(\bX,Z)}$, $\sigma^2_\bX=\V{\tr{\bHcal_\bx(f,\bX,Z)}}$ and $\sigma^2_Z=$\\ $\V{\Hcal_z f(\bX,Z)}$. The remaining constants are given by:
	\begin{align*}
		\sigma^2=&\,R(f)\times\gamma_q \lambda_q(L)^{-4} \int_0^{\infty} r^{\frac{q}{2}-1}\lrb{\int_0^{\infty} \rho^{\frac{q}{2}-1} L(\rho) \varphi_q(r,\rho) \,d\rho}^2\,dr\\
		&\qquad\;\times \int_\R \lrb{\int_\R  K(u)K(u+v)\,du}^2\,dv,\\
		\varphi_q(r,\rho)=&\,\left\{ 
		\begin{array}{ll}
			L\lrp{r+\rho-2(r\rho)^{\frac{1}{2}}}+L\lrp{r+\rho+2(r\rho)^{\frac{1}{2}}}, & q=1,\\
			\int_{-1}^1 \left( 1-\theta^2 \right)^{\frac{q-3}{2}} L\lrp{r+\rho-2\theta(r\rho)^{\frac{1}{2}}}\,d\theta, & q\geq2,\\
		\end{array}
		\right.\\
		\gamma_q=&\,\left\{ 
		\begin{array}{ll}
			2^{-\frac{1}{2}}, & q=1,\\
			\om{q-1}\om{q-2}^2 2^{\frac{3q}{2}-3}, & q\geq2.
		\end{array}
		\right.
	\end{align*}
	The same limit distributions hold in \ref{gofdens:theo:clt:i}--\ref{gofdens:theo:clt:iii} if $\E{I_n}$ is replaced by
	\[
	\Iqr{\lrp{\E{\hat f_{h,g}(\bx,z)}-f(\bx,z)}^2}{\bx}{z}+\frac{\lambda_q(L^2)\lambda_q(L)^{-2}R(K)}{nh^qg}.
	\]
\end{theo}
Bearing in mind the CLT result in \cite{Hall1984} for the linear case, a bandwidth-free rate of convergence should be expected in \ref{gofdens:theo:clt:iii}. Nevertheless, when $n\phi(h,g)\allowbreak h^qg\to\delta$, the analytical difficulty of joining the two rates of convergence of the dominant terms forces the normalizing rate to be $n(h^qg)^{\frac{1}{2}}$, although the sequence of bandwidths is restricted to satisfy the constraint $n\phi(h,g)h^qg\to\delta$. To clarify this point, a corollary presents a special case with proportional bandwidth sequences where the rate of convergence can be analytically stated in a bandwidth-free form.
\begin{coro}
	\label{gofdens:coro:clt}
	Under \ref{gofdens:assump:a1}--\ref{gofdens:assump:a3}, and assuming $g_n=\beta h_n$ for a fixed $\beta>0$ and $0<\delta<\infty$,
	\begin{enumerate}[label=\roman{*}., ref=\textit{\roman{*}}]
		\item $n^{\frac{1}{2}}h^{-2}\lrp{I_n-\E{I_n}}\stackrel{d}{\longrightarrow}\mathcal{N}(0,\phi(1,\beta))$, if $nh^{q+5}\to\infty$,\label{gofdens:coro:clt:i}
		\item $nh^{\frac{q+1}{2}}\lrp{I_n-\E{I_n}}\stackrel{d}{\longrightarrow}\mathcal{N}\lrp{0,2\sigma^2}$, if $nh^{q+5}\to0$,\label{gofdens:coro:clt:ii}
		\item $n^{\frac{q+9}{2(q+5)}}\lrp{I_n-\E{I_n}}\stackrel{d}{\longrightarrow} \mathcal{N}\Big(0,\phi(1,\beta)\delta^\frac{4}{q+5}+2\sigma^2\delta^{-\frac{q+1}{q+5}}\Big)$, if $nh^{q+5}\to\delta$.\label{gofdens:coro:clt:iii}
	\end{enumerate}
\end{coro}

%-------------------------------------------------%
\subsection{Extensions of Theorem \ref{gofdens:theo:clt}}
\label{gofdens:subsec:clt:exten}
%-------------------------------------------------%

The previous results can be adapted to other contexts involving directional variables, such as directional-directional or directional-multivariate random vectors. Once the common structure and the effects of each component are determined, it is easy to reproduce the computations duplicating a certain component or modifying it. This will be used to derive the directional-directional versions of the most relevant results along the paper. By considering a single bandwidth for the estimator defined in $\R^p$ (as in \cite{Hall1984}, for example), Theorem \ref{gofdens:theo:clt} can be easily\nopagebreak[4] adapted to account for a multivariate component.\\

Considering the directional-directional estimator $\hat f_{h_1,h_2}$, the corresponding analogues of conditions \ref{gofdens:assump:a1}--\ref{gofdens:assump:a3} are obtained (extending $f$ from $\Om{q_1}\times\Om{q_2}$ to $\{(\bx,\by)\in\R^{q_1+q_2+2}: \bx\neq\mathbf{0},\,\by\neq\mathbf{0}\}$ and assuming $nh_{1,n}^{q_1}h_{2,n}^{q_2}\to\infty$). Then, it is possible to derive a directional-directional version of Theorem\nolinebreak[4] \ref{gofdens:theo:clt}.
\begin{coro}[CLT for the directional-directional ISE]
	\label{gofdens:coro:clt:dirdir}
	Denote the ISE of $\hat f_{h_1,h_2}$ by $I_n=\int_{\Om{q_1}\times\Om{q_2}}$ $(\hat f_{h_1,h_2}(\bx,\by)-f(\bx,\by))^2\,\om{q_2}(\by)\,\om{q_1}(\bx)$. Then, under the directional-directional analogues of \ref{gofdens:assump:a1}--\ref{gofdens:assump:a3}, 
	\begin{enumerate}[label=\roman{*}., ref=\textit{\roman{*}}]
		\item $n^{\frac{1}{2}}\phi(h_1,h_2)^{-\frac{1}{2}}\lrp{I_n-\E{I_n}}\stackrel{d}{\longrightarrow}Z$,  if $n\phi(h_1,h_2)h_1^{q_1}h_2^{q_2}\to\infty$,\label{gofdens:coro:clt:dirdir:i}
		\item $n(h_1^{q_1}h_2^{q_2})^{\frac{1}{2}}\lrp{I_n-\E{I_n}}\stackrel{d}{\longrightarrow}2^\frac{1}{2}\sigma Z$, if $n\phi(h_1,h_2)h_1^{q_1}h_2^{q_2}\to0$,\label{gofdens:coro:clt:dirdir:ii}
		\item $n(h_1^{q_1}h_2^{q_2})^{\frac{1}{2}}\lrp{I_n-\E{I_n}}\stackrel{d}{\longrightarrow}\lrp{\delta+2\sigma^2}^{\frac{1}{2}} Z$, if $n\phi(h_1,h_2)h_1^{q_1}h_2^{q_2}\to\delta$,\label{gofdens:coro:clt:dirdir:iii}
	\end{enumerate}
	where $0<\delta<\infty$ and 
	\begin{align*}
		\phi(h_1,h_2)=&\,\frac{4b_{q_1}(L_1)^2}{q_1^2}\sigma_\bX^2h_1^4+\frac{4b_{q_2}(L_2)^2}{q_2^2}\sigma_\bY^2h_2^4+\frac{8b_{q_1}(L_1)b_{q_2}(L_2)}{q_1q_2}\sigma_{\bX,\bY}h_1^2h_2^2.\\
		\sigma^2=&\,R(f)\times\gamma_{q_1} \lambda_{q_1}(L_1)^{-4}\! \int_0^{\infty} \!r^{\frac{q_1}{2}-1}\lrb{\int_0^{\infty}\! \rho^{\frac{q_1}{2}-1} L_1(\rho) \varphi_{q_1}(r,\rho) \,d\rho}^2\!\!\,dr\\
		&\qquad\;\times\gamma_{q_2} \lambda_{q_2}(L_2)^{-4} \!\int_0^{\infty} \!r^{\frac{q_2}{2}-1}\lrb{\int_0^{\infty} \!\rho^{\frac{q_2}{2}-1} L_2(\rho) \varphi_{q_2}(r,\rho) \,d\rho}^2\!\!\,dr,
	\end{align*}
	with $\sigma_{\bX,\bY}=\Cov{\tr{\bHcal_\bx(f,\bX,\bY)}}{\tr{\bHcal_\by(f,\bX,\bY)}}$, $\sigma^2_\bX=\V{\tr{\bHcal_\bx(f,\bX,\bY)}}$ and $\sigma^2_\bY=$\\ $\V{\tr{\bHcal_\by(f,\bX,\bY)}}$. The same limit distributions hold in \ref{gofdens:coro:clt:dirdir:i}--\ref{gofdens:coro:clt:dirdir:iii} if $\E{I_n}$ is replaced by
	\[
	\int_{\Om{q_1}\times\Om{q_2}}\!\!\!\!\lrp{\mathbb{E}\lrc{\hat f_{h_1,h_2}(\bx,\by)}\!-\!f(\bx,\by)}^2\!\! \om{q_2}(\by)\,\om{q_1}(\bx)\!+\!\frac{\lambda_{q_1}(L_1^2)\lambda_{q_2}(L_2^2)}{\lambda_{q_1}(L_1)^{2}\lambda_{q_2}(L_2)^{2}nh_1^{q_1}h_2^{q_2}}.
	\]
\end{coro}

%-------------------------------------------------%
\section{Testing independence with directional random variables}
\label{gofdens:sec:indep}
%-------------------------------------------------%

Given a random sample $(\bX_1,Z_1),\ldots,(\bX_n,Z_n)$ from a directional-linear variable $(\bX,Z)$, one may be interested in the assessment of independence between components. If such a hypothesis is rejected, the joint kernel density estimator may give an idea of the dependence structure between them.\\

Let denote by $f_{(\bX,Z)}$ the directional-linear density of $(\bX,Z)$, with $f_\bX$ and $f_Z$ the directional and linear marginal densities. In this setting, the null hypothesis of independence is stated as $H_0: f_{(\bX,Z)}(\bx,z)=f_\bX(\bx)f_Z(z)$, $\forall (\bx,z)\in\Om{q}\times\R$, and the alternative as $H_1: f_{(\bX,Z)}(\bx,z)\neq f_\bX(\bx)f_Z(z)$, for some $(\bx,z)\in\Om{q}\times\R$. A statistic to test $H_0$ can be constructed considering the squared distance between the nonparametric estimator of joint density, denoted in this setting by $\hat f_{(\bX,Z);h,g}$, and the product of the corresponding marginal kernel estimators, denoted by $\hat f_{\bX,h}$ and $\hat f_{Z,g}$,
\[
T_n=\int_{\Om{q}\times\R}\lrp{\hat f_{(\bX,Z);h,g}(\bx,z)-\hat f_{\bX;h}(\bx)\hat f_{Z;g}(z)}^2\,dz\,\om{q}(d\bx).
\]
This type of test was introduced by \cite{Rosenblatt1975} and \cite{Rosenblatt1992} for bivariate random variables, considering the same bandwidths for smoothing both components. The directional-linear context requires an assumption on the degree of smoothness in each component.
\begin{enumerate}[label=\textbf{A\arabic{*}}., ref=\textbf{A\arabic{*}}]
	\setcounter{enumi}{3}
	\item $h_n^{q}g_n^{-1}\rightarrow c$, with $0<c<\infty$, as $n\rightarrow\infty$. \label{gofdens:assump:a4}
\end{enumerate}

\begin{theo}[Directional-linear independence test]
	\label{gofdens:theo:indep}
	Under \ref{gofdens:assump:a1}--\ref{gofdens:assump:a4} and the null hypothesis of independence,
	\[
	n(h^qg)^\frac{1}{2}\lrp{T_n-A_n}\stackrel{d}{\longrightarrow}\mathcal{N}(0,2\sigma_I^2),
	\]
	where 
	\[
	A_n=\frac{\lambda_{q}(L^2)\lambda_{q}(L)^{-2}R(K)}{nh^qg}-\frac{\lambda_{q}(L^2)\lambda_{q}(L)^{-2}R(f_Z)}{nh^q}-\frac{R(K)R(f_\bX)}{ng},
	\]
	and $\sigma_I^2$ is defined as $\sigma^2$ in Theorem \ref{gofdens:theo:clt}, but with $R(f)=R(f_\bX)R(f_Z)$.
\end{theo}

Since the leading term is the same as in Theorem \ref{gofdens:theo:clt} for $n\phi(h,g)h^qg\to0$, the asymptotic variance is also the same. As in the CLT for the ISE, the effect of the components can be disentangled in the asymptotic variance and in the bias term. The a priori complex contribution of the directional part in Theorems \ref{gofdens:theo:clt} and \ref{gofdens:theo:gof} is explained for a particular scenario in the supplementary material, together with some numerical experiments for illustrating Theorem \ref{gofdens:theo:indep}.

%-------------------------------------------------%
\section{Goodness-of-fit test with directional random variables}
\label{gofdens:sec:gof}
%-------------------------------------------------%

Testing methods for a specific parametric directional-linear density (simple $H_0$) or for a parametric family (composite $H_0$) are presented in this section. 

%-------------------------------------------------%
\subsection{Testing a simple null hypothesis}
\label{gofdens:subsec:gof:simple}
%-------------------------------------------------%

Given a random sample $\lrb{(\bX_i,Z_i)}_{i=1}^n$ from an unknown directional-linear density $f$, the simple null hypothesis testing problem is stated as $H_0:f=f_{\btheta_0}$, $\btheta_0\in\boldsymbol\Theta$, where $f_{\btheta_0}$ is a certain parametric density with known parameter $\btheta_0$ belonging to the parameter space $\boldsymbol\Theta\subset\R^p$, with $p\geq 1$. The alternative hypothesis is taken as $H_1: f(\bx,z)\neq f_{\btheta_0}(\bx,z),\text{ for some }(\bx,z)\in\Om{q}\times\R$ in a set of positive measure. The proposed test statistic is
\begin{align}
	R_n=\Iqr{\lrp{\hat f_{h,g}(\bx,z)-LK_{h,g}f_{\btheta_0}(\bx,z)}^2}{\bx}{z}, \label{gofdens:gof:simple}
\end{align}
where $LK_{h,g}f_{\btheta_0}(\bx,z)$ represents the expected value of $\hat f_{h,g}(\bx,z)$ under $H_0$. In general, for a function $f$, this expected value is
\begin{align}
	LK_{h,g} f (\bx,z)=\frac{c_{h,q}(L)}{g}\Iqr{LK\lrp{\frac{1-\bx^T\by}{h^2},\frac{z-t}{g}} f(\by,t)}{\by}{t}.\label{gofdens:smoothing}
\end{align}
Smoothing the parametric density was considered by \cite{Fan1994}, in the linear setting, to avoid the bias effects in the integrand of the square error between the nonparametric estimator under the alternative and the parametric estimate under the null. A modification of the smoothing proposal was used by \cite{Boente2013} for the directional case.
\begin{theo}
	\label{gofdens:theo:gof:simp}
	Under \ref{gofdens:assump:a1}--\ref{gofdens:assump:a3} and the simple null hypothesis $H_0: f=f_{\btheta_0}$, with $\btheta_0\in\Theta$ known,
	\[
	n(h^qg)^\frac{1}{2}\lrp{R_n-\frac{\lambda_q(L^2)\lambda_q(L)^{-2}R(K)}{nh^qg}}\stackrel{d}{\longrightarrow} \mathcal{N}\lrp{0,2\sigma_{\btheta_0}^2},
	\]
	where $\sigma_{\btheta_0}^2$ follows from replacing $f=f_{\btheta_0}$ in $\sigma^2$ from Theorem \ref{gofdens:theo:clt}.
\end{theo}

%-------------------------------------------------%
\subsection{Composite null hypothesis}
\label{gofdens:subsec:gof:comp}
%-------------------------------------------------%

Consider the testing problem $H_0:f\in \mathcal{F}_{\boldsymbol\Theta}=\lrb{f_{\btheta}:\btheta\in\boldsymbol\Theta}$, where $\mathcal{F}_{\boldsymbol\Theta}$ is a class of parametric densities indexed by the $p$-dimensional parameter $\btheta$, vs. $H_1:f\notin\mathcal{F}_{\boldsymbol\Theta}$. Under $H_0$, a parametric density estimator $f_{\hat\btheta}$ can be obtained by Maximum Likelihood (ML). The next conditions are required.
\begin{enumerate}[label=\textbf{A\arabic{*}}., ref=\textbf{A\arabic{*}}]
	\setcounter{enumi}{4}
	
	\item The function $f_{\btheta}$ is twice continuously differentiable with respect to $\btheta$, with derivatives that are bounded and uniformly continuous for $(\bx,z)$.\label{gofdens:assump:a5}
	
	\item There exists $\btheta_1\in\Theta$ such that $\hat \btheta-\btheta_1=\mathcal{O}_{\mathbb{P}}\big(n^{-\frac{1}{2}}\big)$ and if $H_0:f=f_{\btheta_0}$ holds for a $\btheta_0\in\Theta$, then $\btheta_1=\btheta_0$.\label{gofdens:assump:a6}
\end{enumerate}

\ref{gofdens:assump:a5} is a regularity assumption on the parametric density, whereas \ref{gofdens:assump:a6} states that the estimation of the unknown parameter must be $\sqrt{n}$-consistent in order to ensure that the effects of parametric estimation can be neglected. The $\sqrt{n}$-consistency is required under $H_0$ (for Theorem \ref{gofdens:theo:gof}) and $H_1$ (for Theorem \ref{gofdens:theo:boots}), which is satisfied by the ML estimator. The test statistic is an adaptation of (\ref{gofdens:gof:simple}), but plugging-in the estimator of the unknown parameter $\btheta_0$ under $H_0$ in the test statistic expression:
\begin{align}
	R_n=\Iqr{\lrp{\hat f_{h,g}(\bx,z)-LK_{h,g}f_{\hat \btheta}(\bx,z)}^2}{\bx}{z}.\label{gofdens:gof:dirlin}
\end{align}

\begin{theo}[Goodness-of-fit test for directional-linear densities]
	\label{gofdens:theo:gof}
	Under \ref{gofdens:assump:a1}--\ref{gofdens:assump:a3}, \ref{gofdens:assump:a5}--\ref{gofdens:assump:a6} and the composite null hypothesis $H_0:f=f_{\btheta_0}$, with $\btheta_0\in\Theta$ unknown,
	\[
	n(h^qg)^\frac{1}{2}\lrp{R_n-\frac{\lambda_q(L^2)\lambda_q(L)^{-2}R(K)}{nh^qg}}\stackrel{d}{\longrightarrow} \mathcal{N}\lrp{0,2\sigma_{\btheta_0}^2}.
	\]
\end{theo}
Families of Pitman alternatives are a common way to measure power for tests based on kernel smoothers (e.g. \cite{Fan1994}). For the directional-linear case, these alternatives can be written as
\begin{align}
	H_{1P}: f(\bx,z)=f_{\btheta_0}(\bx,z)+\big(nh^\frac{q}{2}g^\frac{1}{2}\big)^{-\frac{1}{2}}\Delta(\bx,z),\label{gofdens:pit}
\end{align}
where $\Delta(\bx,z):\Om{q}\times\R\rightarrow\R$ is such that $\Iqr{\Delta(\bx,z)}{\bx}{z}=0$. A necessary condition to derive the limit distribution of $R_n$ under $H_{1P}$ is that the estimator $\hat\btheta$ is a $\sqrt{n}$-consistent estimator for $\btheta_0$.
\begin{enumerate}[label=\textbf{A\arabic{*}}., ref=\textbf{A\arabic{*}}]
	\setcounter{enumi}{6}
	\item For the family of alternatives (\ref{gofdens:pit}), $\hat \btheta-\btheta_0=\mathcal{O}_{\mathbb{P}}\big(n^{-\frac{1}{2}}\big)$. \label{gofdens:assump:a7}
\end{enumerate}
\begin{theo}[Local power under Pitman alternatives]
	\label{gofdens:theo:gofpit}
	Under \ref{gofdens:assump:a1}--\ref{gofdens:assump:a3}, \ref{gofdens:assump:a5}--\ref{gofdens:assump:a7} and the alternative hypothesis (\ref{gofdens:pit}),
	\[
	n(h^qg)^\frac{1}{2}\lrp{R_n-\frac{\lambda_q(L^2)\lambda_q(L)^{-2}R(K)}{nh^qg}}\stackrel{d}{\longrightarrow} \mathcal{N}\lrp{R\lrp{\Delta},2\sigma_{\btheta_0}^2}.
	\]
\end{theo}

%-------------------------------------------------%
\subsection{Calibration in practise}
\label{gofdens:subsec:gof:calib}
%-------------------------------------------------%

In order to effectively calibrate the proposed test, a parametric bootstrap procedure is investigated. The bootstrap statistic is defined as
\[
R_n^*=\Iqr{\lrp{\hat f_{h,g}^*(\bx,z)-LK_{h,g}f_{\hat\btheta^*}(\bx,z) }^2}{\bx}{z},
\]
where the superscript $^*$ indicates that the estimators are computed from the bootstrap sample $\lrb{\lrp{\bX_i^*,Z_i^*}}_{i=1}^n$ obtained from the density $f_{\hat\btheta}$, with $\hat\btheta$ computed from the original sample. The bootstrap procedure, considering the composite null hypothesis testing problem, is detailed in an algorithm. Calibration for the simple null hypothesis test can be done replacing $\hat\btheta$ and $\hat \btheta^*$ by $\btheta_0$.

\begin{algo}[Testing procedure]
	\label{gofdens:testproc}
	Let $\lrb{\lrp{\bX_i,Z_i}}_{i=1}^n$ be a random sample from $f$. To test $H_0:f=f_{\btheta_0}$, with $\btheta_0\in\Theta$ unknown, proceed as follows.
	\begin{enumerate}[label=\textit{\roman{*}}., ref=\textit{\roman{*}}]
		\item Obtain $\hat \btheta$, a $\sqrt{n}$-consistent estimator of $\btheta_0$.\label{gofdens:testproc:i}
		\item Compute $R_n=\Iqr{\big(\hat f_{h,g} (\bx,z)-LK_{h,g}f_{\hat \btheta}(\bx,z)\big)^2}{\bx}{z}$.\label{gofdens:testproc:ii}
		\item \textit{Bootstrap strategy}. For $b=1,\ldots,B$:\label{gofdens:testproc:iii}
		\begin{enumerate}[label=\textit{(\alph{*})}, ref=\textit{(\alph{*})}]
			\item Obtain a random sample $\lrb{\lrp{\bX_i^*,Z_i^*}}_{i=1}^n$ from $f_{\hat \btheta}$.\label{gofdens:testproc:a}
			\item Compute $\hat \btheta^*$ as in step \ref{gofdens:testproc:i}, from the bootstrap sample in \ref{gofdens:testproc:a}.
			\item Compute $R_n^{*b}=\Iqr{\big(\hat f^*_{h,g} (\bx,z)-LK_{h,g}f_{\hat \btheta^*}(\bx,z)\big)^2}{\bx}{z}$, where $\hat f^*_{h,g}$ is obtained from the bootstrap sample in \ref{gofdens:testproc:a}.
		\end{enumerate}
		\item Approximate the $p$-value of the test as $p\text{-value}\approx\#\big\{R_n\leq R_n^{*b}\big\}/B$.
	\end{enumerate}
\end{algo}
The consistency of this testing procedure is proved here, using the bootstrap analogue of \ref{gofdens:assump:a6}.
\begin{enumerate}[label=\textbf{A\arabic{*}}., ref=\textbf{A\arabic{*}}]
	\setcounter{enumi}{7}
	\item $\hat \btheta^*-\hat\btheta=\mathcal{O}_{\mathbb{P}^*}\big(n^{-\frac{1}{2}}\big)$, where $\mathbb{P}^*$ represents the probability of $(\bX^*,Z^*)$ conditioned on the sample $\lrb{\lrp{\bX_i,Z_i}}_{i=1}^n$. \label{gofdens:assump:a8}
\end{enumerate}
\begin{theo}[Bootstrap consistency]
	\label{gofdens:theo:boots}
	Under \ref{gofdens:assump:a1}--\ref{gofdens:assump:a3}, \ref{gofdens:assump:a5}--\ref{gofdens:assump:a6} and \ref{gofdens:assump:a8}, and conditionally on the sample $\lrb{\lrp{\bX_i,Z_i}}_{i=1}^n$,
	\[
	n(h^qg)^\frac{1}{2}\lrp{R_n^*-\frac{\lambda_q(L^2)\lambda_q(L)^{-2}R(K)}{nh^qg}}\stackrel{d}{\longrightarrow} \mathcal{N}\lrp{0,2\sigma_{\btheta_1}^2} \text{in probability}.
	\]
\end{theo}
Then, the probability distribution function (pdf) of $R_n^*$ conditioned on the sample converges 
in probability to a Gaussian pdf, regardless of whether $H_0$ holds or not. The asymptotic distribution coincides with the one of $R_n$ if $H_0$ holds ($\btheta_1=\btheta_0$).

%-------------------------------------------------%
\subsection{Extensions to directional-directional models}
%-------------------------------------------------%

The directional-di\-rec\-tional versions of the previous results follow under analogous assumptions (modifying \ref{gofdens:assump:a5}, (\ref{gofdens:smoothing}) and (\ref{gofdens:pit}) accordingly). The directional-directional test statistic for the composite hypothesis testing problem is
\[
R_n=\Iqq{\lrp{\hat f_{h_1,h_2}(\bx,\by)-L_1L_{2,h_1,h_2}f_{\hat \btheta}(\bx,\by)}^2}{\bx}{\by}.
\]
\begin{coro}[Goodness-of-fit test for directional-directional densities]
	\label{gofdens:coro:gof:dirdir}
	Under the directional-direc\-tion\-al analogues of \ref{gofdens:assump:a1}--\ref{gofdens:assump:a3}, \ref{gofdens:assump:a5}--\ref{gofdens:assump:a6} and the composite null hypothesis $H_0: f=f_{\btheta_0}$, with $\btheta_0\in\Theta$ unknown,
	\[
	n(h_1^{q_1}h_2^{q_2})^\frac{1}{2}\lrp{R_n-\frac{\lambda_{q_1}(L_1^2)\lambda_{q_1}(L_1)^{-2}\lambda_{q_2}(L_2^2)\lambda_{q_2}(L_2)^{-2}}{nh_1^{q_1}h_2^{q_2}}}\stackrel{d}{\longrightarrow} \mathcal{N}\lrp{0,2\sigma_{\btheta_0}^2}.
	\]
\end{coro}

%-------------------------------------------------%
\section{Simulation study}
\label{gofdens:sec:sim}
%-------------------------------------------------%

The finite sample performance of the directional-linear and directional-directional goodness-of-fit tests is illustrated in this section for a variety of models, sample sizes, and bandwidth choices. The study considers circular-linear and circular-circular scenarios, although these tests can be easily applied in higher dimensions, such as spherical-linear or spherical-circular, due to their general definition and resampling procedures. Details on simulated models and further results are included as supplementary material.\\

Circular-Linear (CL) and Circular-Circular (CC) parametric scenarios are considered. Figures \ref{gofdens:fig:cl} and \ref{gofdens:fig:cc} show the density contours in the cylinder (CL) and in the torus (CC) for the different models. The detailed description of each model is given in the supplementary material. Deviations from the composite null hypothesis $H_0:f\in\mathcal{F}_{\Theta}$ are obtained by mixing the true density $f_{\btheta_0}$ with a density $\Delta$ such that the resulting density does not belong to $\mathcal{F}_{\Theta}$:
$H_{\delta}: f=(1-\delta)f_{\btheta_0}+\delta\Delta$, $0\leq\delta\leq1$. The goodness-of-fit tests are applied using the bootstrap strategy, for the whole collection of models, sample sizes $n=100,500,1000$ and deviations $\delta=0,0.10,0.15$ ($\delta=0$ for the null hypothesis). The number of bootstrap and Monte Carlo replicates is $1000$.\\

In each case (model, sample size and deviation), the performance of the goodness-of-fit test is shown for a fixed pair of bandwidths, obtained from the median of $1000$ simulated Likelihood Cross Validation (LCV) bandwidths: 
\begin{align}
\begin{array}{rl}
(h,g)_{\mathrm{LCV}}\!\!\!\!&=\arg\max_{h,g>0} \sum_{i=1}^n\log \hat f_{h,g}^{-i}(\bX_i,Z_i),\\
(h_1,h_2)_{\mathrm{LCV}}\!\!\!\!&=\arg\max_{h_1,h_2>0} \sum_{i=1}^n\log \hat f_{h_1,h_2}^{-i}(\bX_i,\bY_i),
\end{array} \label{gofdens:band:lcv}
\end{align}
where $\hat f^{-i}_{\ldots}$ denotes the kernel estimator computed without the $i$-th datum. A deeper insight on the bandwidth effect is provided for some scenarios, where percentage of rejections are plotted for a grid 

\begin{figure}[H]
	\centering
	\includegraphics[height=0.145\textheight]{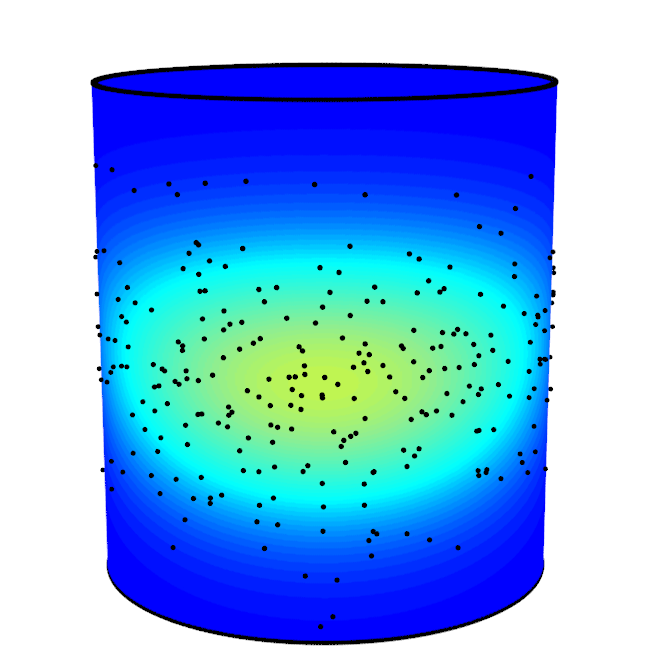}
	\includegraphics[height=0.145\textheight]{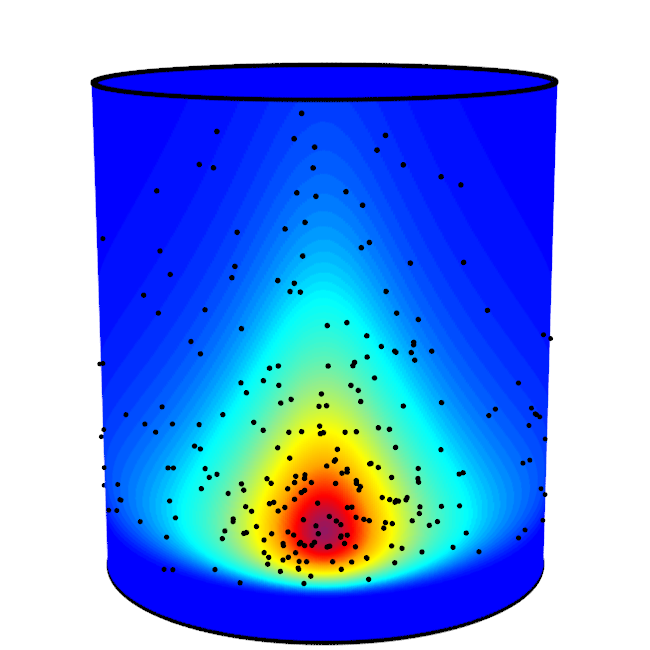}
	\includegraphics[height=0.145\textheight]{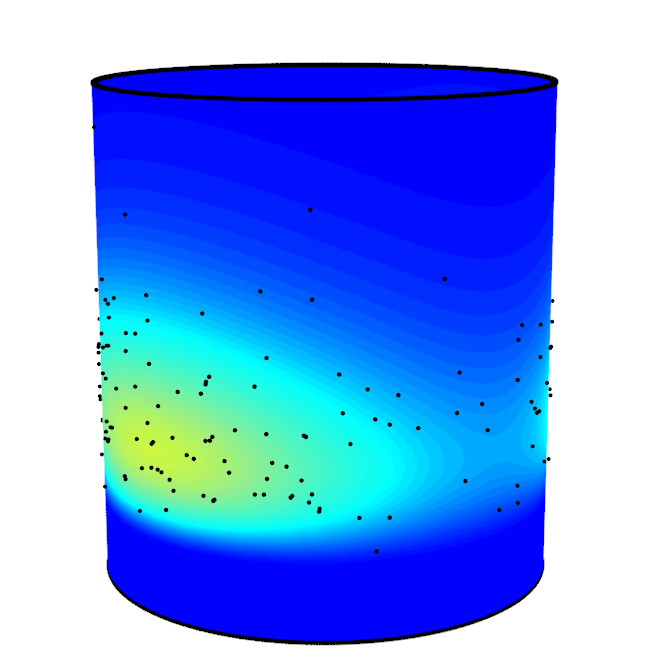}
	\includegraphics[height=0.145\textheight]{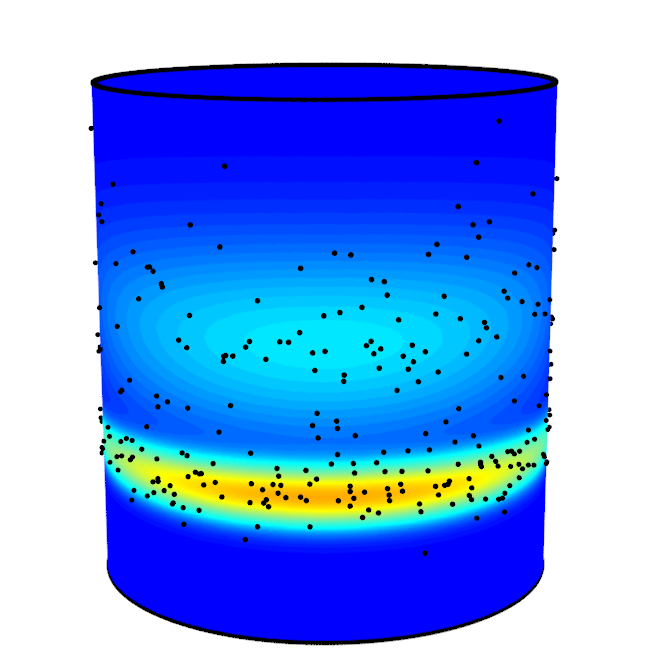}
	\includegraphics[height=0.145\textheight]{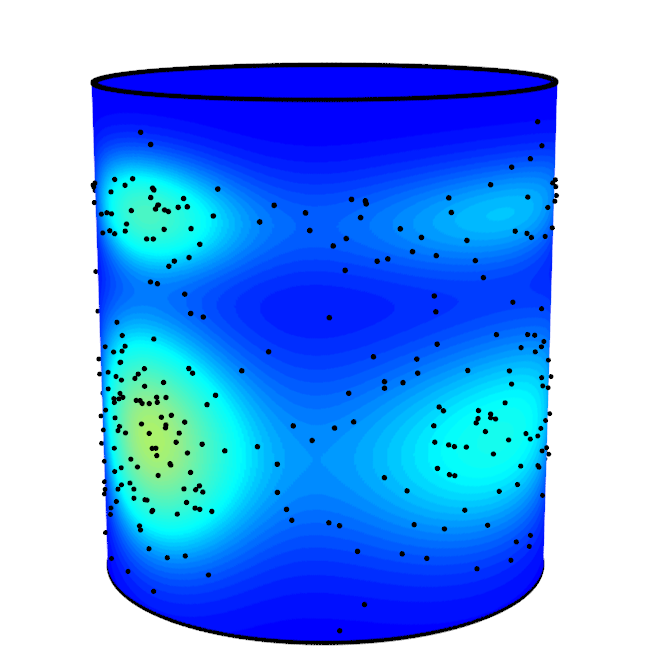}
	\includegraphics[height=0.145\textheight]{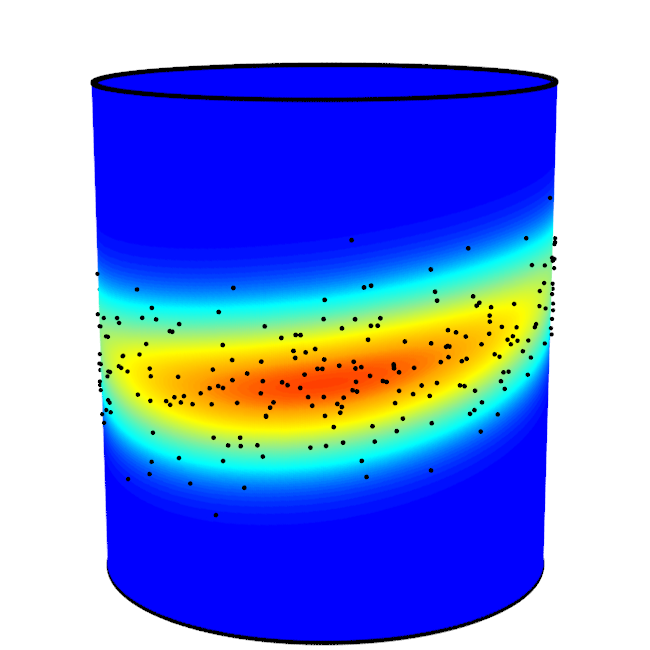}
	\includegraphics[height=0.145\textheight]{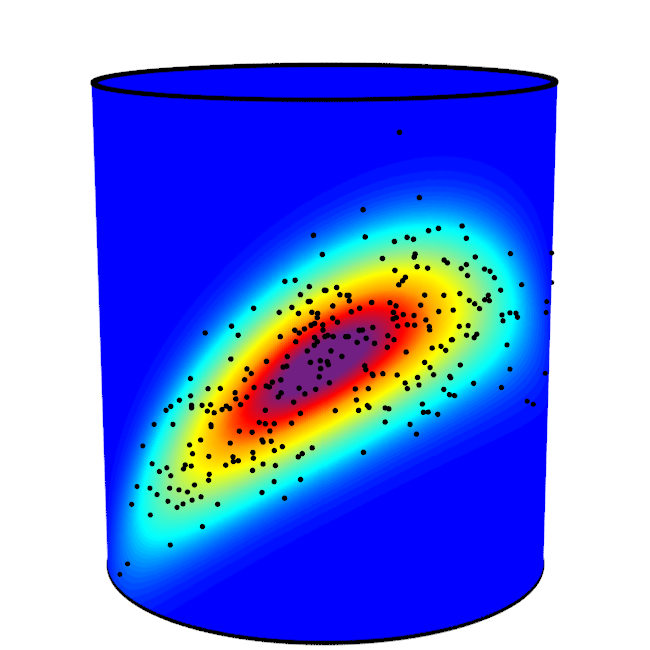}
	\includegraphics[height=0.145\textheight]{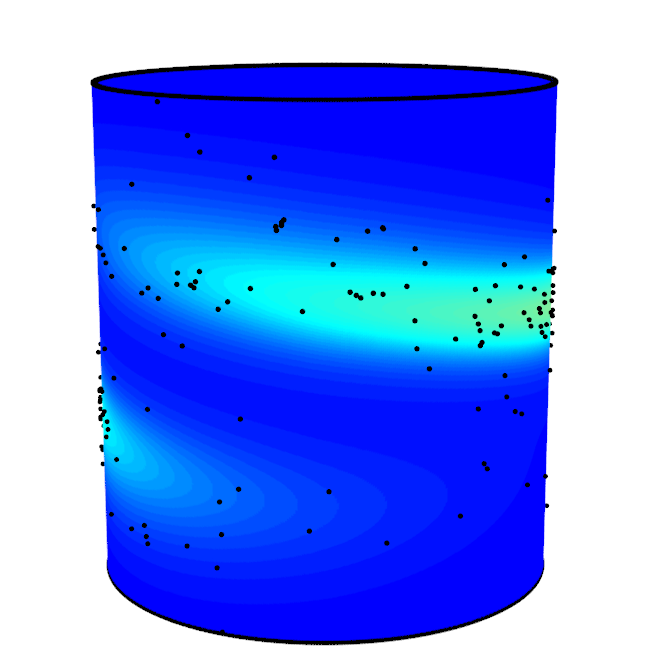}\\
	\includegraphics[height=0.145\textheight]{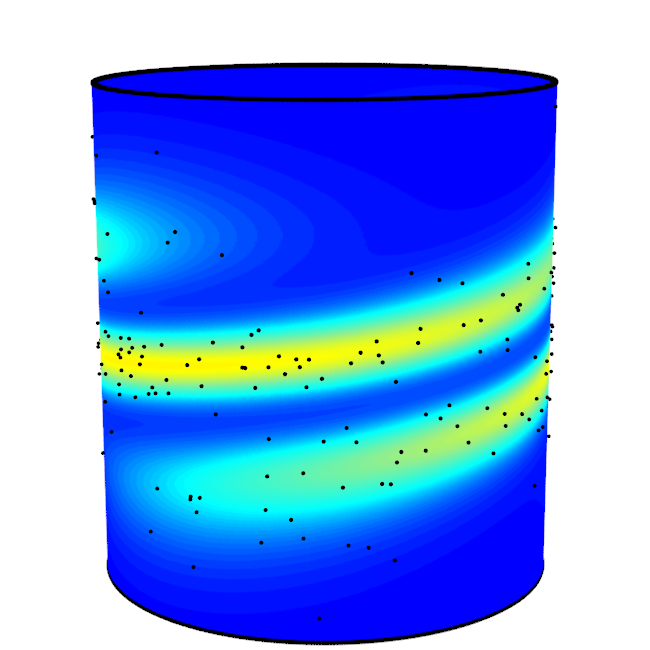}
	\includegraphics[height=0.145\textheight]{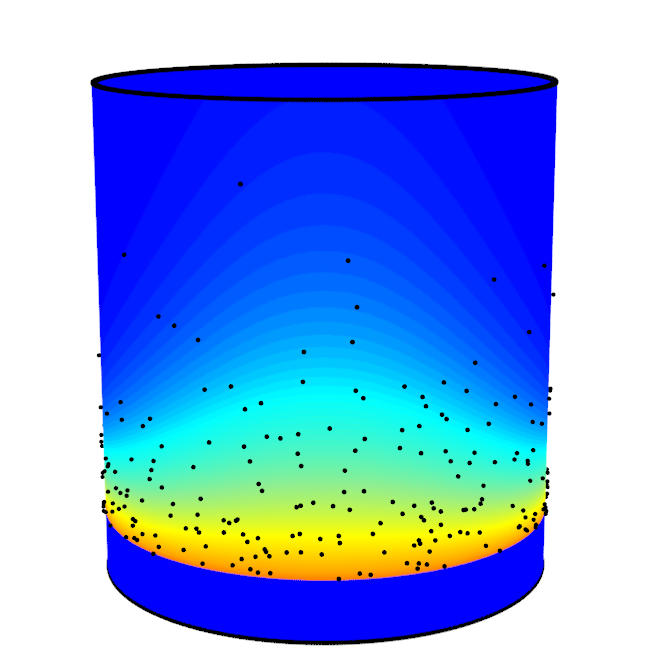}
	\includegraphics[height=0.145\textheight]{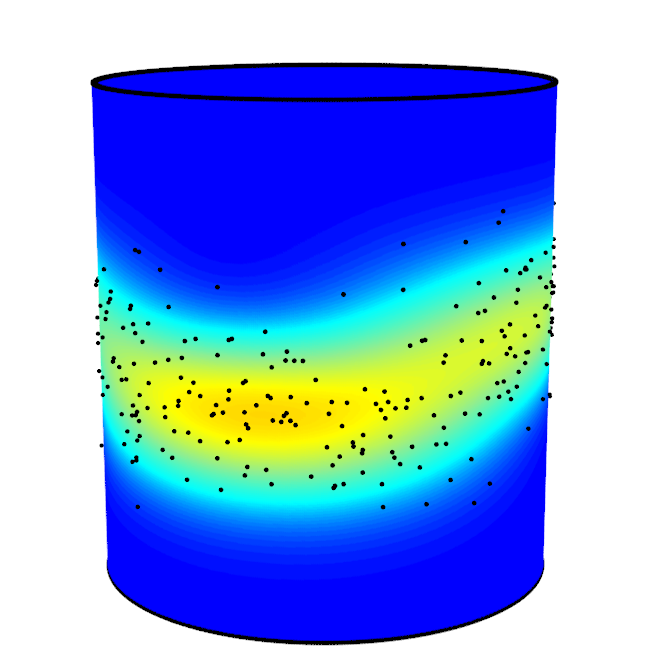}
	\includegraphics[height=0.145\textheight]{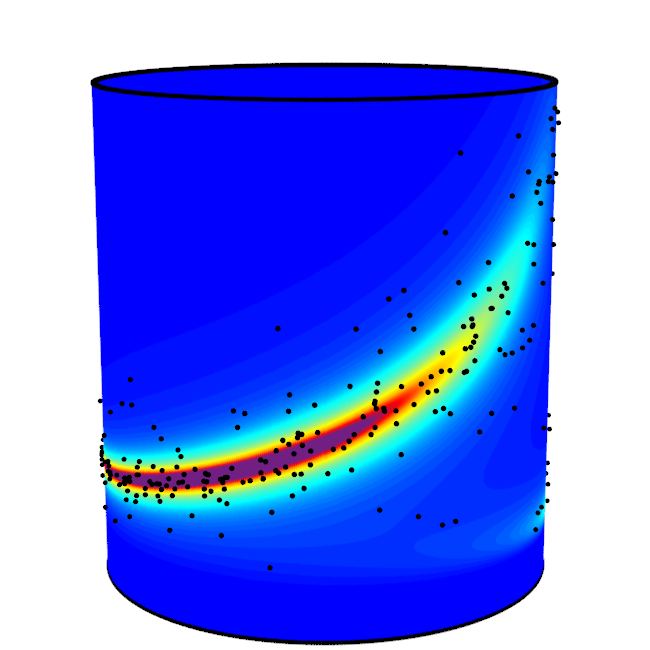}
	\caption{\small Density models for the simulation study in the circular-linear case. From left to right and up to down, models CL1 to CL12. \label{gofdens:fig:cl}}
	\includegraphics[height=0.145\textheight]{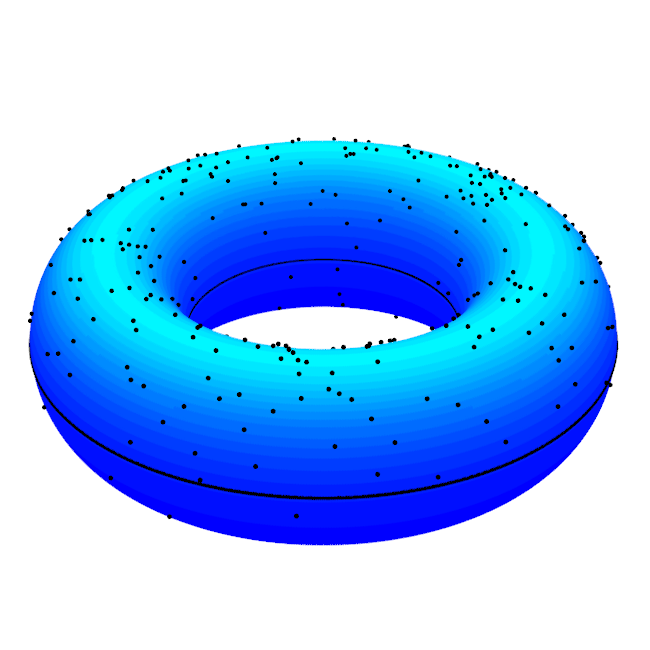}
	\includegraphics[height=0.145\textheight]{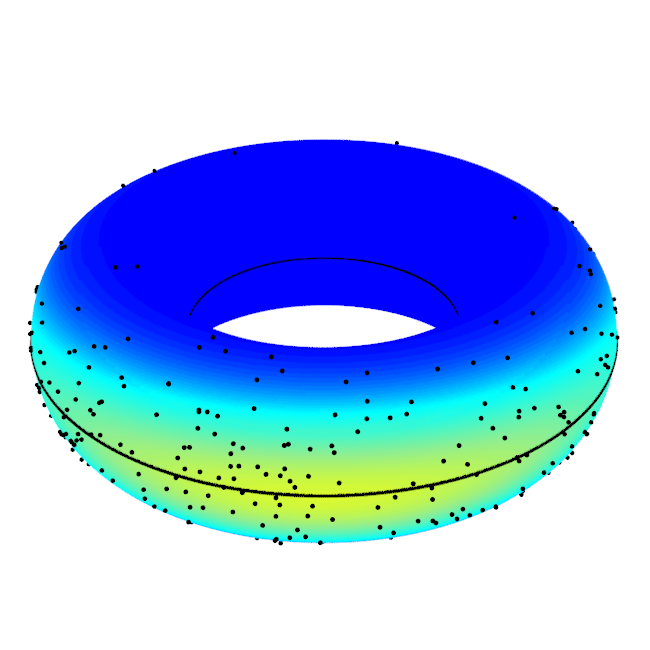}
	\includegraphics[height=0.145\textheight]{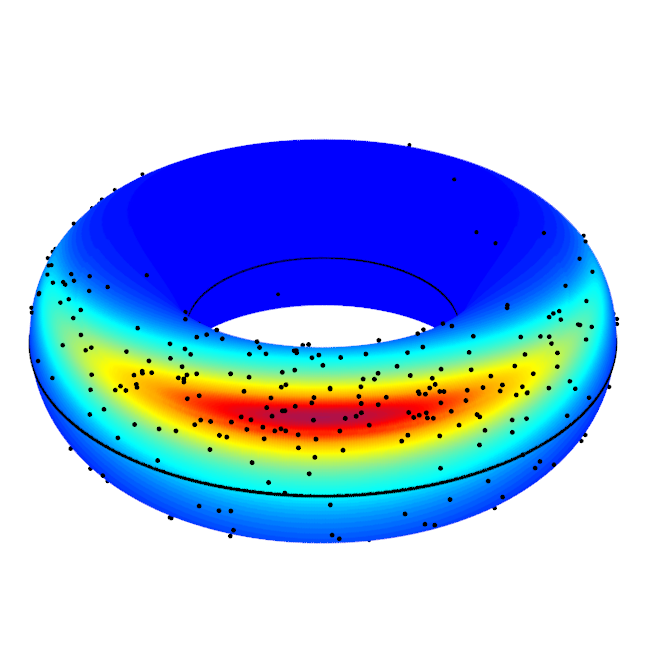}
	\includegraphics[height=0.145\textheight]{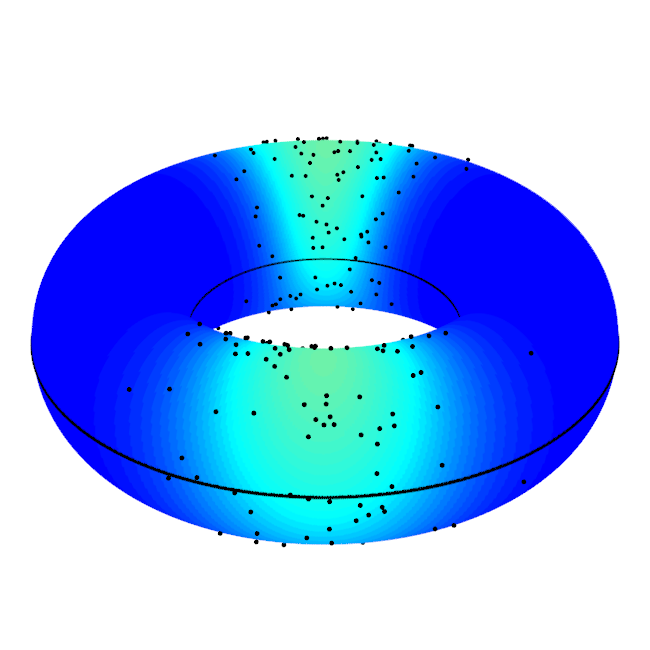}\\
	\includegraphics[height=0.145\textheight]{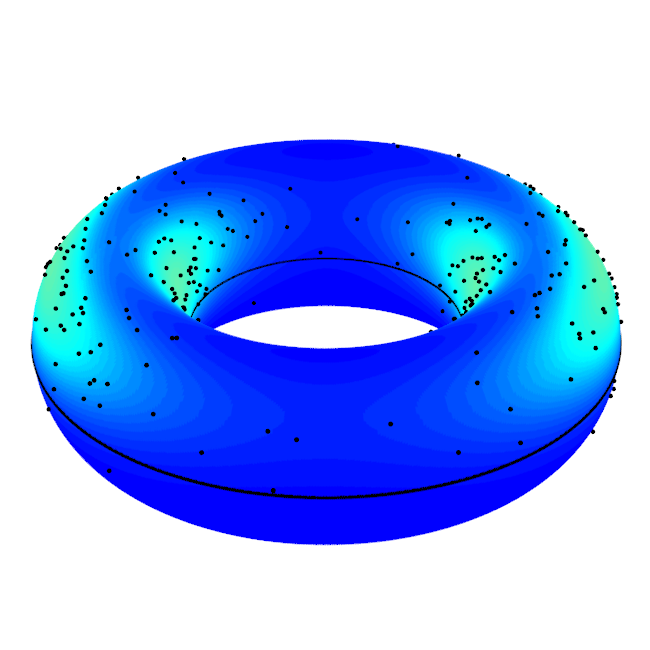}
	\includegraphics[height=0.145\textheight]{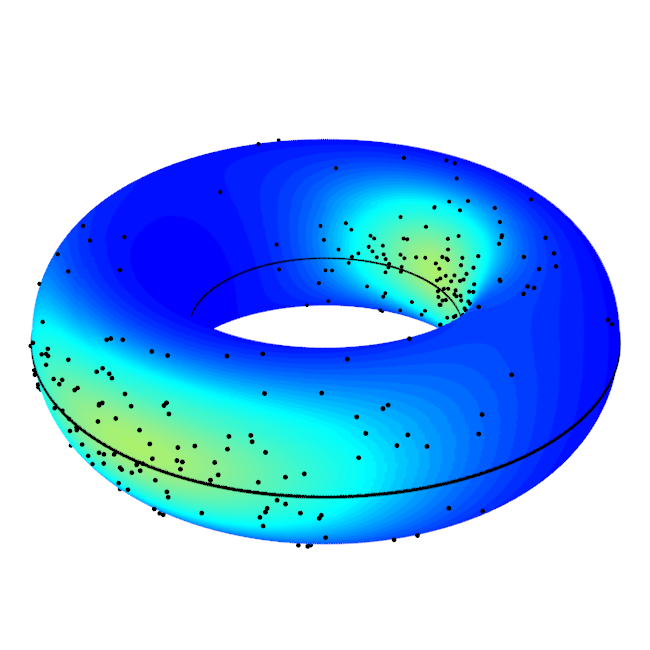}
	\includegraphics[height=0.145\textheight]{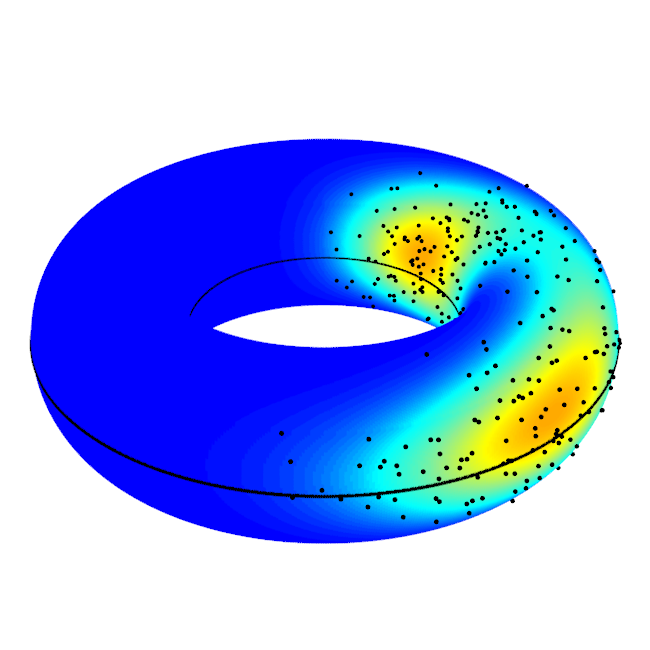}
	\includegraphics[height=0.145\textheight]{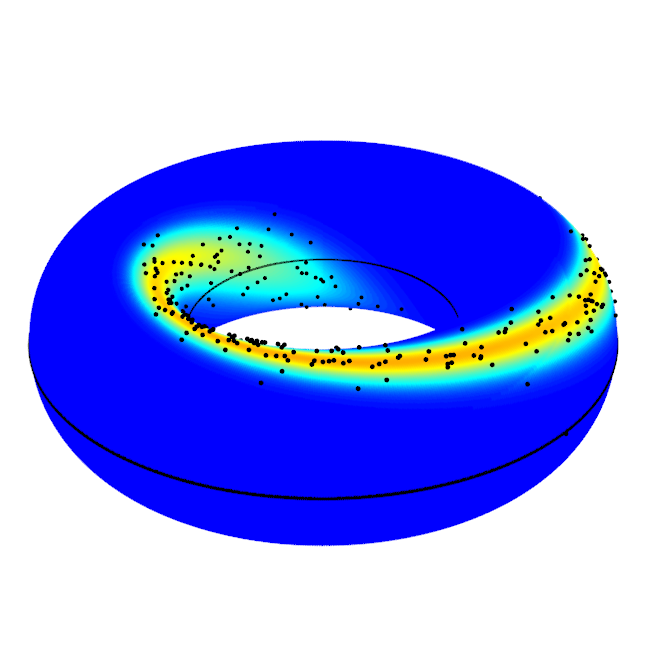}\\
	\includegraphics[height=0.145\textheight]{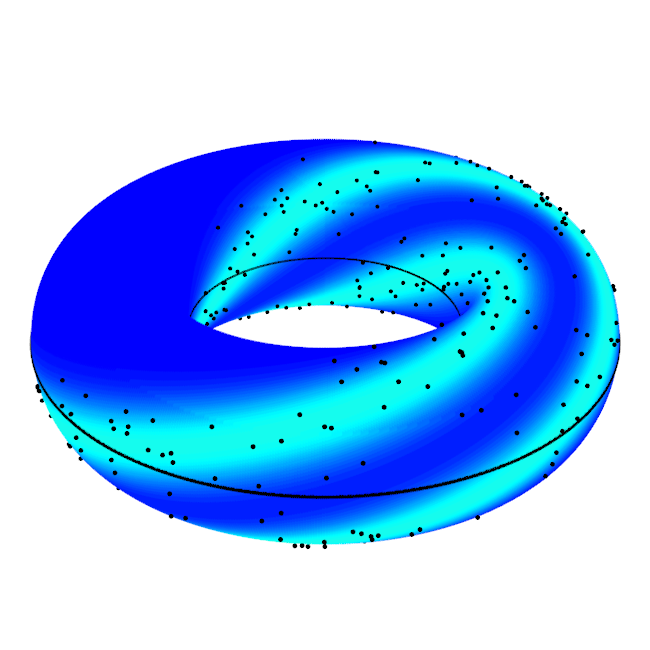}
	\includegraphics[height=0.145\textheight]{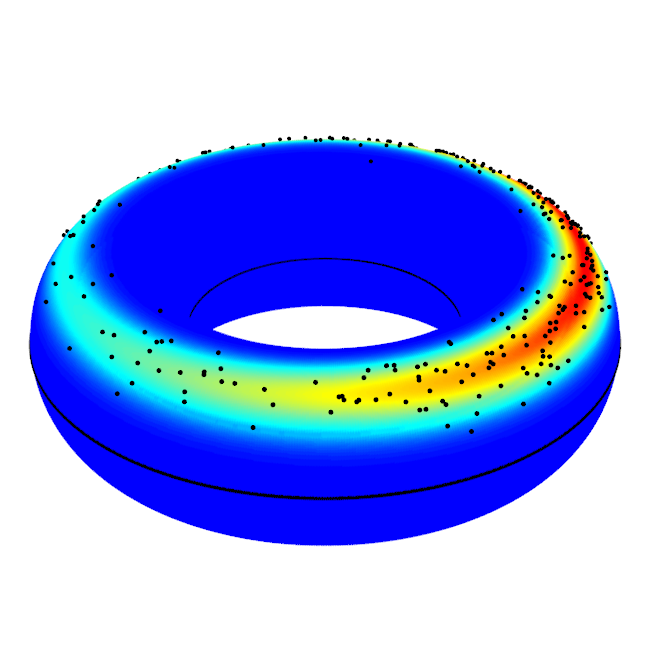}
	\includegraphics[height=0.145\textheight]{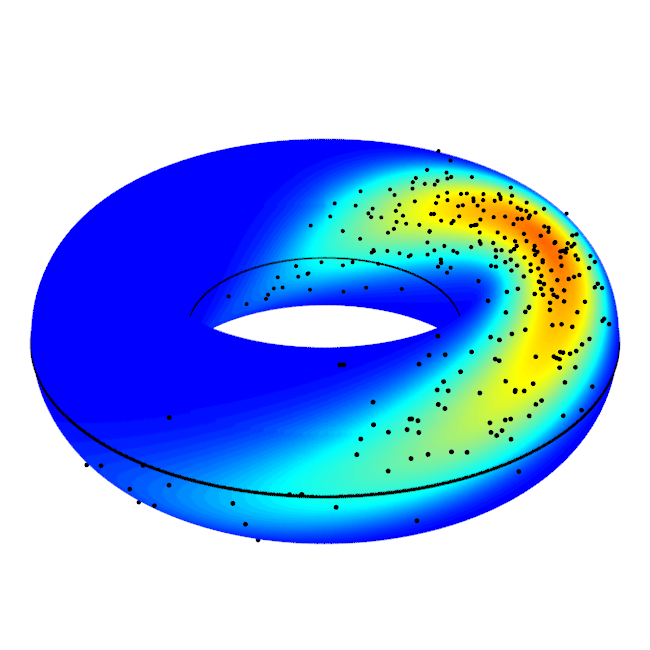}
	\includegraphics[height=0.145\textheight]{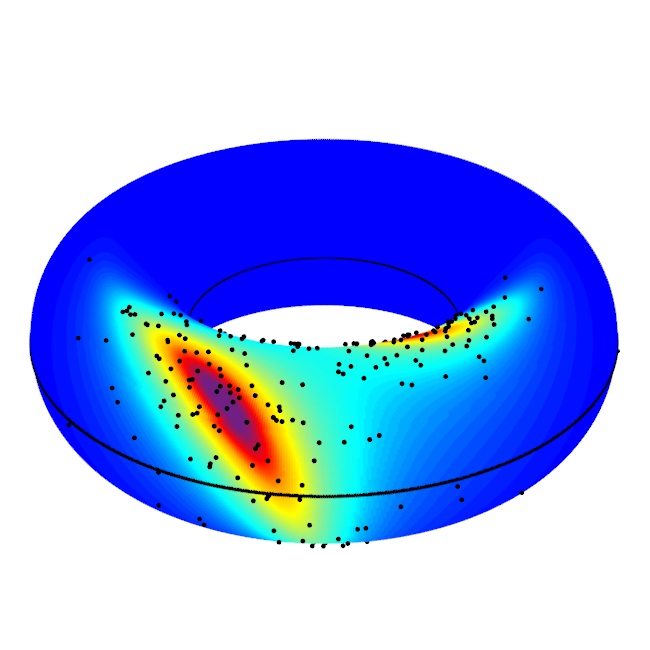}
	\caption{\small Density models for the simulation study in the circular-circular case. From left to right and up to down, models CC1 to CC12.\label{gofdens:fig:cc}}
\end{figure}

\begin{table}[t]
	\centering
	\footnotesize
	\begin{tabular}{l|ccc|ccc|ccc}
		\toprule\toprule
		\multirow{4}{*}{Model} & \multicolumn{9}{c}{Sample size $n$ and deviation $\delta$} \\\cmidrule(lr){2-10}
		& \multicolumn{3}{c|}{$n=100$} & \multicolumn{3}{c|}{$n=500$} & \multicolumn{3}{c}{$n=1000$} \\\cmidrule(lr){2-4} \cmidrule(lr){5-7} \cmidrule(lr){8-10}
		& \phantom{0}$\delta$=0\phantom{0} & $\delta$=0.10 & $\delta$=0.15 & \phantom{0}$\delta$=0\phantom{0} & $\delta$=0.10 & $\delta$=0.15 & \phantom{0}$\delta$=0\phantom{0} & $\delta$=0.10 & $\delta$=0.15 \\
		\midrule
		CL1  & $0.051$ & $0.552$ & $0.997$ & $0.052$ & $0.822$ & $1.000$ & $0.048$ & $1.000$ & $1.000$ \\
		CL2 & $0.051$ & $0.244$ & $0.805$ & $0.049$ & $0.525$ & $0.997$ & $0.050$ & $0.982$ & $1.000$ \\
		CL3 & $0.048$ & $0.107$ & $0.362$ & $0.046$ & $0.163$ & $0.682$ & $0.050$ & $0.659$ & $0.940$ \\
		CL4 & $0.045$ & $0.172$ & $0.568$ & $0.039$ & $0.297$ & $0.869$ & $0.045$ & $0.868$ & $0.993$ \\
		CL5 & $0.049$ & $0.272$ & $0.972$ & $0.049$ & $0.514$ & $0.999$ & $0.041$ & $1.000$ & $1.000$ \\
		CL6 & $0.039$ & $0.996$ & $1.000$ & $0.043$ & $1.000$ & $1.000$ & $0.050$ & $1.000$ & $1.000$ \\
		CL7 & $0.042$ & $1.000$ & $1.000$ & $0.043$ & $1.000$ & $1.000$ & $0.049$ & $1.000$ & $1.000$ \\
		CL8 & $0.049$ & $0.204$ & $0.893$ & $0.050$ & $0.379$ & $0.997$ & $0.044$ & $1.000$ & $1.000$ \\
		CL9 & $0.062$ & $0.914$ & $1.000$ & $0.043$ & $0.989$ & $1.000$ & $0.064$ & $1.000$ & $1.000$ \\
		CL10 & $0.045$ & $0.218$ & $0.723$ & $0.056$ & $0.378$ & $0.975$ & $0.045$ & $0.944$ & $1.000$ \\
		CL11 & $0.059$ & $0.510$ & $0.993$ & $0.056$ & $0.763$ & $1.000$ & $0.056$ & $1.000$ & $1.000$ \\
		CL12 & $0.073$ & $0.152$ & $0.655$ & $0.054$ & $0.254$ & $0.967$ & $0.051$ & $0.969$ & $1.000$ \\\midrule
		CC1  & $0.061$ & $0.456$ & $0.751$ & $0.047$ & $0.995$ & $1.000$ & $0.048$ & $1.000$ & $1.000$\\       
		CC2  & $0.054$ & $0.506$ & $0.798$ & $0.043$ & $0.994$ & $1.000$ & $0.056$ & $1.000$ & $1.000$\\       
		CC3  & $0.061$ & $0.706$ & $0.932$ & $0.042$ & $1.000$ & $1.000$ & $0.058$ & $1.000$ & $1.000$\\       
		CC4  & $0.049$ & $0.837$ & $0.958$ & $0.048$ & $1.000$ & $1.000$ & $0.052$ & $1.000$ & $1.000$\\       
		CC5  & $0.059$ & $0.431$ & $0.720$ & $0.050$ & $1.000$ & $1.000$ & $0.051$ & $1.000$ & $1.000$\\       
		CC6  & $0.069$ & $0.123$ & $0.270$ & $0.045$ & $0.759$ & $0.960$ & $0.034$ & $0.958$ & $0.993$\\        
		CC7  & $0.048$ & $0.112$ & $0.201$ & $0.059$ & $0.724$ & $0.976$ & $0.044$ & $0.989$ & $1.000$\\        
		CC8  & $0.043$ & $0.693$ & $0.945$ & $0.054$ & $1.000$ & $1.000$ & $0.050$ & $1.000$ & $1.000$\\        
		CC9  & $0.043$ & $0.325$ & $0.600$ & $0.057$ & $1.000$ & $1.000$ & $0.042$ & $1.000$ & $1.000$\\        
		CC10 & $0.047$ & $1.000$ & $1.000$ & $0.041$ & $1.000$ & $1.000$ & $0.042$ & $1.000$ & $1.000$\\        
		CC11 & $0.041$ & $0.973$ & $1.000$ & $0.047$ & $1.000$ & $1.000$ & $0.053$ & $1.000$ & $1.000$\\        
		CC12 & $0.062$ & $0.899$ & $0.993$ & $0.058$ & $1.000$ & $1.000$ & $0.048$ & $1.000$ & $1.000$\\\bottomrule\bottomrule
	\end{tabular}
	\caption{\small Empirical size and power of the circular-linear and circular-circular goodness-of-fit tests for models CL1--CL12 and CC1--CC12 (respectively) with significance level $\alpha=0.05$ and different sample sizes and deviations. \label{gofdens:tab:results:short}}
\end{table}
\begin{figure}[H]
	\centering

	\includegraphics[trim=3cm 0cm 7cm 7cm,clip=true,scale=0.25]{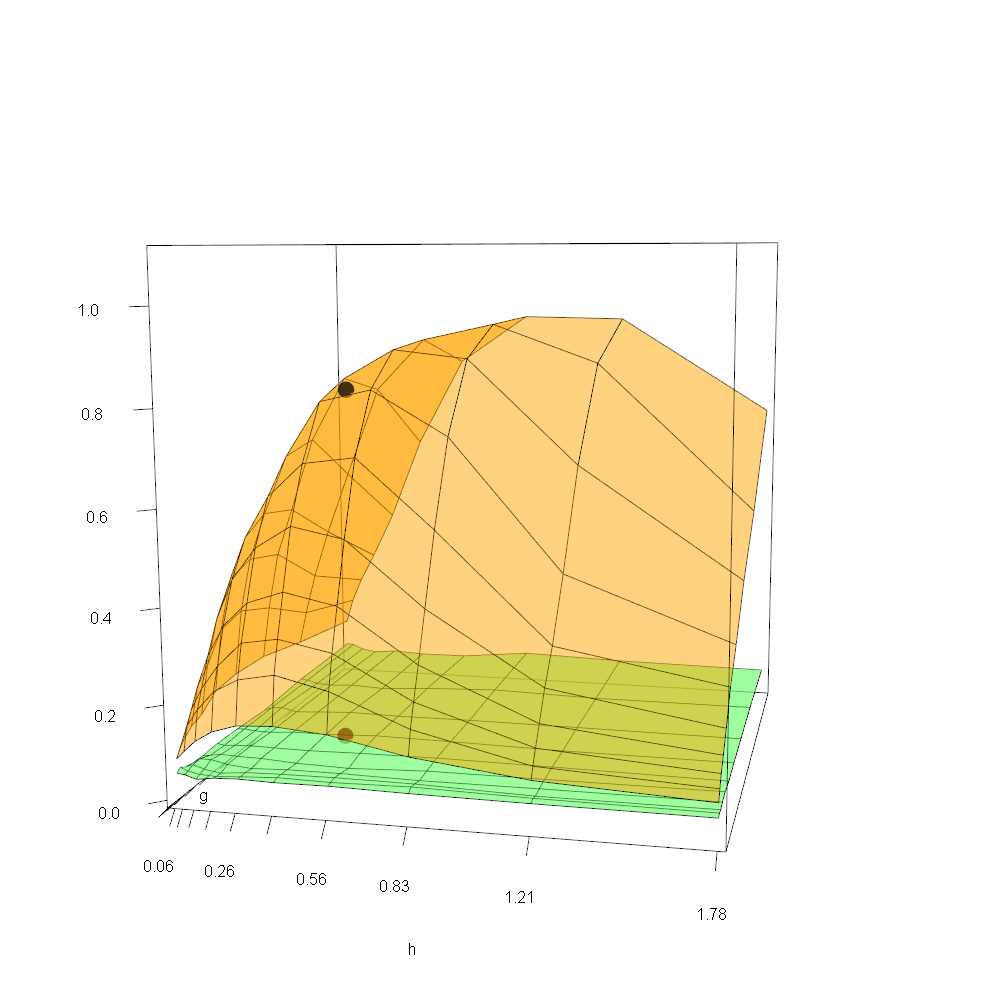}\hspace{1cm}\includegraphics[trim=3cm 0cm 7cm 7cm,clip=true,scale=0.25]{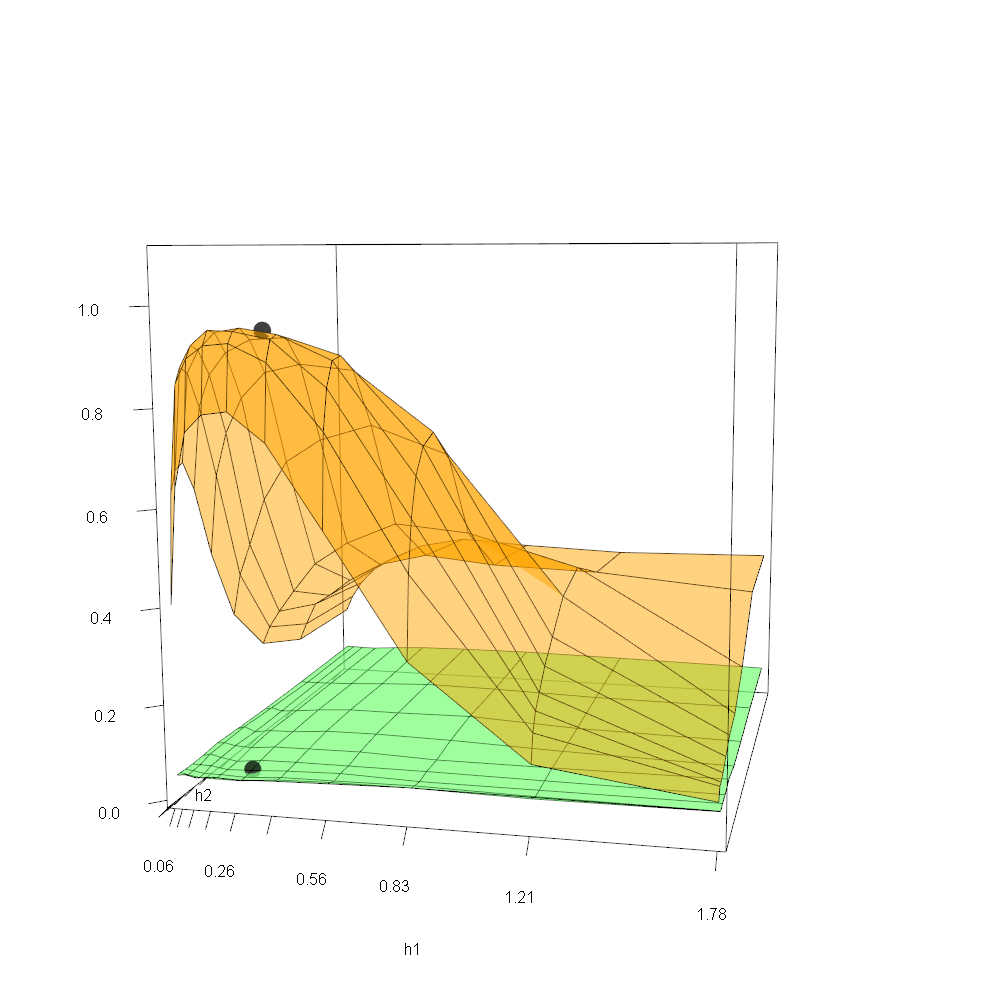}
	\caption{\small Empirical size and power of the circular-linear (left, model CL1) and circular-circular (right, model CC8) goodness-of-fit tests for a $10\times10$ logarithmic spaced grid. Lower surface represents the empirical rejection rate under $H_{0.00}$ and upper surface under $H_{0.15}$. Green colour indicates that the percentage of rejections is in the $95\%$ confidence interval of $\alpha=0.05$, blue that is smaller and orange that is larger. Black points represent the empirical size and power obtained with the median of the LCV bandwidths. \label{gofdens:fig:band}}
\end{figure}

\newpage				
of bandwidths (see Figure \ref{gofdens:fig:band} for two cases, and supplementary material for extended results). The kernels considered are the von Mises and the normal ones.\\

Table \ref{gofdens:tab:results:short} collects the results of the simulation study for each combination of model (CL or CC), deviation ($\delta$) and sample size ($n$). When the null hypothesis holds, significance levels are correctly attained for $\alpha=0.05$ (see supplementary material for $\alpha=0.10,0.01$), for all sample sizes, models and deviations. When the null hypothesis does not hold, the tests perform satisfactorily, having in both cases a quick detection of the alternative when only a $10\%$ and a $15\%$ of the data come from a density out of the parametric family. As expected, the rejection rates grow as the sample size and the deviation from the alternative do.\\

Finally, the effect of the bandwidths is explored in Figure \ref{gofdens:fig:band}. For models CL1 and CC8, the empirical size and power of the tests are computed on a bivariate grid of bandwidths, for sample size $n=100$ and deviations $\delta=0$ (green surface, null hypothesis) and $\delta=0.15$ (orange surface). As it can be seen, the tests are correctly calibrated regardless of the choice of the bandwidths. However, the power is notably affected by the bandwidths, with different behaviours depending on the model and the alternative. Reasonable choices of the bandwidths, such as the median of the LCV bandwidths (\ref{gofdens:band:lcv}), present a competitive power. Further results supporting the same conclusions are available in the supplementary material.

%-------------------------------------------------%
\section{Data application}
\label{gofdens:sec:data}
%-------------------------------------------------%

The proposed goodness-of-fit tests are applied to study two datasets (see supplementary material for further details). The first dataset comes from forestry and contains orientations and log-burnt areas of $26870$ wildfires occurred in Portugal between 1985 and 2005. Data was aggregated in watersheds, giving $102$ observations of the circular mean orientation and mean log-burnt area for each watershed (circular-linear example). Further details on the data acquisition procedure, measurement of fires orientation and watershed delimitation can be seen in \cite{Barros:2012p1135} and \cite{Garcia-Portugues:testindep}. The model proposed by \cite{Mardia1978} was tested for this dataset (Figure \ref{gofdens:fig:data:mardia}, left) using the LCV bandwidths and $B=1000$ bootstrap replicates, resulting a $p$-value of $0.156$, showing no evidence against the null hypothesis.

\begin{figure}[!htb]
	\centering
	\vspace{-0.25cm}
	\includegraphics[scale=0.375]{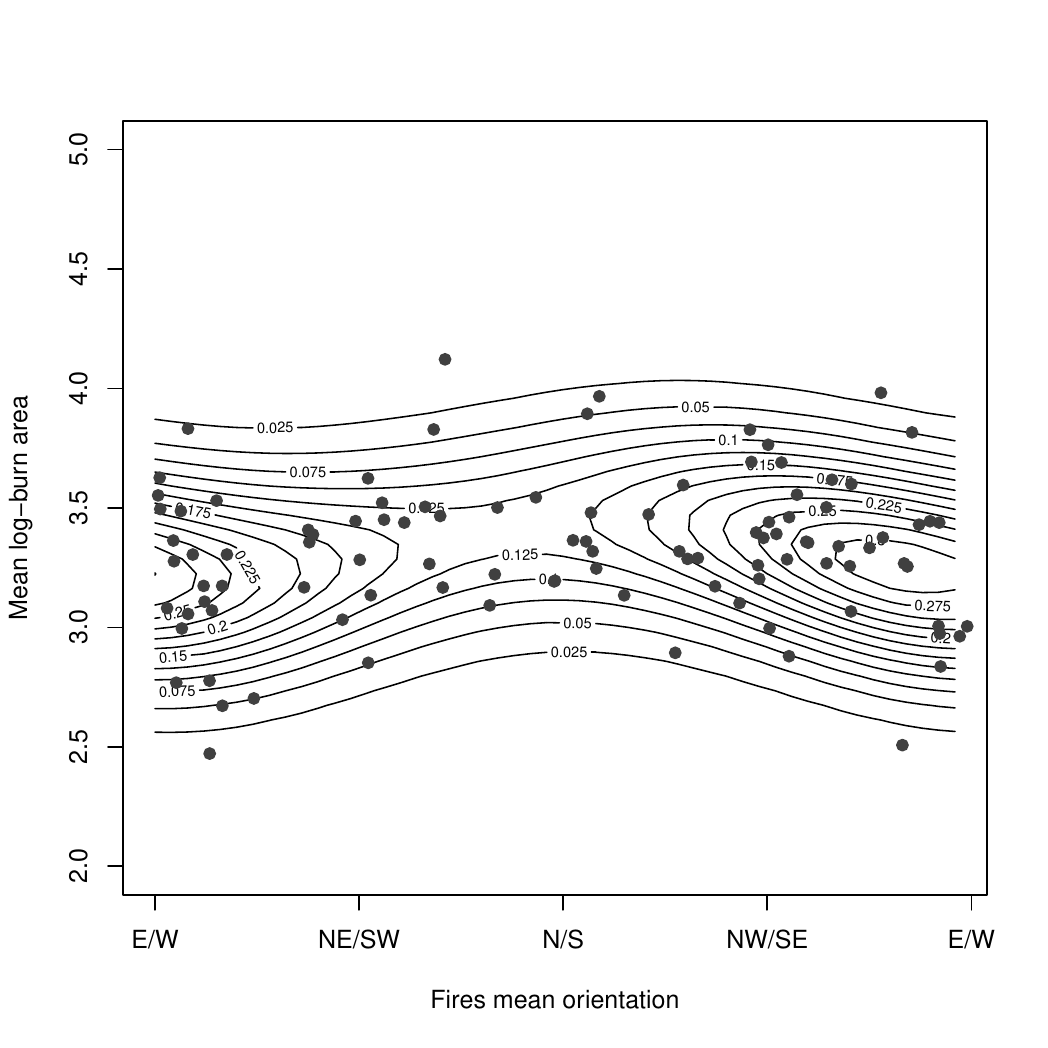}\hspace{1cm}\includegraphics[scale=0.375]{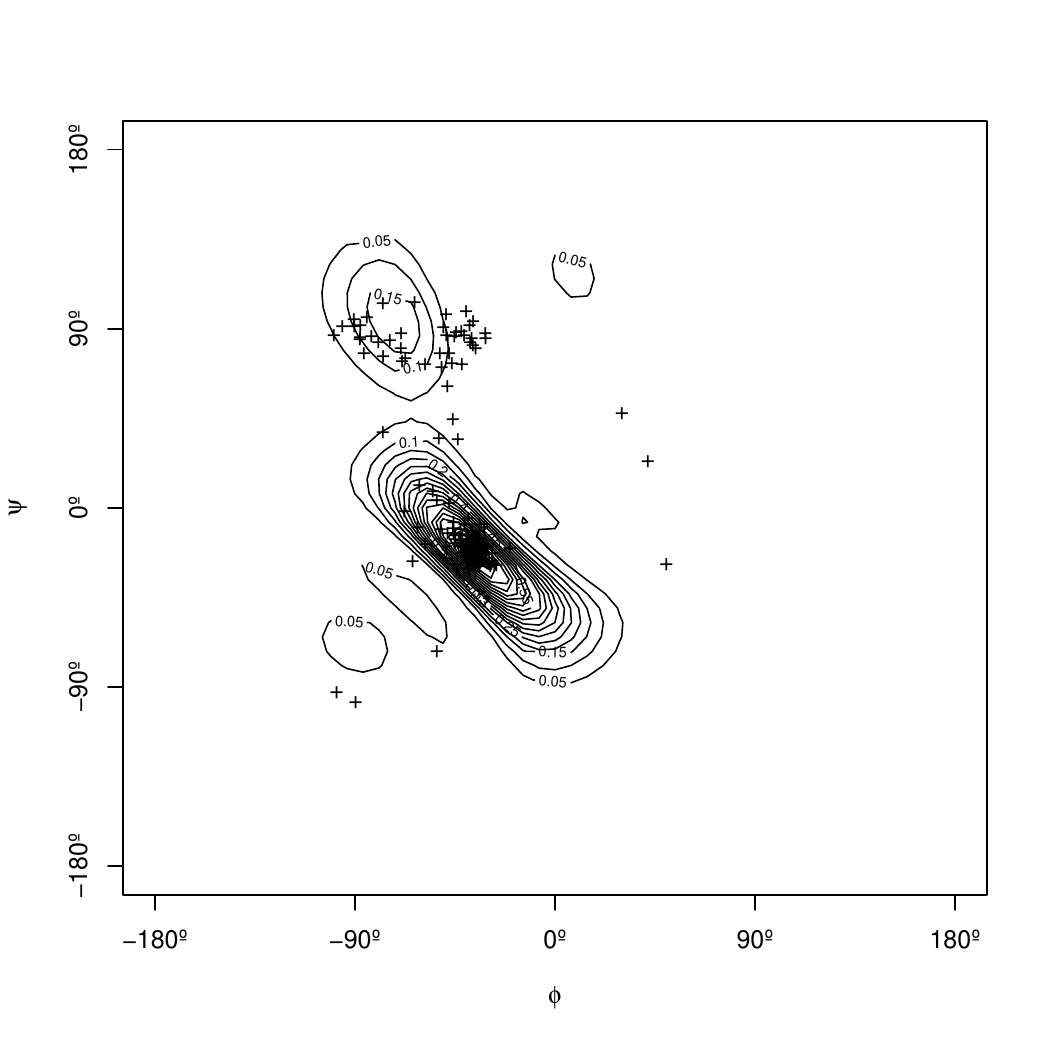} \\[-0.25cm]
	\caption{\small Left: parametric fit (model from \cite{Mardia1978}) to the circular mean orientation and mean log-burnt area of the fires in each of the $102$ watersheds of Portugal. Right: parametric fit (model from \cite{Fern'andez-Dur'an2007}) for the dihedral angles of the alanine-alanine-alanine segments. \label{gofdens:fig:data:mardia}}
\end{figure}
	
The second dataset contains pairs of dihedral angles of segments of the type alanine-alanine-alanine in alanine amino acids in $1932$ proteins. The dataset, formed by $233$ pairs of angles (circular-circular), was studied by \cite{Fern'andez-Dur'an2007} using Nonnegative Trigonometric Sums (NTSS) for the marginal and link function of the model of \cite{Wehrly1980}. The best model in terms of BIC described in \cite{Fern'andez-Dur'an2007} was implemented using a two-step Maximum Likelihood Estimation (MLE) procedure and the tools of the \texttt{CircNNTSR} package \cite{CircNNTSR} for fitting the NTSS parametric densities (Figure \ref{gofdens:fig:data:mardia}, right). The resulting $p$-value with the LCV bandwidths is $0.000$, indicating that the dependence model of \cite{Wehrly1980} is not flexible enough to capture the dependence structure between the two angles. The reason for this lack of fit may be explained by a poor fit in a secondary cluster of data around $\Psi=90^{\circ}$, as can be seen in the contour plot in Figure \ref{gofdens:fig:data:mardia}.

%-------------------------------------------------%
\section*{Supplement}
%-------------------------------------------------%

The supplement contains the detailed proofs of the technical lemmas used to prove the main results, describes in detail the simulation study and shows deeper insights on the real data application. 

%-------------------------------------------------%
\section*{Acknowledgements}
%-------------------------------------------------%

This research has been supported by project MTM2008-03010 from the Spanish Ministry of Science and StuDyS network, from the Interuniversity Attraction Poles Programme (IAP-network P7/06), Belgian Science Policy Office. First author's work has been supported by FPU grant AP2010-0957 from the Spanish Ministry of Education. Authors acknowledge the computational resources used at the SVG cluster of the CESGA Supercomputing Center. The editors and two anonymous referees are acknowledged for their contributions. 

\appendix

%-------------------------------------------------%
\section{Sketches of the main proofs}
\label{gofdens:ap:main}
%-------------------------------------------------%

This section contains the sketches of the main proofs. Proofs for technical lemmas, complete numerical experiments and simulation results, and further details on data analysis are given in the supplementary material.

%-------------------------------------------------%
\subsection{CLT for the integrated squared error}
%-------------------------------------------------%

\begin{proof}[Proof of Theorem \ref{gofdens:theo:clt}]
	The ISE can be decomposed into four addends, $I_n=I_{n,1}+I_{n,2}+I_{n,3}+I_{n,4}$:
	\begin{align*}
		I_{n,1}=&\,2\frac{c_{h,q}(L)}{ng}\sum_{i=1}^n \Iqr{\!\!LK_n\lrp{(\bx,z),(\bX_i,Z_i)}\lrp{\E{\hat f_{h,g}(\bx,z)}-f(\bx,z)}}{\bx}{z},\\
		I_{n,2}=&\,\frac{c_{h,q}(L)^2}{n^2g^2}\sum_{i=1}^n\Iqr{ LK_n^2\lrp{(\bx,z),(\bX_i,Z_i)}}{\bx}{z},\\
		I_{n,3}=&\,\frac{c_{h,q}(L)^2}{n^2g^2}\sum_{1\leq i<j\leq n}\!\!\Iqr{\!\! LK_n\lrp{(\bx,z),(\bX_i,Z_i)} LK_n\lrp{(\bx,z),(\bX_j,Z_j)}}{\bx}{z},\\
		I_{n,4}=&\,\Iqr{\lrp{\E{\hat f_{h,g}(\bx,z)}-f(\bx,z)}^2}{\bx}{z},
	\end{align*}
	where $LK_n\lrp{(\bx,z),(\by,t)}=LK\big(\frac{1-\bx^T\by}{h^2},\frac{z-t}{g}\big)-\mathbb{E}\big[LK\lrp{\frac{1-\bx^T\bX}{h^2},\frac{z-Z}{g}}\big]$. \\
	
	Except for the fourth term, which is deterministic, the CLT for the ISE is derived by examining the asymptotic behaviour of each addend. The first two can be written as $I_{n,1}=\sum_{i=1}^n I_{n,1}^{(i)}$ and $I_{n,2}=\frac{c_{h,q}(L)^2}{n^2g^2}\sum_{i=1}^nI_{n,2}^{(i)}$, where $I_{n,1}^{(i)}$ and $I_{n,2}^{(i)}$ can be directly extracted from the previous expressions. Then, by Lemma \ref{gofdens:lemma:In1},
	\begin{align}
		n^{\frac{1}{2}}\phi(h,g)^{-\frac{1}{2}}I_{n,1}\stackrel{d}{\longrightarrow}\mathcal{N}(0,1)\label{gofdens:res:In1}
	\end{align}
	and by Lemma \ref{gofdens:lemma:In2},
	\begin{align}
		I_{n,2}=\frac{\lambda_{q}(L^2)\lambda_{q}(L)^{-2}R(K)}{nh^qg}+\Orderp{n^{-\frac{3}{2}}h^{-q}g^{-1}}.\label{gofdens:res:In2}
	\end{align}
	
	The third term can be written as
	\begin{align}
		I_{n,3}=2\frac{c_{h,q}(L)^2}{n^2g^2}\sum_{1\leq i<j\leq n} H_n\lrp{(\bX_i,Z_i),(\bX_j,Z_j)}=2\frac{c_{h,q}(L)^2}{n^2g^2}U_n,\label{gofdens:In3:Un}
	\end{align}
	where $U_n$ is an $U$-statistic with kernel function $H_n$ given in Lemma \ref{gofdens:lemma:Hn2}. $U_n$ is degenerate since $\E{LK_n\lrp{(\bx,z),(\bX,Z)}} %
	=0$.\\
	
	In order to properly apply Lemma \ref{gofdens:lemma:extendHn} for obtaining the asymptotic distribution of $U_n$ in (\ref{gofdens:In3:Un}), Lemma \ref{gofdens:lemma:Hn2} provides the explicit expressions for the required elements. Then, considering $\varphi_n\equiv0$ in Lemma \ref{gofdens:lemma:extendHn}, condition $A_nB_n^{-2}\to0$ is satisfied by \ref{gofdens:assump:a3} and, as a consequence, $B_n^{-\frac{1}{2}}U_n\stackrel{d}{\rightarrow}\mathcal{N}(0,1)$. Since the variance of $I_{n,3}$ is
	\begin{align}
		\V{I_{n,3}}=4\frac{c_{h,q}(L)^4}{n^{4}g^4}\V{U_n}=2\frac{\sigma^2}{n^2h^qg}(1+\order{1}),\label{gofdens:var:In3}
	\end{align}
	by Slutsky's theorem, (\ref{gofdens:In3:Un}) and (\ref{gofdens:normalizing}),
	\begin{align}
		n\lrp{h^qg}^\frac{1}{2}I_{n,3}\stackrel{d}{\longrightarrow}\mathcal{N}\lrp{0,2\sigma^2}.\label{gofdens:res:In3}
	\end{align}
	
	From (\ref{gofdens:res:In1}), (\ref{gofdens:res:In2}) and (\ref{gofdens:res:In3}), it follows that:
	\begin{align}
		I_n-\E{I_n} %
		=&\,n^{-\frac{1}{2}}\phi(h,g)^{\frac{1}{2}}N_{n,1}+\Orderp{n^{-\frac{3}{2}}h^{-q}g^{-1}}+2^\frac{1}{2}\sigma n^{-1}(h^qg)^{-\frac{1}{2}}N_{n,3},\label{gofdens:res:In}
	\end{align}
	where $N_{n,1},N_{n,3}\stackrel{d}{\longrightarrow}\mathcal{N}(0,1)$. By \ref{gofdens:assump:a3},  $\big(n^\frac{3}{2}h^{q}g\big)^{-1}=\mathpzc{o}\big((nh^\frac{q}{2}g^\frac{1}{2})^{-1}\big)$ and the second addend $I_{n,2}$ is asymptotically negligible compared with $I_{n,3}$. In order to determine dominance between $I_{n,1}$ and $I_{n,3}$, the squared quotient between their orders is examined, being of order $n\phi(h,g)h^qg$. Then if $n\phi(h,g)h^qg\to\infty$ the last term on (\ref{gofdens:res:In}) is asymptotically negligible in comparison with the first, while if $n\phi(h,g)h^qg\to0$, the first term is negligible in comparison with the last. By (\ref{gofdens:res:In2}), (\ref{gofdens:res:In}) can be stated as
	\begin{align*}
		I_n-\bigg(\Iqr{&\lrp{\E{\hat f_{h,g}(\bx,z)}-f(\bx,z)}^2}{\bx}{z}+\frac{\lambda_q(L^2)\lambda_q(L)^{-2}R(K)}{nh^qg}\bigg)\\
		=&\,n^{-\frac{1}{2}}\phi(h,g)^{\frac{1}{2}}N_{n,1}+\Orderp{n^{-\frac{3}{2}}h^{-q}g^{-1}}+2^\frac{1}{2}\sigma n^{-1}(h^qg)^{-\frac{1}{2}}N_{n,3}.
	\end{align*}
	
	The case where $n\phi(h,g)h^qg\to\delta$, $0<\delta<\infty$, needs a special treatment because none of the terms can be neglected. In this case,
	\begin{align*}
		I_n-\E{I_n}=&\, n^{-\frac{1}{2}}\phi(h,g)^\frac{1}{2}N_{n,1}+2^\frac{1}{2}\sigma n^{-1}(h^qg)^{-\frac{1}{2}}N_{n,3}+\Orderp{n^{-\frac{3}{2}}h^{-q}g^{-1}}\\
		=&\, n^{-1}(h^qg)^{-\frac{1}{2}}\lrp{\delta^\frac{1}{2} N_{n,1}+2^\frac{1}{2}\sigma N_{n,3}}+\Orderp{n^{-\frac{3}{2}}h^{-q}g^{-1}}.
	\end{align*}
	In order to apply Lemma \ref{gofdens:lemma:extendHn}, set $\widetilde U_n=I_{n,1}+I_{n,3}$ with
	\begin{align*}
		\widetilde U_n=&\,\sum_{i=1}^n \varphi_n(\bX_i,Z_i)+\sum_{1\leq i<j\leq n}\widetilde H_n\lrp{(\bX_i,Z_i),(\bX_j,Z_j)},
	\end{align*}
	where $\varphi_n(\bX_1,Z_1)=I_{n,1}^{(1)}$, %
	$\widetilde H_n\lrp{(\bx,z),(\by,t)}=2\frac{c_{h,q}(L)^2}{n^2g^2}H_n\lrp{(\bx,z),(\by,t)}$, and
	$\widetilde G_n\lrp{(\bx,z),(\by,t)}=$\\ $\E{\widetilde H_n\lrp{(\bX,Z),(\bx,z)}\widetilde H_n\lrp{(\bX,Z),(\by,t)}}$.\\
	
	By Lemma \ref{gofdens:lemma:Hn2} and the definitions of $\widetilde H_n$, $\widetilde G_n$, $\varphi_n$, and $M_n$, %
	\begin{align*}
		\E{\widetilde H_n^{2}\lrp{(\bX_1,Z_1),(\bX_2,Z_2)}}=&\,4n^{-4}h^{-q}g^{-1}\sigma^2\lrp{1+\order{1}},\\ %
		\E{\widetilde H_n^{4}\lrp{(\bX_1,Z_1),(\bX_2,Z_2)}}=&\,\Order{n^{-8}h^{-3q}g^{-3}},
		\\
		\E{\widetilde G_n^{2}\lrp{(\bX_1,Z_1),(\bX_2,Z_2)}}=&\,\Order{n^{-8}h^{-q}g^{-1}},\\
		\E{\varphi_n^2(\bX_1,Z_1)}=&\,n^{-2}\phi(h,g)\lrp{1+\order{1}},\\ %
		\E{\varphi_n^4(\bX_1,Z_1)}=&\,\Order{n^{-4}(h^8+g^8)},\\
		\E{M_n^2(\bX_1,Z_1)}=&\,\Order{n^{-6}(h^4+g^4)h^{-\frac{3q}{2}}g^{-\frac{3}{2}}}.
	\end{align*}
	Applying these orders and using $n\phi(h,g)h^qg\to\delta$,
	\begin{align*}
		\frac{A_n}{B_n^2} %
		= \Order{n^{-1}}+\Order{(nh^qg)^{-1}h^\frac{q}{2}g^\frac{1} {2}}+\Order{(nh^{q}g)^{-1}}+\Order{h^qg}.
	\end{align*}
	Then, by \ref{gofdens:assump:a3}, the four previous orders tend to zero and therefore $B_n^{-\frac{1}{2}}\widetilde U_n\stackrel{d}{\longrightarrow}\mathcal{N}(0,1)$, where
	$B_n %
	\sim n^{-1}\phi(h,g)+2n^{-2}(h^qg)^{-1}\sigma^2\sim n^{-2}(h^qg)^{-1}\lrp{\delta+2\sigma^2}$. Finally, $n\lrp{h^qg}^\frac{1}{2} 2^\frac{1}{2} \lrp{\delta+2\sigma^2}^{-\frac{1}{2}}$ $\lrp{I_{n,1}+I_{n,3}}$ $\stackrel{d}{\longrightarrow}\mathcal{N}(0,1)$ by Slutsky's theorem.
\end{proof}

\begin{proof}[Proof of Corollary \ref{gofdens:coro:clt}]
	
	As $g=\beta h$, for a fixed $\beta>0$, $n\phi(h,g)h^qg\allowbreak=\Order{nh^{q+5}}$ and the cases in Theorem \ref{gofdens:theo:clt} are given by the asymptotic behaviour of this sequence.  When $nh^{q+5}\to\infty$ and $nh^{q+5}\to0$, the result is obtained immediately, whereas for $nh^{q+5}\to\delta$, $0<\delta<\infty$, Lemma \ref{gofdens:lemma:extendHn} gives
	\begin{align*}
		B_n\sim \phi(1,\beta)n^{-1}h^{4}+2\sigma^2n^{-2}h^{-(q+1)}
		\sim n^{-\frac{q+9}{q+5}}\lrp{\phi(1,\beta)\delta^\frac{4}{q+5}+2\sigma^2\delta^{-\frac{q+1}{q+5}}}.
	\end{align*}
	Therefore, $n^{\frac{q+9}{2(q+5)}}\Big(\phi(1,\beta)\delta^\frac{4}{q+5}+2\sigma^2\delta^{-\frac{q+1}{q+5}}\Big)^{-\frac{1}{2}}\lrp{I_{n,1}+I_{n,3}}\stackrel{d}{\longrightarrow}\mathcal{N}(0,1)$.
\end{proof}

\begin{proof}[Proof of Corollary \ref{gofdens:coro:clt:dirdir}]
	The proof follows from an adaptation of the proof of Theorem \ref{gofdens:theo:clt} to the directional-directional context.
\end{proof}

%-------------------------------------------------%
\subsection{Testing independence with directional data}
%-------------------------------------------------%

\begin{proof}[Proof of Theorem \ref{gofdens:theo:indep}]
	
	The test statistic is decomposed as $T_n=T_{n,1}+T_{n,2}+T_{n,3}$ taking into account that, under independence, $\mathbb{E}\big[\hat f_{h,g}(\bx,z)\big]=\mathbb{E}\big[\hat f_{h}(\bx)\big]\mathbb{E}\big[\hat f_{g}(z)\big]$:
	\begin{align*}
		T_{n,1}=&\,\Iqr{\lrp{\hat f_{h,g}(\bx,z)-\E{\hat f_{h,g}(\bx,z)}}^2}{\bx}{z},\\
		T_{n,2}=&\,\Iqr{\lrp{\hat f_{h}(\bx) \hat f_{g}(z)-\E{\hat f_{h}(\bx)}\E{\hat f_{g}(z)}}^2}{\bx}{z},\\
		T_{n,3}=&\,-2\Iqr{\lrp{\hat f_{h,g}(\bx,z)-\E{\hat f_{h,g}(\bx,z)}}\lrp{\hat f_{h}(\bx) \hat f_{g}(z)-\E{\hat f_{h}(\bx)}\E{ \hat f_{g}(z)}}}{\bx}{z}.
	\end{align*}
	
	By Chebychev's inequality and Lemmas \ref{gofdens:lem:indep:2} and \ref{gofdens:lem:indep:3}, %
	the sum of the second and third addends is $-\E{T_{n,2}}+\mathcal{O}_\mathbb{P}\big(n^{-1}(h^{-q}+g^{-1})^\frac{1}{2}\big)$. Considering the test statistic decomposition and using Lemma \ref{gofdens:lem:indep:1} yields
	\begin{align*}
		T_n=&\,T_{n,1}-\E{T_{n,2}}+\Orderp{n^{-1}(h^{-q}+g^{-1})^\frac{1}{2}}\\
		=&\,\frac{\lambda_{q}(L^2)\lambda_{q}(L)^{-2}R(K)}{nh^qg}+2^\frac{1}{2}\sigma n^{-1}(h^qg)^{-\frac{1}{2}}N_{n}-\frac{\lambda_{q}(L^2)\lambda_{q}(L)^{-2}R(f_Z)}{nh^q},\\%
		&-\frac{R(K)R(f_\bX)}{ng}+\order{n^{-1}(h^{-q}+g^{-1})}+\Orderp{n^{-1}(h^{-q}+g^{-1})^\frac{1}{2}}.%
	\end{align*}
	Now, $\mathcal{O}_\mathbb{P}\big(n^{-1}(h^{-q}+g^{-1})^\frac{1}{2}\big)$ is negligible in comparison with the second addend by \ref{gofdens:assump:a3} 
	and the deterministic order $\mathpzc{o}\big(n^{-1}(h^{-q}+g^{-1})\big)$ is also negligible by \ref{gofdens:assump:a3} and \ref{gofdens:assump:a4}. %
	Therefore, $n(h^qg)^\frac{1}{2}(T_n-$ $A_n)\stackrel{d}{\longrightarrow}\mathcal{N}(0,2\sigma_I^2)$.%
\end{proof}

%-------------------------------------------------%
\subsection{Goodness-of-fit test for models with directional data}
%-------------------------------------------------%

\begin{proof}[Proof of Theorem \ref{gofdens:theo:gof:simp}]
	Under $H_0: f=f_{\btheta_0}$, the test statistic $R_n=I_{n,2}+I_{n,3}$, where $I_{n,2}$ and $I_{n,3}$ are given by (\ref{gofdens:res:In2}) and (\ref{gofdens:res:In3}) in the proof of Theorem \ref{gofdens:theo:clt}, so 
	\begin{align}
		n(h^qg)^\frac{1}{2}\lrp{I_{n,2}+I_{n,3}-\frac{\lambda_q(L^2)\lambda_q(L)^{-2}R(K)}{nh^qg}}\stackrel{d}{\longrightarrow}\mathcal{N}\lrp{0,2\sigma_{\btheta_0}^2}.\label{gofdens:asymp:I23}
	\end{align}
\end{proof}

\begin{proof}[Proof of Theorem \ref{gofdens:theo:gof}]
	
	The test statistic is decomposed as $R_n=R_{n,1}+I_{n,2}+I_{n,3}+R_{n,4}$ by adding and subtracting $\mathbb{E}\big[\hat f_{h,g}(\bx,z)\big]=LK_{h,g} f(\bx,z)$, with
	\begin{align*}
		R_{n,1}=&\,2\Iqr{\lrp{\hat f_{h,g}(\bx,z)-LK_{h,g} f(\bx,z)} LK_{h,g}\lrp{f(\bx,z)-f_{ \hat \btheta}(\bx,z)}}{\bx}{z},\\
		R_{n,4}=&\,\Iqr{\lrp{LK_{h,g}\lrp{f(\bx,z)-f_{ \hat \btheta}(\bx,z)}}^2}{\bx}{z}.
	\end{align*}
	
	The limit of $I_{n,2}+I_{n,3}$ is given by (\ref{gofdens:asymp:I23}) whereas, by Lemma \ref{gofdens:lem:gof:1}, $R_{n,1}$ and $R_{n,4}$ are negligible in probability. Then, the limit distribution of $R_n$ is determined by $I_{n,2}+I_{n,3}$.
\end{proof}

\begin{proof}[Proof of Theorem \ref{gofdens:theo:gofpit}]
	As in the proof of Theorem \ref{gofdens:theo:gof}, $R_n=\widetilde R_{n,1}+I_{n,2}+I_{n,3}+\widetilde R_{n,4}$, where $I_{n,2}+I_{n,3}$ behaves as (\ref{gofdens:asymp:I23}).
	The asymptotic variance remains $\sigma_{\btheta_0}^2$ since %
	\begin{align*}
		R(f)=R(f_{\btheta_0})+\frac{R(\Delta)}{nh^\frac{q}{2}g^\frac{1}{2}}+\frac{\Iqr{f(\bx,z)\Delta(\bx,z)}{\bx}{z}}{n^\frac{1}{2}h^\frac{q}{4}g^\frac{1}{4}}
	\end{align*}
	and then the second and third addends are negligible with respect to the first by \ref{gofdens:assump:a3}, leaving the same asymptotic variance. The terms $\widetilde R_{n,1}=R_{n,1}+\widetilde R_{n,1}^{(1)}$ and $\widetilde R_{n,4}=R_{n,4}+\widetilde R_{n,4}^{(1)}+\widetilde R_{n,4}^{(2)}$ are decomposed~as
	\begin{align*}
		\widetilde R_{n,1}^{(1)}=&\,\frac{2}{\sqrt{nh^\frac{q}{2}g^\frac{1}{2}}}\Iqr{\lrp{\hat f_{h,g}(\bx,z)-LK_{h,g} f(\bx,z)} LK_{h,g} \Delta(\bx,z)}{\bx}{z},\\
		\widetilde R_{n,4}^{(1)}=&\,\frac{1}{nh^\frac{q}{2}g^\frac{1}{2}}\Iqr{\lrp{LK_{h,g}\Delta(\bx,z)}^2}{\bx}{z},\\
		\widetilde R_{n,4}^{(2)}=&\,\frac{2}{\sqrt{nh^\frac{q}{2}g^\frac{1}{2}}}\Iqr{LK_{h,g}\lrp{f(\bx,z)-f_{ \hat \btheta}(\bx,z)}LK_{h,g}\Delta(\bx,z)}{\bx}{z}.
	\end{align*}
	The remaining terms follow from Lemma \ref{gofdens:lem:gof:2}.
\end{proof}

\begin{proof}[Proof of Theorem \ref{gofdens:theo:boots}]
	Similar to the proof of Theorem \ref{gofdens:theo:gof}, $R_n^*=R_{n,1}^*+I_{n,2}^*+I_{n,3}^*+R_{n,4}^*$, where the terms involved are the bootstrap versions of the ones defined in the aforementioned proof:
	\begin{align*}
		R_{n,1}^*=&\,2\Iqr{\!\!\lrp{\hat f^*_{h,g}(\bx,z)-LK_{h,g} f_{\hat \btheta}(\bx,z)} LK_{h,g}\lrp{f_{\hat \btheta}(\bx,z)-f_{ \hat \btheta^*}(\bx,z)}}{\bx}{z},\\
		I_{n,2}^*=&\,\frac{c_{h,q}(L)^2}{n^2g^2}\sum_{i=1}^n\Iqr{ \lrp{LK_n^*\lrp{(\bx,z),(\bX_i^*,Z_i^*)}}^2 }{\bx}{z},\\
		I_{n,3}^*=&\,\frac{c_{h,q}(L)^2}{n^2g^2}\sum_{1\leq i<j\leq n}\Iqr{LK_n^*\lrp{(\bx,z),(\bX_i^*,Z_i^*)}LK_n^*((\bx,z),(\bX_j^*,Z_j^*))}{\bx}{z},\\
		R_{n,4}^*=&\,\Iqr{\lrp{LK_{h,g}\lrp{f_{\hat \btheta}(\bx,z)-f_{ \hat \btheta^*}(\bx,z)}}^2}{\bx}{z},
	\end{align*}
	with $LK_n^*((\bx,z),(\by,t))=LK\big(\frac{1-\bx^T\by}{h^2},\frac{z-t}{g}\big)-\mathbb{E}^*\big[LK\big(\frac{1-\bx^T\bX^*}{h^2},\frac{z-Z^*}{g}\big)\big]$ and where $\mathbb{E}^*$ represents the expectation with respect to $f_{\hat{\btheta}}$, which is obtained from the original sample.\\
	
	Using the same arguments as in Lemma \ref{gofdens:lem:gof:1}, but replacing \ref{gofdens:assump:a6} by \ref{gofdens:assump:a8}, it follows that $n(h^qg)^\frac{1}{2}R_{n,1}^*$ and $n(h^qg)^\frac{1}{2}R_{n,4}^*$ converge to zero conditionally on the sample, that is, in probability $\mathbb{P}^*$. On the other hand, the terms $I^*_{n,2}$ and $I^*_{n,3}$ follow from considering similar arguments to the ones used for deriving (\ref{gofdens:res:In2}) and (\ref{gofdens:res:In3}), but conditionally on the sample. Specifically, it follows that
	$I_{n,2}^*=\frac{\lambda_q(L^2)\lambda_q(L)^{-2}R(K)}{nh^qg}+\mathcal{O}_{\mathbb{P}^*}\big(n^{-\frac{3}{2}}h^{-q}g^{-1}\big)$ and, for a certain $\btheta_1\in\Theta$, 
	$(nh^qg)^\frac{1}{2}I_{n,3}^*\stackrel{d}{\longrightarrow}\mathcal{N}(0,2\sigma_{\btheta_1}^2)$. The main difference with the proof of Theorem \ref{gofdens:theo:gof} concerns the asymptotic variance given by $n(h^qg)^\frac{1}{2}I^*_{n,3}$: $\mathbb{V}\text{ar}^*\big[n(h^qg)^\frac{1}{2}I_{n,3}^*\big]\stackrel{p}{\longrightarrow}2\sigma_{\btheta_1}^2$,
	since by \ref{gofdens:assump:a5}, $R(f_{\hat \btheta})=R(f_{ \btheta_1})+\mathcal{O}_\mathbb{P}\big(n^{-\frac{1}{2}}\big)$. Hence,
	\begin{align*}
		n(h^qg)^\frac{1}{2}\bigg(R_n^*-\frac{\lambda_q(L^2)\lambda_q(L)^{-2}R(K)}{nh^qg}\bigg)=\mathpzc{o}_{\mathbb{P}^*}(1)+\mathcal{O}_{\mathbb{P}^*}\big((nh^{q}g)^{-\frac{1}{2}}\big)+2^\frac{1}{2}\sigma_{\btheta_1}N_n+\mathpzc{o}_{\mathbb{P}^*}(1)
	\end{align*}
	and bootstrap consistency follows.
\end{proof}

\begin{proof}[Proof of Corollary \ref{gofdens:coro:gof:dirdir}]
	The proof follows by adapting the proof of Theorem \ref{gofdens:theo:gof}.
\end{proof}

\fi

\ifsupplement

\newpage

%---------- TITLE ---------------%
\title{Supplement to ``Central limit theorems for directional and linear random variables with applications''}
\setlength{\droptitle}{-1cm}
\predate{}%
\postdate{}%
%---------- AUTHORS ---------------%
\date{}

\author{Eduardo Garc\'ia-Portugu\'es$^{1,2}$, Rosa M. Crujeiras$^{1}$, and Wenceslao Gonz\'alez-Manteiga$^{1}$}

\footnotetext[1]{
Department of Statistics and Operations Research, University of Santiago de Compostela (Spain).}
\footnotetext[2]{Corresponding author. e-mail: \href{mailto:eduardo.garcia@usc.es}{eduardo.garcia@usc.es}.}

\maketitle

\begin{abstract}
This supplement is organized as follows. Section \ref{gofdens:su:lem} contains the detailed proofs of the required technical lemmas used to prove the main results in the paper. The section is divided into four subsections to classify the lemmas used in the CLT of the ISE, the independence test and the goodness-of-fit test, with an extra subsection for general purpose lemmas. Section \ref{gofdens:su:num} presents closed expressions that can be used in the independence test, the extension of the results to the directional-directional situation and some numerical experiments to illustrate the convergence to the asymptotic distribution. Section \ref{gofdens:su:simus} describes in detail the simulation study of the goodness-of-fit test to allow its reproducibility: parametric models employed, estimation and simulation methods, the construction of the alternatives, the bandwidth choice and further results omitted in the paper. Finally, Section \ref{gofdens:su:data} shows deeper insights on the real data application. 
\end{abstract}

\begin{flushleft}
\small
\textbf{Keywords:} Directional data; Goodness-of-fit; Independence test; Kernel density estimation; Limit distribution.
\end{flushleft}

%-------------------------------------------------%
\section{Technical lemmas}
\label{gofdens:su:lem}
%-------------------------------------------------%

%-------------------------------------------------%
\subsection{CLT for the ISE}
%-------------------------------------------------%

Lemma \ref{gofdens:lemma:extendHn} presents a generalization of Theorem 1 in \cite{Hall1984} for degenerate $U$-statistics that, up to the authors' knowledge, was first stated by \cite{Zhao2001} under different conditions, but without providing a formal proof. This lemma, written under a general notation, is used to prove asymptotic convergence of the ISE when the variance is large relative to the bias ($n\phi(h,g)h^qg\to0$) and when the bias is balanced with the variance ($n\phi(h,g)h^qg\to\delta$).

\begin{lem}
	\label{gofdens:lemma:extendHn}
	Let $\lrb{X_i}_{i=1}^n$ be a sequence of independent and identically distributed random variables. Assume that $H_n(x,y)$ is symmetric in $x$ and $y$,
	\begin{align}
	\E{H_n\lrp{X_1,X_2}\vert X_1}=0\text{ almost surely and }\E{H_n^4(X_1,X_2)}<\infty,\,\forall n. \label{gofdens:lemma:extendHn:1}
	\end{align}
	Define $G_n\lrp{x,y}=\E{H_n\lrp{x,X_1}H_n\lrp{y,X_1}}$ and $\varphi_n$, satisfying $\E{\varphi_n(X_1)}=0$ and $\E{\varphi^4_n(X_1)}<\infty$. Define also:
	\begin{align*}
	M_n(X_1)=&\,\E{\varphi_n(X_2)H_n(X_1,X_2)\vert X_1},\\
	A_n=&\,n\E{\varphi_n^4(X_1)}\!+n^2\E{M_n^2(X_1)}\!+n^3\E{H_n^4(X_1,X_2)}\!+n^4\E{G_n^2(X_1,X_2)},\\
	B_n=&\,n\E{\varphi_n^2(X_1)}+\frac{1}{2}n^2\E{H_n^2(X_1,X_2)}.
	\end{align*}
	If $A_nB_n^{-2}\to 0$ as $n\to\infty$ and $U_n=\sum_{i=1}^n \varphi_n(X_i)+\sum_{1\leq i<j\leq n} H_n\lrp{X_i,X_j}$,
	\[
	B_n^{-\frac{1}{2}}U_n\stackrel{d}{\longrightarrow}\mathcal{N}(0,1).
	\]
\end{lem}
Note that when $\varphi_n\equiv0$, $U_n$ is an $U$-statistic and Theorem 1 in \cite{Hall1984} is a particular case of Lemma \ref{gofdens:lemma:extendHn}.

\begin{proof}[Proof of Lemma \ref{gofdens:lemma:extendHn}]
	To begin with, let consider the sequence of random variables $\lrb{Y_{n_i}}_{i=1}^n$, defined\nolinebreak[4] by
	\begin{align*}
	Y_{n_i}=\left\{
	\begin{array}{ll}
	\varphi_n(X_1),&i=1,\\
	\varphi_n(X_i)+\sum_{j=1}^{i-1}H_n(X_i,X_j),&2\leq i\leq n.
	\end{array}
	\right.
	\end{align*}
	This sequence generates a martingale $S_i=\sum_{j=1}^i Y_{n_j}$, $1\leq i\leq n$ with respect to the sequence of random variables $\lrb{X_i}_{i=1}^n$, with differences $Y_{n_i}$ and with $S_n=U_n$. To see that $S_i=\sum_{j=1}^i Y_{n_j}$, $1\leq i\leq n$ is indeed a martingale with respect to $\lrb{X_i}_{i=1}^n$, recall that
	\begin{align*}
	\E{S_{i+1}|X_1,\ldots,X_{i}}=&\,\sum_{j=1}^{i+1}\E{\varphi_n(X_j)|X_1,\ldots,X_{i}}+\sum_{j=1}^{i+1}\sum_{k=1}^{j-1}\E{H_n(X_j,X_k)|X_1,\ldots,X_{i}}\\
	=&\,\sum_{j=1}^{i}\varphi_n(X_j)+\sum_{j=1}^i\sum_{k=1}^{j-1}H_n(X_j,X_k)\\
	=&\,S_i
	\end{align*}
	because of the null expectations of $\E{\varphi_n(X)}$ and $\E{H_n\lrp{X_1,X_2}\vert X_1}$.\\
	
	The main idea of the proof is to apply the martingale CLT of \cite{Brown1971} (see also Theorem 3.2 of \cite{Hall1980}), in the same way as \cite{Hall1984} did for the particular case where $\varphi_n\equiv0$. Theorem 2 of \cite{Brown1971} ensures that if the conditions 
	\begin{enumerate}[label=\textbf{C\arabic{*}}., ref=\textbf{C\arabic{*}}]
		\item $\displaystyle \lim_{n\to\infty}s_n^{-2}\sum_{i=1}^n \label{gofdens:assump:c1} \E{Y_{n_i}^2\mathbbm{1}_{\lrb{\abs{Y_{n_i}}>\varepsilon s_n}}}=0$, $\forall \varepsilon>0$,
		\item $s_n^{-2}V_n^2\stackrel{p}{\longrightarrow}1$, \label{gofdens:assump:c2}
	\end{enumerate}
	are satisfied, with $s_n^2=\E{U_n^2}$ and $V_n^2=\sum_{i=1}^n \mathbb{E}\big[Y_{n_i}^2|X_1,\ldots,X_{i-1}\big]$, then $s_n^{-1}U_n\stackrel{d}{\longrightarrow}\mathcal{N}(0,1)$. The aim of this proof is to prove separately both conditions. From now on, expectations will be taken with respect to the random variables $X_1,\ldots,X_n$, except otherwise is stated. \\
	
	\textit{Proof of \ref{gofdens:assump:c1}}. The key idea is to give bounds for $\mathbb{E}\big[Y_{n_i}^4\big]$ and prove that $s_n^{-4}\sum_{i=1}^{n}\mathbb{E}\big[Y_{n_i}^4\big]\to0$ as $n\to\infty$. In that case, the Lindenberg's condition \ref{gofdens:assump:c1} follows immediately:
	\begin{align*}
	\lim_{n\to\infty}s_n^{-2}\sum_{i=1}^n\E{Y_{n_i}^2\mathbbm{1}_{\lrb{\abs{Y_{n_i}}>\varepsilon s_n}}}\leq&\lim_{n\to\infty} s_n^{-2}\sum_{i=1}^n\E{Y_{n_i}^4\varepsilon^{-2}s_n^{-2}\times 1}\\
	=&\,\varepsilon^{-2}\lim_{n\to\infty}s_n^{-4}\sum_{i=1}^n\E{Y_{n_i}^4}\\
	=&\,0.
	\end{align*}
	In order to compute $s_n^2=\E{U_n^2}$, it is needed
	\begin{align*}
	\E{Y_{n_i}^2}=\left\{
	\begin{array}{ll}
	\E{\varphi_n^2(X_1)},&i=1,\\
	\E{\varphi_n^2(X_i)}+(i-1)\E{H_n^2(X_1,X_2)},&2\leq i\leq n,
	\end{array}
	\right.
	\end{align*}
	where the second case holds because the independence of the variables, the tower property of the conditional expectation and (\ref{gofdens:lemma:extendHn:1}) ensure that
	\[
	\E{\varphi_n(X_1)H_n(X_1,X_2)}=\E{H_n(X_1,X_2)H_n(X_1,X_3)}=0. %
	\]
	Using these relations and the null expectation of $\varphi_n(X_1)$, it follows that for $j\neq k$,
	\begin{align*}
	\E{Y_{n_j}Y_{n_k}}=&\,\E{\varphi_n(X_j)}\E{\varphi_n(X_k)}+\sum_{l=1}^{k-1}\E{\varphi_n(X_j)H_n(X_k,X_l)}\\
	&+\sum_{m=1}^{j-1}\E{\varphi_n(X_k)H_n(X_j,X_m)}+\sum_{l=1}^{k-1}\sum_{m=1}^{j-1}\E{H_n(X_k,X_l)H_n(X_j,X_m)}\\
	=&\,0.
	\end{align*}
	Then:
	\begin{align}
	s_n^2=n\E{\varphi_n^2(X_1)}+\sum_{j=1}^n(j-1)\E{H_n^2(X_1,X_2)}=\Order{B_n}.\label{gofdens:lemma:extendHn:sn2}
	\end{align}
	On the other hand,
	\begin{align*}
	\E{Y_{n_i}^4}=&\,\mathbb{E}\bigg[\Big(\varphi_n(X_i)+\sum_{j=1}^{i-1} H_n(X_i,X_j)\Big)^4\bigg]\nonumber\\
	=&\, \Order{\E{\varphi^4_n(X_i)}}+\mathcal{O}\bigg(\mathbb{E}\bigg[\Big(\sum_{j=1}^{i-1} H_n(X_i,X_j)\Big)^4\bigg]\bigg)\\
	=&\,\Order{\E{\varphi_n^4(X_1)}}+(i-1)\Order{\E{H_n^4(X_1,X_2)}}\\
	&+3(i-1)(i-2)\Order{\E{H_n^2(X_1,X_2)H_n^2(X_1,X_3)}},
	\end{align*}
	where the equalities are true in virtue of Lemma \ref{gofdens:lemma:orders} and because
	\[
	\mathbb{E}\big[H_n\lrp{X_1,X_2}H_n\lrp{X_1,X_3}H_n\lrp{X_1,X_4}H_n\lrp{X_1,X_5}\big]=\E{H_n^3\lrp{X_1,X_2}H_n\lrp{X_1,X_3}}=0.
	\]
	Finally,
	\begin{align}
	\sum_{i=1}^n \E{Y_{n_i}^4}=&\,n\Order{\E{\varphi_n^4(X_1)}}+\frac{1}{2}n(n-1)\Order{\E{H_n^4(X_1,X_2)}}\nonumber\\
	&+(n^3-n)\Order{\E{G_n^2(X_1,X_2)}}\nonumber\\
	=&\,\Order{A_n}.\label{gofdens:lemma:extendHn:sumyni4}
	\end{align}
	Then, joining (\ref{gofdens:lemma:extendHn:sn2}) and (\ref{gofdens:lemma:extendHn:sumyni4}),
	\[
	s_n^{-4}\sum_{i=1}^n \E{Y_{n_i}^4}=\Order{B_n^{-2}A_n}\xrightarrow[n\to\infty]{}0
	\]
	and \ref{gofdens:assump:c1} is satisfied.\\
	
	\textit{Proof of \ref{gofdens:assump:c2}}. Now it is proved the convergence in squared mean of $s_n^{-2}V_n^2$ to $1$, which implies that $s_n^{-2}V_n^2\stackrel{p}{\longrightarrow}1$, by obtaining bounds for $\E{V_n^4}$. \\
	
	First of all, let denote $V_n^2=\sum_{i=1}^n \nu_{n_i}$, where
	\begin{align*}
	\nu_{n_i}=&\,\E{Y_{n_i}^2|X_1,\ldots,X_{i-1}}\\
	=&\,\mathbb{E}\bigg[\varphi_n^2(X_i)+2\varphi_n(X_i)\sum_{j=1}^{i-1}H_n(X_i,X_j)+\sum_{j=1}^{i-1}\sum_{k=1}^{i-1}H_n(X_i,X_j)H_n(X_i,X_k)\bigg|X_1,\ldots,X_{i-1}\bigg]\\
	=&\,\E{\varphi_n^2(X_i)}+2\sum_{j=1}^{i-1}M_n(X_j)+\sum_{j=1}^{i-1}\sum_{k=1}^{i-1}\E{H_n(X_i,X_j)H_n(X_i,X_k) |X_j,X_k}\\
	=&\,\E{\varphi_n^2(X_1)}+2\sum_{j=1}^{i-1}M_n(X_j)+\sum_{j=1}^{i-1}G_n(X_j,X_j)+2\sum_{1\leq j<k\leq i-1}G_n(X_j,X_k).
	\end{align*}
	Using Lemma \ref{gofdens:lemma:orders}, the Jensen inequality and that for $j_1\leq k_1$, $j_2\leq k_2$,
	\begin{align*}
	\E{G_n(X_{j_1},X_{k_1})G_n(X_{j_2},X_{k_2})}=\left\{
	\begin{array}{ll}
	\E{G_n^2(X_1,X_1)},& j_1=k_1=j_2=k_2,\\
	\E{G_n(X_1,X_1)}^2,& j_1=k_1\neq j_2=k_2,\\
	\E{G_n^2(X_1,X_2)},& j_1=j_2<k_1=k_2,\\
	0,& \text{otherwise},\\
	\end{array}\right.
	\end{align*}
	it follows:
	\begin{align*}
	\E{\nu_{n_i}^2}=&\,\Order{\E{\varphi_n^2(X_1)}^2}+\sum_{j=1}^{i-1}\sum_{k=1}^{i-1}\Order{\E{M_n(X_j)M_n(X_k)}}\\
	&\!+\sum_{j=1}^{i-1}\sum_{k=1}^{i-1}\Order{\E{G_n(X_j,X_j)G_n(X_k,X_k)}}+\!\!\!\!\sum_{\substack{1\leq j_1<k_1\leq i-1\\1\leq j_2<k_2\leq i-1}}\!\!\!\!\Order{\E{G_n(X_{j_1},X_{k_1})G_n(X_{j_2},X_{k_2})}}\\
	=&\,\Order{\E{\varphi_n^4(X_1)}}+(i-1)\Order{\E{M_n^2(X_1)}}+(i-1)(i-2)\Order{\E{M_n(X_1)}^2}\\
	&+(i-1)\Order{\E{G_n^2(X_1,X_1)}}+(i-1)(i-2)\Order{\E{G_n(X_1,X_1)}^2}\\
	&+(i-1)(i-2)\Order{\E{G_n^2(X_1,X_2)}}.
	\end{align*}
	Applying again the Lemma \ref{gofdens:lemma:orders},
	\begin{align*}
	\E{V_n^4}=\mathbb{E}\bigg[\Big(\sum_{i=1}^n \nu_{n_i}\Big)^2\bigg]=\sum_{i=1}^n\Order{\E{\nu_{n_i}^2}}.
	\end{align*}
	By the two previous computations and bearing in mind that $\E{G_n(X_1,X_1)}^2=\Order{\E{H_n^4(X_1,X_2)}}$ (by the Cauchy--Schwartz inequality) and $\E{M_n(X_1)}=0$ (by the tower property), it yields:
	\begin{align*}
	\E{V_n^4}=&\,n\Order{\E{\varphi_n^4(X_1)}}+n(n-1)\Order{\E{M_n^2(X_1)}}\\
	&+n(n-1)(n-3)\Order{\E{H_n^4(X_1,X_2)}}\\
	&+n(n-1)(n-3)\Order{\E{G_n^2(X_1,X_2)}}\\
	=&\,\Order{A_n}.
	\end{align*}
	Then, using the bound for $\E{V_n^4}$, that $s_n^2=B_n$ and that $\E{V_n^2}=s_n^2$, it results
	\begin{align*}
	\E{\big(s_n^{-2}V_n^2-1\big)^2}=s_n^{-4}\E{\big(V_n^2-s_n^{2}\big)^2}=s_n^{-4}\lrp{\E{V_n^4}-s_n^4}\leq s_n^{-4}\E{V_n^4}=\Order{B_n^{-2}A_n}.
	\end{align*}
	Then $s_n^{-2}V_n^2$ converges to $1$ in squared mean, which implies $s_n^{-2}V_n^2\stackrel{p}{\longrightarrow}1$. 
\end{proof}
\begin{lem}
	\label{gofdens:lemma:In1}
	Under \ref{gofdens:assump:a1}--\ref{gofdens:assump:a3}, 
	\begin{align*}
	n^{\frac{1}{2}}\phi(h,g)^{-\frac{1}{2}}I_{n,1}\stackrel{d}{\longrightarrow}\mathcal{N}(0,1). %
	\end{align*}
\end{lem}

\begin{proof}[Proof of Lemma \ref{gofdens:lemma:In1}]
	The asymptotic normality of $I_{n,1}=\sum_{i=1}^n I_{n,1}^{(i)}$ will be derived checking the Lindenberg's condition. To that end, it is needed to prove the following relations: 
	\begin{align*}
	\E{I_{n,1}^{(i)}}=&\,0,\quad\E{\big(I_{n,1}^{(i)}\big)^2}= n^{-2}\phi(h,g)(1+\order{1}),\\
	\E{\big(I_{n,1}^{(i)}\big)^4}=&\,\Order{n^{-4}(h^8+g^8)},\quad s_n^4=\Order{n^{-2}(h^8+g^8)},
	\end{align*}
	where $s_n^2=\sum_{i=1}^n\mathbb{E}\big[\big(I_{n,1}^{(i)}\big)^2\big]$ and $\phi(h,g)$ is defined as in Theorem \ref{gofdens:theo:clt}. If these relations hold, the Lindenberg's condition
	\begin{align*}
	\lim_{n\to\infty}s_n^{-2}\sum_{i=1}^n\E{\big( I_{n,1}^{(i)}\big)^2\mathbbm{1}_{\big\{\big|I_{n,1}^{(i)}\big|>\varepsilon s_n\big\}}}=0,\quad\forall \varepsilon>0
	\end{align*}
	is satisfied:
	\begin{align*}
	s_n^{-2}\sum_{i=1}^n\E{\big(I_{n,1}^{(i)}\big)^2\mathbbm{1}_{\big\{\big|I_{n,1}^{(i)}\big|>\varepsilon s_n\big\}}} %
	\leq&\, \sum_{i=1}^n\E{\big(I_{n,1}^{(i)}\big)^4\varepsilon^{-2}s_n^{-4}\times1}\\
	=&\,\varepsilon^{-2}n\E{\big(I_{n,1}^{(i)}\big)^4}\Order{n^{2}(h^8+g^8)^{-1}}\\
	=&\,\varepsilon^{-2}\Order{n^{-1}}.
	\end{align*}
	Therefore $s_n^{-1}I_{n,1}\stackrel{d}{\longrightarrow}\mathcal{N}(0,1)$, which, by Slutsky's theorem, implies that
	\begin{align*}
	n^{\frac{1}{2}}\phi(h,g)^{-\frac{1}{2}}I_{n,1}\stackrel{d}{\longrightarrow}\mathcal{N}(0,1).
	\end{align*}
	
	In order to prove the moment relations for $I_{n,1}^{(i)}$ and bearing in mind the smoothing operator (\ref{gofdens:smoothing}), let denote
	\begin{align*}
	\widetilde I_{n,1}^{(i)}=&\,2\frac{c_{h,q}(L)}{ng}\Iqr{LK\lrp{\frac{1-\bx^T\bX_i}{h^2},\frac{z-Z_i}{g}}\lrp{\E{\hat f_{h,g}(\bx,z)}-f(\bx,z)}}{\bx}{z},\\
	=&\,2n^{-1}LK_{h,g}\lrp{\E{\hat f_{h,g}(\bX_i,Z_i)}-f(\bX_i,Z_i)}
	\end{align*}
	so that $I_{n,1}^{(i)}=\widetilde I_{n,1}^{(i)}-\mathbb{E}\big[\widetilde I_{n,1}^{(i)}\big]$. Therefore, $\mathbb{E}\big[I_{n,1}^{(i)}\big]=0$ and $\widetilde I_{n,1}^{(i)}$ can be decomposed in two addends by virtue of Lemma \ref{gofdens:lemma:biasvar}:
	\begin{align*}
	\widetilde I_{n,1}^{(i)}=&\,2n^{-1}LK_{h,g}\bigg(\frac{b_q(L)}{q}\tr{\bHcal_\bx f(\bX_i,Z_i)}h^2+\frac{\mu_2(K)}{2}\Hcal_z f(\bX_i,Z_i)g^2\bigg)+\order{n^{-1}(h^2+g^2)}\\
	=&\,\widetilde I_{n,1}^{(i,1)}+\widetilde I_{n,1}^{(i,2)}+\order{n^{-1}(h^2+g^2)},
	\end{align*}
	where $\widetilde I_{n,1}^{(i,j)}=\delta_{j}LK_{h,g} \varphi_j(f,\bX_i,Z_i)$ and
	\begin{align*}
	\varphi_j(f,\bx,z)=&\,\lb\begin{array}{ll}\tr{\bHcal_\bx f(\bx,z)},& j=1,\\\Hcal_z f(\bx,z),& j=2,\end{array}\ri\, \delta_{j}=\lb\begin{array}{ll}\frac{2b_q(L)}{q}h^2n^{-1},& j=1,\\\mu_2(K)g^2n^{-1},& j=2.\end{array}\ri
	\end{align*}
	Note that as the order $\order{h^2+g^2}$ is uniform in $(\bx,z)\in\Om{q}\times\R$, then it is possible to extract it from the integrand of $\widetilde I_{n,1}^{(i)}$. Applying Lemma \ref{gofdens:lemma:convunif} to the functions $\varphi_j(f,\cdot,\cdot)$, that by \ref{gofdens:assump:a1} are uniformly continuous and bounded, it yields
	$LK_{h,g}\varphi_j(f,\by,t)\to \varphi_j(f,\by,t)$
	uniformly in $(\by,t)\in\Om{q}\times\R$ as $n\to\infty$. So, for any integers $k_1$ and $k_2$:
	\begin{align*}
	\lim_{n\to\infty}\delta_{1}^{-k_1}\delta_{2}^{-k_2}&\E{\big(\widetilde I_{n,1}^{(i,1)}\big)^{k_1}\big(\widetilde I_{n,1}^{(i,2)}\big)^{k_2}}\\
	=&\,\lim_{n\to\infty}\Iqr{\lrp{LK_{h,g}\varphi_1(f,\by,t)}^{k_1}\lrp{LK_{h,g}\varphi_2(f,\by,t)}^{k_2} f(\by,t)}{\by}{t}\\
	=&\,\Iqr{\varphi_1(f,\by,t)^{k_1}\varphi_2(f,\by,t)^{k_2} f(\by,t)}{\by}{t}\\
	=&\,\E{\varphi_1(f,\bX,Z)^{k_1}\varphi_2(f,\bX,Z)^{k_2}}.
	\end{align*}
	Here the limit can commute with the integral by the Dominated Convergence Theorem (DCT), since the functions $\lrp{LK_{h,g}\varphi_j(f,\by,t)}^k$ are bounded by \ref{gofdens:assump:a1} and the construction of the smoothing operator (\ref{gofdens:smoothing}), being this dominating function integrable:
	\begin{align*}
	\lrp{LK_{h,g}\varphi_1(f,\by,t)}^{k_1}&\lrp{LK_{h,g}\varphi_2(f,\by,t)}^{k_2} f(\by,t)\leq\sup_{(\bx,z)\in\Om{q}\times\R} \abs{\varphi_1(f,\bx,z)^{k_1}\varphi_2(f,\bx,z)^{k_2}} f(\by,t).
	\end{align*} 
	
	Recapitulating, the relation obtained is:
	\begin{align*}
	\E{\big(\widetilde I_{n,1}^{(i,1)}\big)^{k_1}\big(\widetilde I_{n,1}^{(i,2)}\big)^{k_2}}\sim 2^{k_1}n^{-(k_1+k_2)}\frac{b_q(L)^{k_1}}{q^{k_1}}\mu_2(K)^{k_2}h^{2k_1}g^{2k_2}\E{\tr{\bHcal_\bx(f,\bX,Z)}^{k_1}\Hcal_z f(\bX,Z)^{k_2}}.
	\end{align*}
	Now it is easy to prove:
	\begin{align*}
	\E{\big( I_{n,1}^{(i)}\big)^2}%
	\sim&\,\E{\big( \widetilde I_{n,1}^{(i,1)}+\widetilde I_{n,1}^{(i,2)}\big)^2}-\lrp{\E{\widetilde I_{n,1}^{(i,1)}}+\E{\widetilde I_{n,1}^{(i,2)}}}^2\\
	=&\,\E{\big( \widetilde I_{n,1}^{(i,1)}\big)^2}+\E{\big(\widetilde I_{n,1}^{(i,2)}\big)^2}-2\E{\widetilde I_{n,1}^{(i,1)}\widetilde I_{n,1}^{(i,2)}}\\
	&-\E{\widetilde I_{n,1}^{(i,1)}}^2-\E{\widetilde I_{n,1}^{(i,2)}}^2-2\E{\widetilde I_{n,1}^{(i,1)}}\E{\widetilde I_{n,1}^{(i,2)}}\\
	\sim&\,n^{-2}\Bigg(\frac{4b_q(L)^2}{q^2}\V{\tr{\bHcal_\bx(f,\bX,Z)}}h^4+\mu_2(K)^2\V{\Hcal_z f(\bX,Z)}g^4\\
	&+\frac{4b_q(L)\mu_2(K)}{q} \Cov{\tr{\bHcal_\bx(f,\bX,Z)}}{\Hcal_z f(\bX,Z)}h^2g^2\Bigg)\\
	=&\,n^{-2}\phi(h,g).
	\end{align*}
	With the previous expression, it follows $\mathbb{E}\big[\big(I_{n,1}^{(i)}\big)^2\big]=\Order{n^{-2}(h^4+g^4)}$ (see the first point of Lemma \ref{gofdens:lemma:orders}) and $s_n^2=n^{-1}\phi(h,g)(1+\order{1})=\Order{n^{-1}(h^4+g^4)}$. Then by the fourth point of Lemma \ref{gofdens:lemma:orders}:
	\begin{align*}
	s_n^4=&\,\lrp{s_n^2}^2=\Order{n^{-2}(h^4+g^4)^2}=\Order{n^{-2}(h^8+g^8)},\\
	\E{\big(I_{n,1}^{(i)}\big)^4}=&\,\Order{\E{\big(\widetilde I_{n,1}^{(i)}\big)^4}+\E{\widetilde I_{n,1}^{(i)}}^4}=\Order{\E{\big(\widetilde I_{n,1}^{(i)}\big)^4}},
	\end{align*}
	where
	\begin{align*}
	\E{\big(\widetilde I_{n,1}^{(i)}\big)^4}=&\,\Order{\E{\big(\widetilde I_{n,1}^{(i)}\big)^4}+\E{\big(\widetilde I_{n,1}^{(i)}\big)^4}}=\Order{n^{-4}(h^8+g^8)}.
	\end{align*}
\end{proof}

\begin{lem}
	\label{gofdens:lemma:In2}
	Under \ref{gofdens:assump:a1}--\ref{gofdens:assump:a3},
	\begin{align*}
	I_{n,2}=\E{I_{n,2}}+\Orderp{n^{-\frac{3}{2}}h^{-q}g^{-1}}=\frac{\lambda_{q}(L^2)\lambda_{q}(L)^{-2}R(K)}{nh^qg}+\Orderp{n^{-\frac{3}{2}}h^{-q}g^{-1}}. %
	\end{align*}
\end{lem}

\begin{proof}[Proof of Lemma \ref{gofdens:lemma:In2}]
	To prove the result the Chebychev inequality will be used. To that end, the expectation and variance of $I_{n,2}=\frac{c_{h,q}(L)^2}{n^2g^2}\sum_{i=1}^nI_{n,2}^{(i)}$ have to be computed. But first recall that, by Lemma \ref{gofdens:lemma:convunif} and (\ref{gofdens:normalizing}), for $i$ and $j$ naturals,
	\begin{align}
	\Iqr{LK^j\lrp{\frac{1-\bx^T\by}{h^2},\frac{z-t}{g}}\varphi^i(\by,t)}{\by}{t}\sim h^qg\lambda_q(L^j)\varphi^i(\bx,z),\label{gofdens:lemma:In2:p:1}
	\end{align}
	uniformly in $(\bx,z)\in\Om{q}\times\R$, with $\varphi$ a uniformly continuous and bounded function and $\lambda_{q}(L^j)=\om{q-1}2^{\frac{q}{2}-1}\int_0^\infty L^j(r)r^{\frac{q}{2}-1}\,dr$. The following particular cases of this relation are useful to shorten the next computations:
	\begin{enumerate}[label=\textit{\roman{*}}., ref=\textit{\roman{*}}]
		\item $\mathbb{E}\big[LK\big(\frac{1-\bx^T\bX}{h^2},\frac{z-Z}{g}\big)\big]\sim h^qg\lambda_{q}(L) f(\bx,z)$, \label{gofdens:lemma:In2:p:2}
		\item $\Iqr{LK^2\big(\frac{1-\bx^T\by}{h^2},\frac{z-t}{g}\big)}{\bx}{z}\sim h^qg\lambda_{q}(L^2)R(K)$. \label{gofdens:lemma:In2:p:4}
	\end{enumerate}
	
	\textit{Expectation of $I_{n,2}$}. The expectation is divided in two addends, which can be computed by applying the relations \ref{gofdens:lemma:In2:p:2}--\ref{gofdens:lemma:In2:p:4}:
	\begin{align*}
	\E{I_{n,2}^{(i)}}=&\,\E{\Iqr{LK_n^2\lrp{(\bx,z),(\bX,Z)}}{\bx}{z}}\\
	=&\,\Iqr{\E{LK^2\lrp{\frac{1-\bx^T\bX}{h^2},\frac{z-Z}{g}}}}{\bx}{z}\\
	&-\Iqr{\E{LK\lrp{\frac{1-\bx^T\bX}{h^2},\frac{z-Z}{g}}}^2}{\bx}{z}\\
	=&\,\E{\Iqr{LK^2\lrp{\frac{1-\bx^T\bX}{h^2},\frac{z-Z}{g}}}{\bx}{z}}-h^{2q}g^2\lambda_q(L)^2R(f)(1+\order{1})\\
	=&\,h^qg\lambda_q(L^2)R(K)+\Order{h^{2q}g^2}.
	\end{align*}
	Therefore, the expectation of $I_{n,2}$ is 
	\begin{align*}
	\E{I_{n,2}}=&\,\frac{\lambda_q(L)^{-2}}{nh^{2q}g^2}\lrp{h^qg\lambda_{q}(L^2)R(K)+\Order{h^{2q}g^2}}=\frac{\lambda_{q}(L^2)\lambda_q(L)^{-2}R(K)}{nh^{q}g}+\Order{n^{-1}}.
	\end{align*}
	
	\textit{Variance of $I_{n,2}$}. For the variance it suffices to compute its order, which follows considering the third point of Lemma \ref{gofdens:lemma:orders}:
	\begin{align*}
	\E{\big(I_{n,2}^{(i)}\big)^2}=&\,\Iqr{\bigg\{\Iqr{LK_n^2\lrp{(\bx,z),(\by,t)}}{\bx}{z}\bigg\}^2f(\by,t)}{\by}{t}\\
	=&\,\Order{I_{n,2}^{(i,1)}+I_{n,2}^{(i,2)}},%
	\end{align*}
	where the involved terms are
	\begin{align*}
	I_{n,2}^{(i,1)}=&\,\Iqr{\lrb{\Iqr{LK^2\lrp{\frac{1-\bx^T\by}{h^2},\frac{z-t}{g}}}{\bx}{z}}^2f(\by,t)}{\by}{t},\\
	I_{n,2}^{(i,2)}=&\,\Iqr{\Bigg\{\Iqr{\E{LK\lrp{\frac{1-\bx^T\bX}{h^2},\frac{z-Z}{g}}}^2}{\bx}{z}\Bigg\}^2f(\by,t)}{\by}{t}.
	\end{align*}
	
	Using relations \ref{gofdens:lemma:In2:p:2}--\ref{gofdens:lemma:In2:p:4} the orders of the addends $I_{n,2}^{(i,k)}$, $k=1,2$, follow easily:
	\begin{align*}
	I_{n,2}^{(i,1)}\sim&\,\Iqr{\lrb{h^qg\lambda_{q}(L^2)R(K)}^2f(\by,t)}{\by}{t}\\
	=&\,h^{2q}g^2\lambda_{q}(L^2)^2R(K)^2,\\
	I_{n,2}^{(i,2)}\sim&\,\Iqr{\bigg\{\Iqr{\lrp{h^qg\lambda_{q}(L)f(\bx,z)}^2}{\bx}{z}\bigg\}^2f(\by,t)}{\by}{t}\\
	=&\,h^{4q}g^4\lambda_{q}(L)^4R(f)^2.
	\end{align*}
	Therefore $I_{n,2}^{(i,1)}=\Order{h^{2q}g^2}$, $I_{n,2}^{(i,2)}=\Order{h^{4q}g^4}$ and $\mathbb{E}\big[\big(I_{n,2}^{(i)}\big)^2\big]=\mathcal{O}\big(I_{n,2}^{(i,1)}\big)+\mathcal{O}\big(I_{n,2}^{(i,2)}\big)=\Order{h^{2q}g^2}$. The variance of $I_{n,2}$ is
	\begin{align*}
	\V{I_{n,2}} %
	\leq n^{-4}c_{h,q}(L)^{4}g^{-4}\sum_{i=1}^n \E{\big(I_{n,2}^{(i)}\big)^2}
	=\Order{n^{-3}h^{-2q}g^{-2}},
	\end{align*}
	so by Chebychev's inequality %
	\begin{align*}
	\Prob{\abs{I_{n,2}-\E{I_{n,2}}}\geq kn^{-\frac{3}{2}}h^{-q}g^{-1}}\leq \frac{1}{k^2},\quad\forall k>0,
	\end{align*}
	which, by definition, is
	\begin{align*}
	I_{n,2}=\E{I_{n,2}}+\Orderp{n^{-\frac{3}{2}}h^{-q}g^{-1}}	=\frac{\lambda_{q}(L^2)\lambda_q(L)^{-2}R(K)}{nh^{q}g}+\Orderp{n^{-\frac{3}{2}}h^{-q}g^{-1}},
	\end{align*}
	because $\Order{n^{-1}}=\mathcal{O}_\mathbb{P}\big(n^{-\frac{3}{2}}h^{-q}g^{-1}\big)$.
\end{proof}

\begin{lem}
	\label{gofdens:lemma:Hn2}
	Let be
	\begin{align*}
	H_n\lrp{(\bx,z),(\by,t)}=&\,\Iqr{LK_n\lrp{(\bu,v),(\bx,z)}LK_n\lrp{(\bu,v),(\by,t)}}{\bu}{v},\\
	G_n\lrp{(\bx,z),(\by,t)}=&\,\E{H_n\lrp{(\bX,Z),(\bx,z)}H_n\lrp{(\bX,Z),(\by,t)}},\\
	M_n(\bX_1,Z_1)=&\,2\frac{c_{h,q}(L)^2}{n^2g^2}\E{I_{n,1}^{(2)}H_n\lrp{(\bX_1,Z_1),(\bX_2,Z_2)}|(\bX_1,Z_1)}.
	\end{align*}
	Then, under \ref{gofdens:assump:a1}--\ref{gofdens:assump:a3}, 
	\begin{align}
	\E{H_n^2\lrp{(\bX_1,Z_1),(\bX_2,Z_2)}}=&\,h^{3q}g^3\lambda_q(L)^4\sigma^2\lrp{1+\order{1}},\label{gofdens:lemma:Hn2:1}\\
	\E{H_n^4\lrp{(\bX_1,Z_1),(\bX_2,Z_2)}}=&\,\Order{h^{5q}g^5},\label{gofdens:lemma:Hn2:2}
	\\
	\E{G_n^2\lrp{(\bX_1,Z_1),(\bX_2,Z_2)}}=&\,\Order{h^{7q}g^7},\label{gofdens:lemma:Hn2:3}\\
	\E{M_n^2(\bX_1,Z_1)}=&\,\Order{n^{-6}(h^4+g^4)h^{-\frac{3q}{2}}g^{-\frac{3}{2}}}.\label{gofdens:lemma:Hn2:4}
	\end{align}
\end{lem}

\begin{proof}[Proof of Lemma \ref{gofdens:lemma:Hn2}]
	The proof is divided in four sections.\\
	
	\textit{Proof of (\ref{gofdens:lemma:Hn2:1}).} $\mathbb{E}\big[H_n^2\big(\bX_1,Z_1),(\bX_2,Z_2)\big)\big]$ can be split into three addends:
	\begin{align*}
	\mathbb{E}\big[H_n^2\big(&\bX_1,Z_1),(\bX_2,Z_2)\big)\big]\\
	=&\,\E{\bigg(\Iqr{LK_n\lrp{(\bx,z),(\bX_1,Z_1)}LK_n\lrp{(\bx,z),(\bX_2,Z_2)}}{\bx}{z}\bigg)^2}\\
	=&\,\mathbb{E}\Bigg[\Iqr{\Iqr{LK_n\lrp{(\bx,z),(\bX_1,Z_1)}LK_n\lrp{(\bx,z),(\bX_2,Z_2)}\\
			&\times LK_n\lrp{(\by,t),(\bX_1,Z_1)}LK_n\lrp{(\by,t),(\bX_2,Z_2)}}{\bx}{z}}{\by}{t}\Bigg]\\
	=&\,\Iqr{\Iqr{\E{LK_n\lrp{(\bx,z),(\bX,Z)}LK_n\lrp{(\by,t),(\bX,Z)}}^2}{\bx}{z}}{\by}{t}\\
	=&\,\Iqr{\Iqr{\lrp{E_1((\bx,z),(\by,t))-E_2((\bx,z),(\by,t))}^2}{\bx}{z}}{\by}{t}\\
	=&\,A_1-2A_2+A_3,
	\end{align*}
	where:
	\begin{align*}
	E_1((\bx,z),(\by,t))=&\,\E{LK\lrp{\frac{1-\bx^T\bX}{h^2},\frac{z-Z}{g}}LK\lrp{\frac{1-\by^T\bX}{h^2},\frac{t-Z}{g}}},\\
	E_2((\bx,z),(\by,t))=&\,\E{LK\lrp{\frac{1-\bx^T\bX}{h^2},\frac{z-Z}{g}}}\E{LK\lrp{\frac{1-\by^T\bX}{h^2},\frac{t-Z}{g}}}.
	\end{align*}
	The dominant term of the three is $A_1$, which has order $\Order{h^{3q}g^3}$, as it will be seen. The terms $A_2$ and $A_3$ have order $\Order{h^{4q}g^4}$, which can be seen applying iteratively the relation (\ref{gofdens:lemma:In2:p:1}):
	\begin{align*}
	A_2=&\,\Iqr{\Iqr{E_1((\bx,z),(\by,t))E_2((\bx,z),(\by,t))}{\by}{t}}{\bx}{z}\\
	\sim&\Iqr{\Iqr{\lrp{h^qg\lambda_q(L)LK\lrp{\frac{1-\bx^T\bu}{h^2},\frac{z-t}{g}}f(\by,t)}\\
			&\times\lrp{h^{2q}g^2\lambda_q(L)^2f(\bx,z) f(\by,t)}}{\by}{t}}{\bx}{z}\\
	\sim&\,h^{4q}g^4\lambda_q(L)^4\Iqr{f(\bx,z)^3}{\bx}{z},\\
	A_3=&\,\Iqr{\Iqr{ L^2_2((\bx,z),(\by,t))}{\by}{t}}{\bx}{z}\\
	\sim&\,\Iqr{\Iqr{h^{4q}g^4\lambda_q(L)^4f(\bx,z)^2f(\by,t)^2}{\by}{t}}{\bx}{z}\\
	=&\, h^{4q}g^4\lambda_q(L)^4R(f)^2.
	\end{align*}
	
	Let recall now on the term $A_1$. In order to clarify the following computations, let denote by $(\bx,x)$, $(\by,y)$ and $(\bz,z)$ the three variables in $\Om{q}\times\R$ that play the role of $(\bx,z)$, $(\by,t)$ and $(\bu,v)$, respectively. The addend $A_1$ in this new notation is:
	\begin{align*}
	A_1%
	=&\,\Iqr{\Iqr{\Bigg[\Iqr{LK\lrp{\frac{1-\bx^T\bz}{h^2},\frac{x-z}{g}}LK\lrp{\frac{1-\by^T\bz}{h^2},\frac{y-z}{g}}\\
				&\times f(\bz,z)}{\bz}{z}\Bigg]^2}{\by}{y}}{\bx}{x}.
	\end{align*}
	The computation of $A_1$ will be divided in the cases $q\geq2$ and $q=1$. There are several changes of variables involved, which will be detailed in \ref{gofdens:exp1}--\ref{gofdens:exp5}. To begin with, let suppose $q\geq2$:
	\begin{align*}
	A_1\stackrel{\text{\ref{gofdens:exp1}}}{=}&   \,                      \int_{\Om{q}\times\R}\int_{\Om{q-1}}\int_{-1}^{1}\int_\R \Bigg[\int_{\Om{q-2}}\iint_{t^2+\tau^2<1}\int_\R \\
	&\times LK\lrp{\frac{1-t}{h^2},\frac{x-z}{g}}LK\lrp{\frac{1-st-\tau(1-s^2)^\frac{1}{2}}{h^2},\frac{y-z}{g}}\\ &\times f\lrp{t\bx+\tau \bB_{q}\bxi+\lrp{1-t^2-\tau^2}^\frac{1}{2}\bA_{\bxi}\bbeta,z} (1-t^2-\tau^2)^\frac{q-3}{2}\\
	&\times\,dz\,dt\,d\tau\,\om{q-2}(d\bbeta)\Bigg]^2 (1-s^2)^{\frac{q}{2}-1}\,dy\,ds\,\om{q-1}(d\bxi)\,dx\,\om{q}(d\bx)\\
	\stackrel{\text{\ref{gofdens:exp2}}}{=}&\,                         \int_{\Om{q}\times\R}\int_{\Om{q-1}}\int_{0}^{2h^{-2}}\int_\R \bigg[\int_{\Om{q-2}}\int_0^{2h^{-2}}\int_{-1}^1 \int_\R LK\lrp{\rho,\frac{x-z}{g}}\\
	&\times LK\lrp{r+\rho-h^2r\rho-\theta\lrc{r\rho(2-h^2r)(2-h^2\rho)}^\frac{1}{2},\frac{y-z}{g}}\\
	&\times f\lrp{(1-h^2\rho)\bx+h\lrc{\rho(2-h^2\rho)}^\frac{1}{2}\lrc{\theta\bB_{\bx}\bxi+(1-\theta^2)^\frac{1}{2}\bA_{\bxi}\bbeta},z}\\
	&\times (1-\theta^2)^\frac{q-3}{2}h^{q-3}\lrc{\rho(2-h^2\rho)}^\frac{q-3}{2}  h^3\lrc{\rho(2-h^2\rho)}^\frac{1}{2}\,dz\,d\theta\,d\rho\,\om{q-2}(d\bbeta)\bigg]^2\\
	&\times h^{q-2}r^{\frac{q}{2}-1}(2-h^2r)^{\frac{q}{2}-1}h^2\,dy\,dr\,\om{q-1}(d\bxi)\,dx\,\om{q}(d\bx)\\
	\stackrel{\text{\ref{gofdens:exp3}}}{=}& \,                        h^{3q}g^3\int_{\Om{q}\times\R}\int_{\Om{q-1}}\int_{0}^{2h^{-2}}\int_\R \bigg[\int_{\Om{q-2}}\int_0^{2h^{-2}}\int_{-1}^1 \int_\R  LK\lrp{\rho,u}\\
	&\times LK\lrp{r+\rho-h^2r\rho-\theta\lrc{r\rho(2-h^2r)(2-h^2\rho)}^\frac{1}{2},u+v}\\
	&\times f\lrp{(1-h^2\rho)\bx+h\lrc{\rho(2-h^2\rho)}^\frac{1}{2}\lrc{\theta\bB_{\bx}\bxi+(1-\theta^2)^\frac{1}{2}\bA_{\bxi}\bbeta},x-ug}\\
	&\times (1-\theta^2)^\frac{q-3}{2}\lrc{\rho(2-h^2\rho)}^{\frac{q}{2}-1}\,du\,d\theta\,d\rho\,\om{q-2}(d\bbeta)\bigg]^2 r^{\frac{q}{2}-1}(2-h^2r)^{\frac{q}{2}-1}\\
	&\times\,dv\,dr\,\om{q-1}(d\bxi)\,dx\,\om{q}(d\bx)\\
	\stackrel{\text{\ref{gofdens:exp5}}}{\sim}&\,h^{3q}g^3\int_{\Om{q}\times\R}\int_{\Om{q-1}}\int_{0}^{\infty}\int_\R \bigg[\int_{\Om{q-2}}\int_0^{\infty}\int_{-1}^1 \int_\R LK\lrp{\rho,u}\\
	&\times LK\lrp{r+\rho-2\theta\lrp{r\rho}^\frac{1}{2},u+v} f\lrp{\bx,x} (1-\theta^2)^\frac{q-3}{2}\lrp{2\rho}^{\frac{q}{2}-1}\\
	&\times\,du\,d\theta\,d\rho\,\om{q-2}(d\bbeta)\bigg]^2 (2r)^{\frac{q}{2}-1}\,dv\,dr\,\om{q-1}(d\bxi)\,dx\,\om{q}(d\bx)\\
	=&\,h^{3q}g^3R(f)\om{q-1}\om{q-2}^22^{\frac{3q}{2}-1} \int_0^\infty r^{\frac{q}{2}-1}\int_\R\bigg[\int_\R\int_0^\infty \rho^{\frac{q}{2}-1}LK(\rho,u)\\
	&\times\int_{-1}^1(1-\theta^2)^{\frac{q-3}{2}}LK\lrp{r+\rho-2\theta(r\rho)^\frac{1}{2},u+v}\,d\theta\,d\rho\,du\bigg]^2\,dv\,dr\\
	=&\,h^{3q}g^3\lambda_q(L)^4\sigma^2.
	\end{align*}
	The steps for the computation of the case $q\geq 2$ are the following:
	\begin{enumerate}[label=\textit{\roman{*}}., ref=\textit{\roman{*}}]
		\item Let $\bx$ a fixed point in $\Om{q}$, $q\geq 2$. Let be the change of variables: \label{gofdens:exp1} 
		\begin{align*}
		\by=s\bx+(1-s^2)^\frac{1}{2}\bB_{\bx}\bxi,\quad\om{q}(d\by)=(1-s^2)^{\frac{q}{2}-1}\,ds\,\om{q-1}(d\bxi),
		\end{align*}
		where $s\in(-1,1)$, $\bxi\in\Om{q-1}$ and $\bB_{\bx}=(\bb_1,\ldots,\bb_{q})_{(q+1)\times q}$ is the semi-orthonormal matrix ($\bB_{\bx}^T\bB_{\bx}=\bI_q$ and $\bB_{\bx}\bB_{\bx}^T=\bI_{q+1}-\bx\bx^T$) resulting from the completion of $\bx$ to the orthonormal basis $\lrb{\bx,\bb_1,\ldots,\bb_{q}}$ of $\R^{q+1}$. Here $\bI_q$ represents the identity matrix with dimension $q$. See Lemma 2 of \cite{Garcia-Portugues:dirlin} for further details. Consider also the other change of variables
		\begin{align*}
		\bz=t\bx+\tau\bB_{\bx}\bxi+(1-t^2-\tau^2)^\frac{1}{2}\bA_{\bxi}\bbeta,\quad \om{q}(d\bz)=(1-t^2-\tau^2)^\frac{q-3}{2}\,dt\,d\tau\,\om{q-2}(d\bbeta),
		\end{align*}
		where $t,\tau\in(-1,1)$, $t^2+\tau^2<1$, $\bbeta\in\Om{q-2}$ and $\bA_{\bxi}=(\mathbf{a}_1,\ldots,\mathbf{a}_q)_{(q+1)\times (q-1)}$ is the semi-orthonormal matrix ($\bA_{\bxi}^T\bA_{\bxi}=\bI_{q}$ and $\bA_{\bxi}\allowbreak\bA_{\bxi}^T=\bI_{q+1}-\bx\bx^T-\bB_{\bx}\bxi\bxi^T\bB_{\bx}^T$) resulting from the completion of $\lrb{\bx,\bB_{\bx}\bxi}$ to the orthonormal basis $\lrb{\bx,\bB_{\bx}\bxi,\mathbf{a}_1,\ldots,\mathbf{a}_{q-1}}$ of $\R^{q+1}$. This change of variables can be obtained by replicating the proof of Lemma 2 in \cite{Garcia-Portugues:dirlin} with an extra step for the case $q\geq2$. With these two changes of variables,
		\begin{align*}
		\by^T\bz=st+\tau(1-s^2)^\frac{1}{2},\quad \bx^T(\bB_{\bx}\bxi)=\bx^T(\bA_{\bxi}\bbeta)=(\bB_{\bx}\bxi)^T(\bA_{\bxi}\bbeta)=0.
		\end{align*}
		
		\item Consider first the change of variables $r=\frac{1-s}{h^2}$ and then \label{gofdens:exp2}
		\begin{align*}
		\lb\begin{array}{l}
		\rho=\frac{1-t}{h^2}, \\
		\theta=\frac{\tau}{h\lrc{\rho(2-h^2\rho)}^\frac{1}{2}},
		\end{array}\ri
		\quad \abs{\frac{\partial(t,\tau)}{\partial(\rho,\theta)}}=h^3\lrc{\rho(2-h^2\rho)}^\frac{1}{2}.
		\end{align*}
		With this last change of variables, $\tau=h \theta \lrc{\rho(2-h^2\rho)}^\frac{1}{2}$, $t=1-h^2\rho$ and, as a result:
		\begin{align*}
		1-s^2=&\,h^2r(2-h^2r),\\
		1-t^2=&\,h^2\rho(2-h^2\rho),\\
		1-t^2-\tau^2=&\,(1-\theta^2)h^2\rho(2-h^2\rho),\\
		\frac{1-st-\tau(1-s^2)^\frac{1}{2}}{h^2}=&\,r+\rho-h^2r\rho-\theta\lrc{r\rho(2-h^2r)(2-h^2\rho)}^\frac{1}{2}.
		\end{align*}
		\item Use $u=\frac{x-z}{g}$ and $v=\frac{y-x}{g}$. \label{gofdens:exp3}
		
		\item By expanding the square, $A_1$ can be written as \label{gofdens:exp5}
		\begin{align}
		A_1=&\,h^{3q}g^3\int_{\Om{q}\times\R}\int_{\Om{q-1}}\int_{0}^{\infty}\int_\R\bigg[ \int_{\Om{q-2}}\int_0^{\infty}\int_{-1}^1 \int_\R \int_{\Om{q-2}}\int_0^{\infty}\int_{-1}^1 \int_\R \nonumber\\
		&\times \varphi_n(\bx,x,r,\rho_1,\theta_1,u_1,v,\bxi,\bbeta_1) \varphi_n(\bx,x,r,\rho_2,\theta_2,u_2,v,\bxi,\bbeta_2)\nonumber\\
		&\times
		\,du_1\,d\theta_1\,d\rho_1\,\om{q-2}(d\bbeta_1)
		\,du_2\,d\theta_2\,d\rho_2\,\om{q-2}(d\bbeta_2)\bigg]\nonumber\\
		&\times\,dv\,dr\,\om{q-1}(d\bxi)\,dx\,\om{q}(d\bx),\label{gofdens:A1:aux}
		\end{align}
		where
		\begin{align*}
		\varphi_n(\bx,x,r,\rho_i,&\theta_i,u_i,v,\bxi,\bbeta_i)\\
		=&\,L\lrp{\rho_i} L\lrp{r+\rho_i-h^2r\rho_i-\theta\lrc{r\rho_i(2-h^2r)(2-h^2\rho_i)}^\frac{1}{2}}\\
		&\times K\lrp{u_i} K\lrp{u_i+v} f\lrp{(\bx,x)+\ba_{h,g}} (1-\theta_i^2)^\frac{q-3}{2}\\
		&\times \rho_i^{\frac{q}{2}-1}(2-h^2\rho_i)^{\frac{q}{2}-1} r^{\frac{q}{4}-\frac{1}{2}}(2-h^2r)^{\frac{q}{4}-\frac{1}{2}}\mathbbm{1}_{[0,2h^{-2})}(r)\mathbbm{1}_{[0,2h^{-2})}(\rho_i),
		\end{align*}
		with $\ba_{h,g}=\Big(-h^2\rho_i\bx+h\lrc{\rho_i(2-h^2\rho_i)}^\frac{1}{2}\big[\theta\bB_{\bx}\bxi+(1-\theta_i^2)^\frac{1}{2}\bA_{\bxi}\bbeta_i\big],\allowbreak-u_ig\Big)$ and $i=1,2$. A first step to apply the DCT is to see that by the Taylor's theorem,
		\begin{align*}
		f\lrp{(\bx,x)+\ba_{h,g}}=f(\bx,x)+\Order{\ba_{h,g}^T\bnab f(\bx,x)},
		\end{align*}
		where the remaining order is $\mathcal{O}\big((h^2\rho_i+g^2u_i^2)^\frac{1}{2}\norm{\bnab f(\bx,x)}\big)$ because $\norm{\ba_{h,g}}^2=2h^2\rho_i+g^2u_i^2$. Furthermore, the order is uniform for all points $(\bx,x)$ because of the boundedness assumption of the second derivative given by \ref{gofdens:assump:a1} (see the proof of Lemma \ref{gofdens:lemma:biasvar}). Next, as $h,g\to0$, then the order becomes $\order{(\sqrt{\rho_i}+u_i)\norm{\bnab f(\bx,x)}}$.\\
		
		For bounding the directional kernel $L$, recall that by completing the square,
		\begin{align*}
		(2-h^2r)(2-h^2\rho_i)=&\,4-2h^2(r+\rho_i)+h^4\lrp{(r+\rho_i)/2}^2-h^4\lrp{\lrp{(r+\rho_i)/2)}^2-r\rho_i}\\
		\leq&\,\lrp{2-h^2\frac{r+\rho_i}{2}}^2.
		\end{align*}
		Using this, and the fact that $\theta\in(-1,1)$, for all $r,\rho_i\in[0,2h^{-2})$,
		\begin{align*}
		r+\rho_i-h^2r\rho_i&-\theta\lrc{r\rho_i(2-h^2r)(2-h^2\rho_i)}^\frac{1}{2}\\
		&\geq r+\rho_i-h^2r\rho_i-(r\rho_i)^\frac{1}{2}\lrc{(2-h^2r)(2-h^2\rho_i)}^\frac{1}{2}\\
		&\geq r+\rho_i-h^2r\rho_i-(r\rho_i)^\frac{1}{2}\lrp{2-h^2\frac{r+\rho_i}{2}}\\
		&= r+\rho_i-2(r\rho_i)^\frac{1}{2}+h^2(r\rho_i)^\frac{1}{2}\lrp{\frac{r+\rho_i}{2}-(r\rho_i)^\frac{1}{2}}\\
		&\geq r+\rho_i-2(r\rho_i)^\frac{1}{2},
		\end{align*}
		where the last inequality follows because the last addend is positive by the inequality of the geometric and arithmetic means. As $L$ is a decreasing function by \ref{gofdens:assump:a2},
		\begin{align*}
		L\bigg(r+\rho_i-h^2r\rho_i- \theta\lrc{r\rho_i(2-h^2r)(2-h^2\rho_i)}^\frac{1}{2}\bigg)\leq L\lrp{r+\rho_i-2(r\rho_i)^\frac{1}{2}}.
		\end{align*}
		Then for all the variables in the integration domain of $A_1$,
		\begin{align*}
		\varphi_n(\bx,x,r,\rho_i,\theta_i,u_i,v,\bxi,\bbeta_i)\leq&\,L\lrp{\rho_i} L\lrp{r+\rho_i-2(r\rho_i)^\frac{1}{2}}K\lrp{u_i} K\lrp{u_i+v}\\
		&\times  \lrp{f(\bx,x)+\order{(\sqrt{\rho_i}+u_i)\norm{\bnab f(\bx,x)}}} (1-\theta_i^2)^\frac{q-3}{2}\\
		&\times \rho_i^{\frac{q}{2}-1}2^{\frac{q}{2}-1} r^{\frac{q}{4}-\frac{1}{2}}2^{\frac{q}{4}-\frac{1}{2}}\mathbbm{1}_{[0,\infty)}(r)\mathbbm{1}_{[0,\infty)}(\rho_i)\\
		=&\,\Psi(\bx,x,r,\rho_i,\theta_i,u_i,v).
		\end{align*}
		
		The product of functions $\varphi_n$ in (\ref{gofdens:A1:aux}) is bounded by the respective product of functions $\Psi$. The product is also integrable as a consequence of assumptions \ref{gofdens:assump:a1} (integrability of $f$ and $\bnab f$), \ref{gofdens:assump:a2} (integrability of kernels) and that the product of integrable functions is integrable. To prove it, recall that by the integral definition of the modified Bessel function of order $\frac{q}{2}-1$ (see equation 10.32.2 of \cite{Olver2010}):
		\begin{align*}
		\int_{-1}^{1} (1-\theta^2)^\frac{q-3}{2}\,d\theta=\frac{\sqrt{\pi } \Gamma\lrp{\frac{q-1}{2}}}{\Gamma\lrp{\frac{q}{2}}}<\infty,\quad \forall q\geq2.
		\end{align*}
		The integral of the linear kernel is proved to be finite using the Cauchy--Schwartz inequality and \ref{gofdens:assump:a2}:
		\begin{align*}
		\int_{\R}\int_{\R}\int_{\R}K(u_1)K(u_1+v)&K(u_2)K(u_2+v)\,du_1\,du_2\,dv\\
		=&\,\int_{\R}\int_{\R}K(u_1)K(u_2)\lrc{\int_{\R}K(u_1+v)K(u_2+v)\,dv}\,du_1\,du_2\\
		\leq&\,\int_{\R}\int_{\R}K(u_1)K(u_2)\mu_2(K)^\frac{1}{2}\mu_2(K)^\frac{1}{2}\,du_1\,du_2\\
		=&\,\mu_2(K).
		\end{align*}
		For the directional situation, the following auxiliary result based on \ref{gofdens:assump:a2} is needed: 
		\begin{align*}
		\int_0^\infty L^2\lrp{\big(\sqrt{r}-\sqrt{\rho_i}\big)^2}r^{\frac{q}{2}-1}\,dr\leq &\,\int_0^\infty L^2\lrp{s}\lrp{\sqrt{s}+\sqrt{\rho_i}}^{q-1}s^{-\frac{1}{2}}\,dr\\
		=&\,\int_0^\infty L^2\lrp{s}\sum_{k=0}^{q-1}s^\frac{k-1}{2}\rho_i^\frac{q-1-k}{2}\,dr\\
		=&\,\sum_{k=0}^{q-1}\lambda_{k+1}(L^2)\rho_i^\frac{q-1-k}{2}\\
		=&\,\Order{\rho_i^\frac{q-1}{2}}.
		\end{align*}
		Using this and that $\int_0^\infty L(\rho) \rho^{\frac{3q-5}{4}}\,dr\leq \lambda_{\left\lceil\frac{2q+1}{3}\right\rceil}(L)<\infty$, it follows:
		\begin{align*}
		\int_0^\infty\int_0^\infty&\int_0^\infty L(\rho_1)L(\rho_2) L\lrp{r+\rho_1-2(r\rho_1)^\frac{1}{2}}L\lrp{r+\rho_2-2(r\rho_2)^\frac{1}{2}}\\
		&\times \rho_1^{\frac{q}{2}-1}\rho_2^{\frac{q}{2}-1}r^{\frac{q}{2}-1}\,dr\,d\rho_1\,d\rho_2\\
		=&\,\int_0^\infty\int_0^\infty L(\rho_1)L(\rho_2) \rho_1^{\frac{q}{2}-1}\rho_2^{\frac{q}{2}-1}\\
		&\times\lrc{\int_0^\infty L\lrp{\big(\sqrt{r}-\sqrt{\rho_1}\big)^2}L\lrp{\big(\sqrt{r}-\sqrt{\rho_2}\big)^2} r^{\frac{q}{2}-1}\,dr}\,d\rho_1\,d\rho_2\\
		\leq&\,\int_0^\infty\int_0^\infty L(\rho_1)L(\rho_2) \rho_1^{\frac{q}{2}-1}\rho_2^{\frac{q}{2}-1}\Order{\rho_1^\frac{q-1}{4}}\Order{\rho_2^\frac{q-1}{4}}\,d\rho_1\,d\rho_2\\
		=&\,\Order{1}.
		\end{align*}
		
		Then, by the DCT,
		\begin{align*}
		A_1\sim&\,h^{3q}g^3\int_{\Om{q}\times\R}\int_{\Om{q-1}}\int_{0}^{\infty}\int_\R \bigg[\int_{\Om{q-2}}\int_0^{\infty}\int_{-1}^1 \int_\R \\
		&\times LK\lrp{\rho,u}LK\lrp{r+\rho-2\theta\lrp{r\rho}^\frac{1}{2},u+v}\\
		&\times f\lrp{\bx,x} (1-\theta^2)^\frac{q-3}{2}\lrp{2\rho}^{\frac{q}{2}-1}\,du\,d\theta\,d\rho\,\om{q-2}(d\bbeta)\bigg]^2\\ &\times(2r)^{\frac{q}{2}-1}\,dv\,dr\,\om{q-1}(d\bxi)\,dx\,\om{q}(d\bx),
		\end{align*}
		because all the functions involved are continuous almost everywhere.
	\end{enumerate}
	
	Turn now to the case $q=1$. As before, the details of the case $q=1$ are explained in \ref{gofdens:exp1b}--\ref{gofdens:exp5b}:
	\begin{align*}
	A_1\stackrel{\text{\ref{gofdens:exp1b}}}{=}& \,                        \int_{\Om{1}\times\R}\int_{\Om{0}}\int_{-1}^{1}\int_\R \Bigg[\int_{\Om{0}}\int_{-1}^{1}\int_\R LK\lrp{\frac{1-t}{h^2},\frac{x-z}{g}}\\
	&\times LK\lrp{\frac{1-st-(1-t^2)^\frac{1}{2}(1-s^2)^\frac{1}{2}(\bB_{\bx}\bxi)^T(\bA_{\bx}\bbeta)}{h^2},\frac{y-z}{g}}\\
	&\times f\lrp{t\bx+(1-t^2)^\frac{1}{2}\bA_{\bx}\bbeta,z}
	(1-t^2)^{-\frac{1}{2}}\,dz\,dt\,\om{0}(d\bbeta)\Bigg]^2\\
	&\times  (1-s^2)^{-\frac{1}{2}}\,dy\,ds\,\om{0}(d\bxi)\,dx\,\om{1}(d\bx)\\
	\stackrel{\text{\ref{gofdens:exp2b}}}{=}&  \,                       \int_{\Om{1}\times\R}\int_{\Om{0}}\int_{0}^{2h^{-2}}\int_\R \bigg[\int_{\Om{0}}\int_0^{2h^{-2}} \int_\R LK\lrp{\rho,\frac{x-z}{g}}\\
	&\times  LK\lrp{r+\rho-h^2r\rho-\lrp{r\rho(2-h^2r)(2-h^2\rho)}^\frac{1}{2}(\bB_{\bx}\bxi)^T\bA_{\bx}\bbeta,\frac{y-z}{g}}\\
	&\times f\lrp{(1-h^2\rho)\bx+h\lrc{\rho(2-h^2\rho)}^\frac{1}{2}\bA_{\bx}\bbeta,z} h^{-1}\rho^{-\frac{1}{2}}(2-h^2\rho)^{-\frac{1}{2}}h^2\\
	&\times\,dz\,d\rho\,\om{0}(d\bbeta)\bigg]^2h^{-1}r^{-\frac{1}{2}}(2-h^2r)^{-\frac{1}{2}}h^2\,dy\,dr\,\om{0}(d\bxi)\,dx\,\om{1}(d\bx)\\
	\stackrel{\text{\ref{gofdens:exp3b}}}{=}&\,                         h^3g^3\int_{\Om{1}\times\R}\int_{\Om{0}}\int_{0}^{2h^{-2}}\int_\R \bigg[\int_{\Om{0}}\int_0^{2h^{-2}} \int_\R LK\lrp{\rho,u}\\
	&\times  LK\lrp{r+\rho-h^2r\rho-\lrp{r\rho(2-h^2r)(2-h^2\rho)}^\frac{1}{2}(\bB_{\bx}\bxi)^T\bA_{\bx}\bbeta,u+v}\\
	&\times f\lrp{(1-h^2\rho)\bx+h\lrc{\rho(2-h^2\rho)}^\frac{1}{2}\bA_{\bx}\bbeta,x-ug} \\
	&\times \rho^{-\frac{1}{2}}(2-h^2\rho)^{-\frac{1}{2}}\,du\,d\rho\,\om{0}(d\bbeta)\bigg]^2r^{-\frac{1}{2}}(2-h^2r)^{-\frac{1}{2}}\\
	&\times\,dv\,dr\,\om{0}(d\bxi)\,dx\,\om{1}(d\bx)\\
	\stackrel{\text{\ref{gofdens:exp1b}}}{=}& \,                        h^3g^3\int_{\Om{1}\times\R}\int_{\Om{0}}\int_{0}^{2h^{-2}}\int_\R \bigg[\int_0^{2h^{-2}} \int_\R LK\lrp{\rho,u}\\
	&\times\bigg[ LK\lrp{r+\rho-h^2r\rho+\lrp{r\rho(2-h^2r)(2-h^2\rho)}^\frac{1}{2},u+v}\\
	&\times f\lrp{(1-h^2\rho)\bx+h\lrc{\rho(2-h^2\rho)}^\frac{1}{2}\bB_{\bx}\bxi,x-ug}\\
	&+LK\lrp{r+\rho-h^2r\rho-\lrp{r\rho(2-h^2r)(2-h^2\rho)}^\frac{1}{2},u+v}\\
	&\times f\lrp{(1-h^2\rho)\bx-h\lrc{\rho(2-h^2\rho)}^\frac{1}{2}\bB_{\bx}\bxi,x-ug}\bigg]\\
	&\times \rho^{-\frac{1}{2}}(2-h^2\rho)^{-\frac{1}{2}}\,du\,d\rho\bigg]^2r^{-\frac{1}{2}}(2-h^2r)^{-\frac{1}{2}}\,dv\,dr\,\om{0}(d\bxi)\,dx\,\om{1}(d\bx)\\
	\stackrel{\text{\ref{gofdens:exp5b}}}{\sim}&\,h^3g^3\int_{\Om{1}\times\R}\int_{\Om{0}}\int_{0}^{\infty}\int_\R \bigg[\int_0^{\infty} \int_\R LK\lrp{\rho,u}\\
	&\times\Big[ LK\lrp{r+\rho+2(r\rho)^\frac{1}{2},u+v}+LK\lrp{r+\rho-2(r\rho)^\frac{1}{2},u+v}\Big] \\
	&\times f\lrp{\bx,x}\rho^{-\frac{1}{2}}2^{-\frac{1}{2}}\,du\,d\rho\bigg]^2r^{-\frac{1}{2}}2^{-\frac{1}{2}}\,dv\,dr\,\om{0}(d\bxi)\,dx\,\om{1}(d\bx)\\
	=&\,h^3g^3R(f)2^{-\frac{1}{2}} \int_{0}^{\infty}r^{-\frac{1}{2}}\int_\R \bigg[\int_0^{\infty} \int_\R \rho^{-\frac{1}{2}}LK\lrp{\rho,u}\\
	&\times\Big[ LK\lrp{r+\rho+2(r\rho)^\frac{1}{2},u+v}+LK\lrp{r+\rho-2(r\rho)^\frac{1}{2},u+v}\Big]\\
	&\times \,du\,d\rho\bigg]^2\,dv\,dr\\
	=&\,h^{3q}g^3\lambda_q(L)^4\sigma^2.
	\end{align*}
	
	The steps used for the computation are the following:
	\begin{enumerate}[label=\textit{\roman{*}}., ref=\textit{\roman{*}}]
		\setcounter{enumi}{5}
		
		\item Let $\bx$ a fixed point in $\Om{q}$. \label{gofdens:exp1b}
		For $q=1$, let be the changes of variables
		\begin{align*}
		\by=&\,s\bx+(1-s^2)^\frac{1}{2}\bB_{\bx}\bxi,\quad\om{1}(d\by)=(1-s^2)^{\frac{q}{2}-1}\,ds\,\om{0}(d\bxi),\\
		\bz=&\,t\bx+(1-t^2)^\frac{1}{2}\bA_{\bx}\bbeta,\quad\om{1}(d\bz)=(1-t^2)^{\frac{q}{2}-1}\,dt\,\om{0}(d\bbeta),
		\end{align*}
		where $s,t\in(-1,1)$ and $\bB_{\bx}$ and $\bA_{\bx}$ are two semi-orthonormal matrices whose $q$ columns are  vectors that extend $\bx$ to an orthonormal basis of $\R^{q+1}$. Note that as $q=1$ and $\bx^T (\bB_{\bx}\bxi)=\bx^T (\bA_{\bx}\bbeta)=0$, then necessarily $\bB_{\bx}\bxi=\bA_{\bx}\bbeta$ or $\bB_{\bx}\bxi=-\bA_{\bx}\bbeta$.
		
		\item Let be the changes of variables $\rho=\frac{1-t}{h^2}$ and $r=\frac{1-s}{h^2}$. With this change, $t=1-h^2\rho$ and $s=1-h^2r$. Then $1-s^2=h^2r(2-h^2r)$, $1-t^2=h^2\rho(2-h^2\rho)$ and \label{gofdens:exp2b}
		\begin{align*}	
		&\frac{1-st-(1-s^2)^\frac{1}{2}(1-t^2)^\frac{1}{2}(\bB_{\bx}\bxi)^T\bA_{\bx}\bbeta}{h^2}\\
		&\qquad\qquad\qquad\qquad  =r+\rho-h^2r\rho-\lrp{r\rho(2-h^2r)(2-h^2\rho)}^\frac{1}{2}(\bB_{\bx}\bxi)^T\bA_{\bx}\bbeta.
		\end{align*}
		
		\item Use $u=\frac{x-z}{g}$ and $v=\frac{y-x}{g}$. \label{gofdens:exp3b}
		
		\item $A_1$ can be written as \label{gofdens:exp5b}
		\begin{align*}
		A_1%
		=&\,h^{3q}g^3\int_{\Om{q}\times\R}\int_{\Om{q-1}}\int_{0}^{\infty}\int_\R\bigg[  \int_0^{\infty} \int_\R \int_0^{\infty} \int_\R  \varphi_n(\bx,x,r,\rho_1,u_1,v,\bxi)\\
		&\times \varphi_n(\bx,x,r,\rho_2,u_2,v,\bxi)\,du_1\,d\rho_1\,du_2\,d\rho_2\bigg] \,dv\,dr\,\om{q-1}(d\bxi)\,dx\,\om{q}(d\bx),
		\end{align*}
		where
		\begin{align*}
		\varphi_n(\bx,x,r,\rho_i,u_i,v,\bxi)=&\,L\lrp{\rho_i}K\lrp{u_i} \bigg[L\lrp{r+\rho_i-h^2r\rho_i+\lrc{r\rho_i(2-h^2r)(2-h^2\rho_i)}^\frac{1}{2}}\\
		&\times K\lrp{u_i+v} f\lrp{(\bx,x)+\ba^{(1)}_{h,g}} + K\lrp{u_i+v} f\lrp{(\bx,x)+\ba^{(2)}_{h,g}} \\
		&\times L\lrp{r+\rho_i-h^2r\rho_i-\lrc{r\rho_i(2-h^2r)(2-h^2\rho_i)}^\frac{1}{2}} \bigg]\\
		&\times \rho_i^{-\frac{1}{2}}(2-h^2\rho_i)^{-\frac{1}{2}} r^{-\frac{1}{4}}(2-h^2r)^{-\frac{1}{4}}\mathbbm{1}_{[0,2h^{-2})}(r)\mathbbm{1}_{[0,2h^{-2})}(\rho_i),
		\end{align*}
		with $\ba^{(j)}_{h,g}=\Big(-h^2\rho_i\bx+k_jh\lrc{\rho_i(2-h^2\rho_i)}^\frac{1}{2}\bB_{\bx}\bxi,-u_ig\Big)$ and $k_1=1$, $k_2=-1$. As before, by the Taylor's theorem,
		\begin{align*}
		f\lrp{(\bx,x)+\ba^{(k)}_{h,g}}=f(\bx,x)+\order{(\sqrt{\rho_i}+u_i)\norm{\bnab f(\bx,x)}},
		\end{align*}
		where the order is uniform for all points $(\bx,x)$. By analogous considerations as for the case $q\geq 2$,
		\begin{align*}
		\varphi_n(\bx,x,r,\rho_i,u_i,v,\bxi)\leq&\, 2L\lrp{\rho_i} L\lrp{r+\rho_i-2(r\rho_i)^\frac{1}{2}}K\lrp{u_i} K\lrp{u_i+v}\Big(f(\bx,x)\\
		&+\order{(\sqrt{\rho_i}+u_i)\norm{\bnab f(\bx,x)}}\Big)\rho_i^{-\frac{1}{2}}r^{-\frac{1}{4}}\mathbbm{1}_{[0,\infty)}(r)\mathbbm{1}_{[0,\infty)}(\rho_i)\\
		=&\,\Psi(\bx,x,r,\rho_i,u_i,v),
		\end{align*}
		Then the product of functions $\varphi_n$ is bounded by the respective product of functions $\Psi$, which is integrable, and by the DCT the limit commute with the integrals.
	\end{enumerate}
	
	\textit{Proof of (\ref{gofdens:lemma:Hn2:2}).} $\E{H_n^4\lrp{(\bX_1,Z_1),(\bX_2,Z_2)}}$ can be decomposed in the sum of two terms:
	\begin{align*}
	\mathbb{E}\big[H_n^4\big(\bX_1,&Z_1),(\bX_2,Z_2)\big)\big]\\
	=&\,\E{\bigg(\Iqr{LK_n\lrp{(\bx,z),(\bX_1,Z_1)}LK_n\lrp{(\bx,z),(\bX_2,Z_2)}}{\bx}{z}\bigg)^4}\\
	=&\,\Iqr{\Iqr{\lrp{E_1((\bx,z),(\by,t))-E_2((\bx,z),(\by,t))}^4}{\bx}{z}}{\by}{t}\\
	=&\,\Order{B_1+B_2}.
	\end{align*}
	The computation of the orders of these terms is analogous to the ones of $A_2$ and $A_3$:
	\begin{align*}
	B_1=&\,\Iqr{\Iqr{\lrp{E_1((\bx,z),(\by,t))}^4}{\by}{t}}{\bx}{z}\\
	\sim&\Iqr{\Iqr{\lrp{h^qg\lambda_q(L)LK\lrp{\frac{1-\bx^T\bu}{h^2},\frac{z-t}{g}}f(\by,t)}^4}{\by}{t}}{\bx}{z}\\
	\sim&\,h^{5q}g^5\lambda_q(L)^4\lambda_q(L^4)R(f^2),\\
	B_2=&\,\Iqr{\Iqr{\lrp{E_2((\bx,z),(\by,t))}^4}{\by}{t}}{\bx}{z}\\
	\sim&\,\Iqr{\Iqr{ h^{8q}g^8\lambda_q(L)^8f(\bx,z)^4f(\by,t)^4}{\by}{t}}{\bx}{z}\\
	=&\, h^{8q}g^8\lambda_q(L)^8R(f^2)^2.
	\end{align*}
	Then $\E{H_n^4\lrp{(\bX_1,Z_1),(\bX_2,Z_2)}}=\Order{h^{5q}g^5}$.\\
	
	\textit{Proof of (\ref{gofdens:lemma:Hn2:3}).} The notation $(\bx,x)$, $(\by,y)$, $(\bz,z)$ and $(\bu,u)$ for variables in $\Om{q}\times\R$ will be employed again:
	\begin{align*}
	G_n((\bx,x),(\by,y))=&\,\Iqr{H_n\lrp{(\bz,z),(\bx,x)}H_n\lrp{(\bz,z),(\by,y)}f(\bz,z)}{\bz}{z}\\
	=&\,\Iqr{\bigg\{\Iqr{LK_n((\bu,u),(\bx,x))LK_n((\bu,u),(\bz,z))}{\bu}{u}\bigg\}\\
		&\times\bigg\{\Iqr{LK_n((\bu,u),(\by,y))LK_n((\bu,u),(\bz,z))}{\bu}{u}\bigg\}\\
		&\times f(\bz,z)}{\bz}{z}.
	\end{align*}
	Therefore:
	\begin{align*}
	\mathbb{E}\big[G_n^2((\bX_1,&Z_1),(\bX_2,Z_2))\big]\\%
	=&\,\Iqr{\Iqr{\Bigg\{\Iqr{\lrc{\Iqr{LK_n((\bu,u),(\bx,x))LK_n((\bu,u),(\bz,z))}{\bu}{u}}\\
				&\times\lrc{\Iqr{LK_n((\bu,u),(\by,y))LK_n((\bu,u),(\bz,z))}{\bu}{u}} f(\bz,z)}{\bz}{z}\Bigg\}^2\\
				&\times f(\by,y)f(\bx,x)}{\by}{y}}{\bx}{x}.
	\end{align*}
	
	Then, according to the expression of $LK_n$, $\E{G_n^2\lrp{(\bX_1,Z_1),(\bX_2,Z_2)}}$ can be decomposed in 16 summands, which, in view of the symmetric roles of $(\bx,x)$ and $(\by,y)$ can be reduced to 9 different summands. The first of all, $C_1$, is the dominant and has order $\Order{h^{7q}g^7}$. Again, the orders are computed using (\ref{gofdens:lemma:In2:p:1}) iteratively:
	\begin{align*}
	C_1=&\,\Iqr{\Iqr{\Bigg\{\Iqr{\\
				&\times\lrc{\Iqr{LK\lrp{\frac{1-\bu^T\bx}{h^2},\frac{u-x}{g}}LK\lrp{\frac{1-\bu^T\bz}{h^2},\frac{u-z}{g}}}{\bu}{u}}\\
				&\times\lrc{\Iqr{LK\lrp{\frac{1-\bu^T\by}{h^2},\frac{u-y}{g}}LK\lrp{\frac{1-\bu^T\bx}{h^2},\frac{u-x}{g}}}{\bu}{u}}\\
				&\times  f(\bz,z)}{\bz}{z}\Bigg\}^2f(\by,y)f(\bx,x)}{\by}{y}}{\bx}{x}\\
	\sim&\,\Iqr{\Iqr{\Bigg\{\Iqr{\lrc{\lambda_q(L)h^qg LK\lrp{\frac{1-\bx^T\bz}{h^2},\frac{x-z}{g}}}\\
				&\times \lrc{\lambda_q(L)h^qg LK\lrp{\frac{1-\by^T\bz}{h^2},\frac{y-z}{g}}}f(\bz,z)}{\bz}{z}\Bigg\}^2\\
				&\times f(\by,y)f(\bx,x)}{\by}{y}}{\bx}{x}\\
	\sim&\,\lambda_q(L)^4h^{4q}g^4 \Iqr{\Iqr{\Bigg\{ \lambda_q(L)h^qg LK\lrp{\frac{1-\by^T\bx}{h^2},\frac{y-x}{g}}f(\bx,x)\Bigg\}^2\\
			&\times f(\by,y)f(\bx,x) }{\by}{y}}{\bx}{x}\\
	\sim&\,\lambda_q(L)^6h^{6q}g^6 \Iqr{\lambda_q(L^2)R(K)h^qg f(\bx,x)f(\bx,x)^3}{\bx}{x}\\
	=&\,\lambda_q(L)^6\lambda_q(L^2)R(K)h^{7q}g^7 R(f^2).
	\end{align*}
	The rest of them have order $\Order{h^{8q}g^8}$, something which can be seen by iteratively applying the Lemma \ref{gofdens:lemma:orders} as before.\\
	
	\textit{Proof of (\ref{gofdens:lemma:Hn2:4}).} It suffices to apply the tower property, the Cauchy--Schwartz inequality, result $\mathbb{E}\big[\big(I_{n,1}^{(2)}\big)^4\big]=\Order{n^{-4}(h^8+g^8)}$ from Lemma \ref{gofdens:lemma:In1} and (\ref{gofdens:lemma:Hn2:2}):
	\begin{align*}
	\E{M_n^2(\bX_1,Z_1)}=&\,4\frac{c_{h,q}(L)^4}{n^4g^4}\E{\E{I_{n,1}^{(2)}H_n\lrp{(\bX_1,Z_1),(\bX_2,Z_2)}\big|(\bX_1,Z_1)}^2}\\
	\leq&\,4\frac{c_{h,q}(L)^4}{n^4g^4}\E{\big(I_{n,1}^{(2)}\big)^2H_n^2\lrp{(\bX_1,Z_1),(\bX_2,Z_2)}}\\
	\leq&\,4\frac{c_{h,q}(L)^4}{n^4g^4} \E{\big(I_{n,1}^{(2)}\big)^4}^\frac{1}{2}\E{H_n^4\lrp{(\bX_1,Z_1),(\bX_2,Z_2)}}^\frac{1}{2}\\
	=&\,\Order{\lrp{nh^qg}^{-4}}\Order{n^{-4}(h^8+g^8)}^\frac{1}{2}\Order{h^{5q}g^5}^\frac{1}{2}\\
	=&\,\Order{n^{-6}(h^4+g^4)h^{-\frac{3q}{2}}g^{-\frac{3}{2}}}.
	\end{align*}
\end{proof}

%-------------------------------------------------%
\subsection{Testing independence with directional data}
%-------------------------------------------------%

\begin{lem}
	\label{gofdens:lem:indep:1}
	Under \ref{gofdens:assump:a1}--\ref{gofdens:assump:a3}, 
	\[
	n(h^qg)^{\frac{1}{2}}\lrp{T_{n,1}-\frac{R(K)\lambda_{q}(L^2)\lambda_{q}(L)^{-2}}{nh^qg}}\stackrel{d}{\longrightarrow}\mathcal{N}\lrp{0,2\sigma^2}.%
	\]
\end{lem}

\begin{proof}[Proof of Lemma \ref{gofdens:lem:indep:1}]
	
	By the decomposition of $I_n$ in the proof of the Theorem \ref{gofdens:theo:clt}, $T_{n,1}=I_{n,2}+I_{n,3}$ and therefore by (\ref{gofdens:res:In2}) and (\ref{gofdens:res:In3}), 
	\begin{align*}
	T_{n,1}=&\,\E{I_{n,2}}+\Orderp{n^{-\frac{3}{2}}h^{-q}g^{-1}}+2^\frac{1}{2}\sigma n^{-1}(h^qg)^{-\frac{1}{2}}N_{n},
	\end{align*}
	where $N_n$ is asymptotically a normal. On the other hand, by (\ref{gofdens:res:In2}),
	\begin{align*}
	\E{I_{n,2}}=\frac{\lambda_{q}(L^2)\lambda_{q}(L)^{-2}R(K)}{nh^qg}+\Order{n^{-1}}
	\end{align*}
	and then
	\begin{align*}
	T_{n,1}=&\,\frac{\lambda_{q}(L^2)\lambda_{q}(L)^{-2}R(K)}{nh^qg}+2^\frac{1}{2}\sigma n^{-1}(h^qg)^{-\frac{1}{2}}N_{n}+\Orderp{n^{-\frac{3}{2}}h^{-q}g^{-1}}, %
	\end{align*}
	because $\big(n^\frac{3}{2}h^{q}g\big)^{-1}=\mathpzc{o}\big((nh^\frac{q}{2}g^\frac{1}{2})^{-1}\big)$. As the last addend is asymptotically negligible compared with the second,
	\[
	n(h^qg)^{\frac{1}{2}}\lrp{T_{n,1}-\frac{R(K)\lambda_{q}(L^2)\lambda_{q}(L)^{-2}}{nh^qg}}\stackrel{d}{\longrightarrow}\mathcal{N}\lrp{0,2\sigma^2}.%
	\]
\end{proof}

\begin{lem}
	\label{gofdens:lem:indep:2}
	Under independence and \ref{gofdens:assump:a1}--\ref{gofdens:assump:a3}, 
	\begin{align*}
	\E{T_{n,2}}=&\,\frac{\lambda_{q}(L^2)\lambda_{q}(L)^{-2}R(f_Z)}{nh^q}+\frac{R(K)R(f_\bX)}{ng}+\order{n^{-1}(h^{-q}+g^{-1})},\\
	\V{T_{n,2}}=&\,\Order{n^{-2}(h^{-q}+g^{-1})}.
	\end{align*}
\end{lem}

\begin{proof}[Proof of Lemma \ref{gofdens:lem:indep:2}]
	
	The term $T_{n,2}$ can be decomposed using the relation
	\begin{align*}
	\hat f_{h}(\bx) \hat f_{g}(z)-\E{\hat f_{h}(\bx)}\E{\hat f_{g}(z)}=S_1(\bx,z)+S_2(\bx,z)+S_3(\bx,z),
	\end{align*}
	where:
	\begin{align*}
	S_1(\bx,z)=&\,\lrp{\hat f_{h}(\bx)-\E{\hat f_{h}(\bx)}}\lrp{\hat f_{g}(z)-\E{\hat f_{g}(z)}},\\
	S_2(\bx,z)=&\,\lrp{\hat f_{h}(\bx)-\E{\hat f_{h}(\bx)}}\E{\hat f_{g}(z)},\\
	S_3(\bx,z)=&\,\lrp{\hat f_{g}(z)-\E{\hat f_{g}(z)}}\E{\hat f_{h}(\bx)}.
	\end{align*}
	Hence,
	\begin{align*}
	T_{n,2}=&\,\Iqr{S_1^2(\bx,z)}{\bx}{z}+\Iqr{S_2^2(\bx,z)}{\bx}{z}\\
	&+\Iqr{S_3^2(\bx,z)}{\bx}{z}+2\Iqr{S_1(\bx,z)S_2(\bx,z)}{\bx}{z}\\
	&+2\Iqr{S_1(\bx,z)S_3(\bx,z)}{\bx}{z}+2\Iqr{S_2(\bx,z)S_3(\bx,z)}{\bx}{z}\\
	=&\,T^{(1)}_{n,2}+T^{(2)}_{n,2}+T^{(3)}_{n,2}+T^{(4)}_{n,2}+T^{(5)}_{n,2}+T^{(6)}_{n,2}.
	\end{align*}
	
	To compute the expectation of each addend under independence, use the variance and expectation expansions for the directional and linear estimator (see for example \cite{Garcia-Portugues:dirlin} for both) and relation (\ref{gofdens:normalizing}). Recall that due to \ref{gofdens:assump:a1} it is possible to consider Taylor expansions on the marginal densities that have uniform remaining orders.
	\begin{align*}
	\E{T^{(1)}_{n,2}}=&\,\Iq{\V{\hat f_{h}(\bx)}}{\bx}\Ir{\V{\hat f_{g}(z)}}{z}\\
	=&\,\Order{(n^2h^qg)^{-1}},\\
	\E{T^{(2)}_{n,2}}=&\,\Iq{\V{\hat f_{h}(\bx)}}{\bx}\Ir{\E{\hat f_{g}(z)}^2}{z}\\
	=&\,\lrc{\frac{\lambda_{q}(L^2)\lambda_{q}(L)^{-2}}{nh^q}+\Order{n^{-1}}}\lrc{R(f_Z)+\order{1}}\\
	=&\,\frac{\lambda_{q}(L^2)\lambda_{q}(L)^{-2}R(f_Z)}{nh^q}+\order{\lrp{nh^q}^{-1}},\\
	\E{T^{(3)}_{n,2}}=&\,\Ir{\V{\hat f_{g}(z)}}{z}\Iq{\E{\hat f_{h}(\bx)}^2}{\bx}\\
	=&\,\lrc{\frac{R(K)}{ng}+\Order{n^{-1}}}\lrc{R(f_\bX)+\order{1}}\\
	=&\,\frac{R(K)R(f_\bX)}{ng}+\order{\lrp{ng}^{-1}}.
	\end{align*}
	The expectation of $T^{(4)}_{n,2}$, $T^{(5)}_{n,2}$ and $T^{(6)}_{n,2}$ is zero because of the separability of the directional and linear components. Joining these results,
	\begin{align*}
	\E{T_{n,2}}=\frac{\lambda_{q}(L^2)\lambda_{q}(L)^{-2}R(f_Z)}{nh^q}+\frac{R(K)R(f_\bX)}{ng}+\order{n^{-1}(h^{-q}+g^{-1})},
	\end{align*}
	because $\lrp{n^2h^qg}^{-1}=\order{n^{-1}\lrp{h^{-q}+g^{-1}}}$.\\
	
	Computing the variance is not so straightforward as the expectation and some extra results are needed. First of all, recall that by the formula of the variance of the sum, the Cauchy--Schwartz inequality and Lemma \ref{gofdens:lemma:orders},
	\begin{align*}
	\V{T_{n,2}}=\V{\sum_{i=1}^6 T^{(i)}_{n,2}}=\sum_{i=1}^6\Order{\V{T^{(i)}_{n,2}}}.
	\end{align*}
	Then the variance of each addend will be computed separately. For that purpose, recall that by the decomposition of the ISE given in Theorem \ref{gofdens:theo:clt},
	\[
	\Iqr{\lrp{\hat f_{h,g}(\bx,z)-\E{\hat f_{h,g}(\bx,z)}}^2}{\bx}{z}=I_{n,2}+I_{n,3},
	\]
	so by equations (\ref{gofdens:res:In2}) and (\ref{gofdens:var:In3}),
	\begin{align*}
	\V{I_{n,2}+I_{n,3}}=&\,\Order{\V{I_{n,2}}+\V{I_{n,3}}}\\
	=&\,\Order{(n^3h^qg)^{-1}+(n^2h^qg)^{-1}}\\
	=&\,\Order{(n^2h^qg)^{-1}},\\
	\E{\lrp{I_{n,2}+I_{n,3}}^2}=&\,\V{I_{n,2}+I_{n,3}}+\E{I_{n,2}+I_{n,3}}^2\\
	=&\,\Order{(n^2h^qg)^{-1}+(nh^qg)^{-2}}\\
	=&\,\Order{(nh^qg)^{-2}}.
	\end{align*}
	The marginal directional and linear versions of these relations will be required:
	\begin{align*}
	\E{\bigg(\Iq{\lrp{\hat f_h(\bx)-\E{\hat f_h(\bx)}}^2}{\bx}\bigg)^2}=&\,\Order{(nh^q)^{-2}},\\
	\E{\lrp{\Ir{\lrp{\hat f_g(z)-\E{\hat f_g(z)}}^2}{z}}^2}=&\,\Order{(ng)^{-2}},\\
	\V{\Iq{\lrp{\hat f_h(\bx)-\E{\hat f_h(\bx)}}^2}{\bx}}=&\,\Order{(n^2h^q)^{-1}},\\
	\V{\Ir{\lrp{\hat f_g(z)-\E{\hat f_g(z)}}^2}{z}}=&\,\Order{(n^2g)^{-1}}.
	\end{align*}
	Then:
	\begin{align*}
	\V{T^{(1)}_{n,2}}\leq&\, \E{\big(T^{(1)}_{n,2}\big)^2}\\
	=&\,\E{\bigg(\Iq{\lrp{\hat f_h(\bx)-\E{\hat f_h(\bx)}}^2}{\bx}\bigg)^2} \E{\lrp{\Ir{\lrp{\hat f_g(z)-\E{\hat f_g(z)}}^2}{z}}^2}\\
	=&\,\Order{(nh^q)^{-2}}\Order{(ng)^{-2}}\\
	=&\,\Order{n^{-4}h^{-2q}g^{-2}},\\
	\V{T^{(2)}_{n,2}}=&\,\V{\!\bigg(\!\Iq{\!\lrp{\hat f_h(\bx)-\E{\hat f_h(\bx)}}^2\!}{\bx}\bigg)\!\bigg(\Ir{\E{\hat f_g(z)}^2\!}{z}\bigg)}\\
	=&\,\lrp{\Ir{\E{\hat f_g(z)}^2}{z}}^2\V{\Iq{\lrp{\hat f_h(\bx)-\E{\hat f_h(\bx)}}^2}{\bx}}\\
	=&\,\Order{1}\Order{(n^2h^q)^{-1}}\\
	=&\,\Order{(n^2h^q)^{-1}},\\
	\V{T^{(3)}_{n,2}}=&\,\V{\!\bigg(\!\Ir{\lrp{\hat f_g(z)-\E{\hat f_g(z)}}^2\!}{z}\bigg)\!\bigg(\Iq{\!\E{\hat f_h(\bx)}^2\!}{\bx}\bigg)}\\
	=&\,\bigg(\Iq{\E{\hat f_h(\bx)}^2}{\bx}\bigg)^2\V{\Ir{\lrp{\hat f_g(z)-\E{\hat f_g(z)}}^2}{z}}\\
	=&\,\Order{1}\Order{(n^2g)^{-1}}\\
	=&\,\Order{(n^2g)^{-1}}.
	\end{align*}
	The next results follows from applying iteratively Cauchy--Schwartz and the previous orders:
	\begin{align*}
	\V{T^{(4)}_{n,2}}\leq&\, \E{\big(T^{(4)}_{n,2}\big)^2}\\
	\leq&\,\E{\bigg(\Iqr{S_1^2(\bx,z)}{\bx}{z}\bigg)\bigg(\Iqr{S_2^2(\by,t)}{\by}{t}\bigg)}\\
	\leq&\,\E{\big(T^{(1)}_{n,2}\big)^2}^\frac{1}{2}\E{\big(T^{(2)}_{n,2}\big)^2}^\frac{1}{2}\\
	=&\,\Order{n^{-3}h^{-2q}},\\
	\V{T^{(5)}_{n,2}}\leq&\,\E{\big(T^{(1)}_{n,2}\big)^2}^\frac{1}{2}\E{\big(T^{(3)}_{n,2}\big)^2}^\frac{1}{2}\\
	=&\,\Order{n^{-3}g^{-2}},\\
	\V{T^{(6)}_{n,2}}\leq&\,\E{\big(T^{(2)}_{n,2}\big)^2}^\frac{1}{2}\E{\big(T^{(3)}_{n,2}\big)^2}^\frac{1}{2}\\
	=&\,\Order{n^{-2}}.
	\end{align*}
	Therefore, the order of $\V{T_{n,2}}$ is $\Order{n^{-2}(h^{-q}+g^{-1})}$ since it dominates $\Order{n^{-4}h^{-2q}g^{-2}}$,\linebreak $\Order{n^{-3}(h^{-2q}+g^{-2})}$ and $\Order{n^{-2}}$ by \ref{gofdens:assump:a3}.
\end{proof}
\begin{lem}
	\label{gofdens:lem:indep:3}
	Under independence and \ref{gofdens:assump:a1}--\ref{gofdens:assump:a3}, 
	$\E{T_{n,3}}=-2\E{T_{n,2}}$ and $\V{T_{n,3}}=\mathcal{O}\big(n^{-2}(h^{-q}\allowbreak+g^{-1})\big)$.
\end{lem}
\begin{proof}[Proof of Lemma \ref{gofdens:lem:indep:3}]
	The term $T_{n,3}$ can be split in a similar fashion to $T_{n,2}$. Let denote
	\begin{align*}
	S_4(\bx,z)=\hat f_{h,g}(\bx,z)-\E{\hat f_{h,g}(\bx,z)}.
	\end{align*}
	Then:
	\begin{align*}
	T_{n,3}=&\,-2\Iqr{S_4(\bx,z)\lrp{S_1(\bx,z)+S_2(\bx,z)+S_3(\bx,z)}}{\bx}{z}\\
	=&\,-2\lrp{T^{(1)}_{n,3}+T^{(2)}_{n,3}+T^{(3)}_{n,3}}.
	\end{align*}
	The key idea now is to use that, under independence,
	\begin{align}
	LK_n\lrp{(\bx,z),(\bX,Z)}=&\,L_n\lrp{\bx,\bX}K_n\lrp{z,Z}+L_n\lrp{\bx,\bX}\E{K\lrp{\frac{z-Z}{g}}}\nonumber\\
	&+K_n\lrp{z,Z}\E{L\lrp{\frac{1-\bx^T\bX}{h^2}}},\label{gofdens:lem:indep:3:1}
	\end{align}
	where $L_n$ and $K_n$ are the marginal versions of $LK_n$:
	\begin{align*}
	L_n\lrp{\bx,\by}=L\lrp{\frac{1-\bx^T\by}{h^2}}-\E{L\lrp{\frac{1-\bx^T\bX}{h^2}}},\quad K_n\lrp{z,t}=K\lrp{\frac{z-t}{g}}-\E{K\lrp{\frac{z-Z}{g}}}.
	\end{align*}
	By repeated use of (\ref{gofdens:lem:indep:3:1}) in the integrands of $T_{n,3}$ and applying the Fubini theorem, it follows:
	\begin{align*}
	\E{T^{(1)}_{n,3}}%
	=&\,\frac{c_{h,q}(L)^2}{n^2g^2}\Iqr{\E{LK_n\lrp{(\bx,z),(\bX,Z)}L_n\lrp{\bx,\bX}K_n\lrp{z,Z}}}{\bx}{z}\\
	=&\,\frac{c_{h,q}(L)^2}{n^2g^2}\Iqr{\mathbb{E} \Bigg[L_n\lrp{\bx,\bX}^2K_n\lrp{z,Z}^2+L_n\lrp{\bx,\bX}^2K_n\lrp{z,Z}\E{K\lrp{\frac{z-Z}{g}}}\\
		&+L_n\lrp{\bx,\bX}K_n\lrp{z,Z}^2\E{L\lrp{\frac{1-\bx^T\bX}{h^2}}}\Bigg]}{\bx}{z}\\
	=&\,\frac{c_{h,q}(L)^2}{n^2g^2}\Iqr{\E{L_n\lrp{\bx,\bX}^2}\E{K_n\lrp{z,Z}^2}}{\bx}{z}\\
	=&\,\Iqr{\E{S_1(\bx,z)^2}}{\bx}{z}\\
	=&\,\E{T^{(1)}_{n,2}},\\
	\E{T^{(2)}_{n,3}}%
	=&\,\frac{c_{h,q}(L)^2}{n^2g^2}\Iqr{\mathbb{E}\bigg[LK_n\lrp{(\bx,z),(\bX,Z)}L_n\lrp{\bx,\bX}\E{K\lrp{\frac{z-Z}{g}}}\bigg]}{\bx}{z}\\
	=&\,\frac{c_{h,q}(L)^2}{n^2g^2}\Iqr{\mathbb{E} \Bigg[L_n\lrp{\bx,\bX}^2K_n\lrp{z,Z}\E{K\lrp{\frac{z-Z}{g}}}\\
		&+L_n\lrp{\bx,\bX}K_n\lrp{z,Z} \E{K\lrp{\frac{z-Z}{g}}}\E{L\lrp{\frac{1-\bx^T\bX}{h^2}}}\\
		&+L_n\lrp{\bx,\bX}^2\E{K\lrp{\frac{z-Z}{g}}}^2\Bigg]}{\bx}{z}\\
	=&\,\frac{c_{h,q}(L)^2}{n^2g^2}\Iqr{\E{L_n\lrp{\bx,\bX}^2}\E{K\lrp{\frac{z-Z}{g}}}^2}{\bx}{z}\\
	=&\,\Iqr{\E{S_2(\bx,z)^2}}{\bx}{z}\\
	=&\,\E{T^{(2)}_{n,2}},\\
	\E{T^{(3)}_{n,3}}%
	=&\,\frac{c_{h,q}(L)^2}{n^2g^2}\Iqr{\mathbb{E}\Bigg[LK_n\lrp{(\bx,z),(\bX,Z)}K_n\lrp{z,Z}\E{L\lrp{\frac{1-\bx^T\bX}{h^2}}}\Bigg]}{\bx}{z}\\
	=&\,\frac{c_{h,q}(L)^2}{n^2g^2}\Iqr{\mathbb{E} \Bigg[L_n\lrp{\bx,\bX}K_n\lrp{z,Z}\E{L\lrp{\frac{1-\bx^T\bX}{h^2}}}\\
		&+L_n\lrp{\bx,\bX}K_n\lrp{z,Z}\E{L\lrp{\frac{1-\bx^T\bX}{h^2}}}\E{K\lrp{\frac{z-Z}{g}}}\\
		&+K_n\lrp{z,Z}^2 \E{L\lrp{\frac{1-\bx^T\bX}{h^2}}}^2\Bigg]}{\bx}{z}\\
	=&\,\frac{c_{h,q}(L)^2}{n^2g^2}\Iqr{\E{K_n\lrp{z,Z}^2}\E{L\lrp{\frac{1-\bx^T\bX}{h^2}}}^2}{\bx}{z}\\
	=&\,\Iqr{\E{S_3(\bx,z)^2}}{\bx}{z}\\
	=&\,\E{T^{(3)}_{n,2}}.
	\end{align*}
	Then $\E{T_{n,3}}=-2\E{T_{n,2}}$.\\
	
	Computing the variance is much more tedious: the order obtained by bounding the variances by repeated use of the Cauchy--Schwartz inequality is not enough. Instead of, a laborious decomposition of the term $T_{n,3}$ has to be done in order to compute separately the variance of each addend, by following the steps of \cite{Rosenblatt1992}. The first step is to split the variance using the Cauchy--Schwartz inequality and Lemma \ref{gofdens:lemma:orders}:
	\begin{align*}
	\V{T_{n,3}}=&\,\Order{\V{T^{(1)}_{n,3}}+\V{T^{(2)}_{n,3}}+\V{T^{(3)}_{n,3}}}.
	\end{align*}
	Each of the three terms will be also decomposed into other addends. To simplify their computation the following notation will be employed:
	\begin{align*}
	CLK_n((\bx_1,&z_1),(\bX_1,Z_1);(\bx_2,z_2),(\bX_2,Z_2))\\
	=&\,\Cov{LK\lrp{\frac{1-\bx_1^T\bX_1}{h^2},\frac{z_1-Z_1}{g}}}{LK\lrp{\frac{1-\bx_2^T\bX_2}{h^2},\frac{z_2-Z_2}{g}}}
	\end{align*}
	and also its marginal versions:
	\begin{align*}
	CL_n(\bx_1,\bX_1;\bx_2,\bX_2)=&\,\Cov{L\lrp{\frac{1-\bx_1^T\bX_1}{h^2}}}{L\lrp{\frac{1-\bx_2^T\bX_2}{h^2}}},\\
	CK_n(z_1,Z_1;z_2,Z_2)=&\,\Cov{K\lrp{\frac{z_1-Z_1}{g}}}{K\lrp{\frac{z_2-Z_2}{g}}}.
	\end{align*}
	
	\textit{Term $T^{(2)}_{n,3}$.} To begin with, let examine $T^{(2)}_{n,3}$ using the notation of $LK_n$, $L_n$ and $K_n$:
	\begin{align*}
	T^{(2)}_{n,3}=&\,\frac{c_{h,q}(L)^2}{n^{2}g^2}\sum_{i=1}^n\sum_{j=1}^n\Iqr{LK_n((\bx,z),(\bX_i,Z_i))L_n(\bx,\bX_j)\E{K\lrp{\frac{z-Z}{g}}}}{\bx}{z}.
	\end{align*}
	where the double summation can be split into two summations (a single sum plus the sum of the cross terms). Then, \begin{align*}
	\V{T^{(2)}_{n,3}}=\frac{c_{h,q}(L)^4}{n^{4}g^4}\Order{n\V{T^{(2,1)}_{n,3}}+n^2\V{T^{(2,2)}_{n,3}}},
	\end{align*}
	where:
	\begin{align*}
	T^{(2,1)}_{n,3}=&\,\Iqr{LK_n((\bx,z),(\bX,Z))L_n(\bx,\bX)\E{K\lrp{\frac{z-Z}{g}}}}{\bx}{z},\\
	T^{(2,2)}_{n,3}=&\,\Iqr{LK_n((\bx,z),(\bX_1,Z_1))L_n(\bx,\bX_2)\E{K\lrp{\frac{z-Z}{g}}}}{\bx}{z}.
	\end{align*}
	The first term is computed by 
	\begin{align*}
	\V{T^{(2,1)}_{n,3}}\leq&\,\E{\big(T^{(2,1)}_{n,3}\big)^2}\\
	=&\,\Order{g^2}\E{\bigg(\Iqr{LK_n((\bx,z),(\bX,Z))L_n(\bx,\bX)}{\bx}{z}\bigg)^2}\\
	=&\,\Order{g^2}\mathbb{E}\left[\Bigg(\Iqr{\lrc{LK\lrp{\frac{1-\bx^T\bX}{h^2},\frac{z-Z}{g}}-\Order{h^qg}}^{\mbox{}}\right.\\
		&\times\left.\lrc{L\lrp{\frac{1-\bx^T\bX}{h^2}}-\Order{h^q}} }{\bx}{z}\Bigg)^2\,\right]\\
	=&\,\Order{g^2} \bigg(\int_{\Om{q-1}}\int_{0}^{2h^{-2}}\Ir{\lrc{LK\lrp{r,t}-\Order{h^qg}}\lrc{L\lrp{r}-\Order{h^q}}\\
		&\times h^q(2-h^2r)^{\frac{q}{2}-1}r^{\frac{q}{2}-1}g}{z}\,dr\,\om{q-1}(d\bxi)\bigg)^2 \\
	=&\,\Order{h^{2q}g^4},
	\end{align*}
	where the second equality follows from $\mathbb{E}\big[LK\big(\frac{1-\bx^T\bX}{h^2},\frac{z-Z}{g}\big)\big]=\Order{h^qg}$ and $\mathbb{E}\big[L\big(\frac{1-\bx^T\bX}{h^2}\big)\big]=\Order{h^q}$, and the third from applying the changes of variables of the proof of Lemma \ref{gofdens:lemma:Hn2}. The second addend\nolinebreak[4] is
	\begin{align*}
	\V{T^{(2,2)}_{n,3}}\leq&\,\mathbb{E}\Bigg[\Iqr{\Iqr{LK_n((\bx_1,z_1),(\bX_1,Z_1))L_n(\bx_1,\bX_2)\\
			&\times\E{K\lrp{\frac{z_1-Z}{g}}} LK_n((\bx_2,z_2),(\bX_1,Z_1))L_n(\bx_2,\bX_2)\\
			&\times\E{K\lrp{\frac{z_2-Z}{g}}} }{\bx_1}{z_1}}{\bx_2}{z_2}\Bigg]\\
	=&\,\Iqr{\Iqr{ CLK_n((\bx_1,z_1),(\bX_1,Z_1);(\bx_2,z_2),(\bX_1,Z_1))\\
			&\times CL_n(\bx_1,\bX_2;\bx_2,\bX_2) \E{K\lrp{\frac{z_1-Z}{g}}}\E{K\lrp{\frac{z_2-Z}{g}}}\\
			&\times  }{\bx_1}{z_1}}{\bx_2}{z_2}\\
	\leq &\, \bigg(\Iqr{\Iqr{ CLK_n((\bx_1,z_1),(\bX_1,Z_1);(\bx_2,z_2),(\bX_1,Z_1))\\
			&\times}{\bx_1}{z_1}}{\bx_2}{z_2}\bigg)\Order{h^qg^2},
	\end{align*}
	because $CL_n(\bx_1,\bX_2;\bx_2,\bX_2)=\Order{h^q}$ by Cauchy--Schwartz and the directional version of Lemma \ref{gofdens:lemma:biasvar}, and $\mathbb{E}\big[K\big(\frac{z-Z}{g}\big)\big]=\Order{g}$. Also, the integral of the covariance is
	\begin{align*}
	\Iqr{\Iqr{&CLK_n((\bx_1,z_1),(\bX_1,Z_1);(\bx_2,z_2),(\bX_1,Z_1))}{\bx_1}{z_1}}{\bx_2}{z_2}\\
	=&\,\Iqr{\Iqr{\mathbb{E}\Bigg[LK\lrp{\frac{1-\bx_1^T\bX_1}{h^2},\frac{z_1-Z_1}{g}} LK\lrp{\frac{1-\bx_2^T\bX_1}{h^2},\frac{z_2-Z_1}{g}}\Bigg]\\
			&\times}{\bx_1}{z_1}}{\bx_2}{z_2}-\Order{h^{2q}g^2}\\
	=&\,\Iqr{\Iqr{\Iqr{LK\lrp{\frac{1-\bx_1^T\by}{h^2},\frac{z_1-t}{g}} LK\lrp{\frac{1-\bx_2^T\by}{h^2},\frac{z_2-t}{g}}\\
				&\times f(\by,t)}{\by}{t}}{\bx_1}{z_1}}{\bx_2}{z_2}-\Order{h^{2q}g^2}\\
	=&\,\Order{h^{2q}g^2},
	\end{align*}
	as it follows that the order of the first addend is $\Order{h^{2q}g^2}$ by applying \ref{gofdens:exp1}--\ref{gofdens:exp5b} in the same way as in the computation of $A_1$ in Lemma \ref{gofdens:lemma:Hn2} (recall that the square in $A_1$ is not present here and therefore the order is larger). Then $\mathbb{V}\mathrm{ar}\big[T^{(2,2)}_{n,3}\big]=\Order{h^{3q}g^4}$ and as a consequence,
	\begin{align}
	\V{T^{(2)}_{n,3}}=\frac{c_{h,q}(L)^4}{n^{4}g^4}\Order{nh^{2q}g^4+n^2h^{3q}g^4}=\Order{n^{-2}h^{-q}}.\label{gofdens:VarTn3_2}
	\end{align}
	
	\textit{Term $T^{(3)}_{n,3}$.} This addend follows analogously from $T^{(2)}_{n,3}$, as the only difference is the swapping of the roles of the directional and linear components: 
	\begin{align*}
	T^{(3)}_{n,3}=&\,\frac{c_{h,q}(L)^2}{n^{2}g^2}\sum_{i=1}^n\sum_{j=1}^n\Iqr{LK_n((\bx,z),(\bX_i,Z_i))K_n(z,Z_j)\E{L\lrp{\frac{1-\bx^T\bX}{h^2}}}}{\bx}{z},
	\end{align*}
	with the same decomposition that gives
	\begin{align*}
	\V{T^{(3)}_{n,3}}=\frac{c_{h,q}(L)^4}{n^{4}g^4}\Order{n\V{T^{(3,1)}_{n,3}}+n^2\V{T^{(3,2)}_{n,3}}},
	\end{align*}
	where:
	\begin{align*}
	T^{(3,1)}_{n,3}=&\,\Iqr{\!LK_n((\bx,z),(\bX,Z))K_n(z,Z) \E{L\lrp{\frac{1-\bx^T\bX}{h^2}}}\!}{\bx}{z},\\
	T^{(3,2)}_{n,3}=&\,\Iqr{\!LK_n((\bx,z),(\bX_1,Z_1))K_n(z,Z_2) \E{L\lrp{\frac{1-\bx^T\bX}{h^2}}}\!}{\bx}{z}.
	\end{align*}
	Then, by similar computations to those of $T^{(3)}_{n,3}$, $\mathbb{V}\mathrm{ar}\big[T^{(3,1)}_{n,3}\big]=\Order{h^{4q}g^2}$,
	$\mathbb{V}\mathrm{ar}\big[T^{(3,2)}_{n,3}\big]=\Order{h^{4q}g^3}$ and
	\begin{align}
	\V{T^{(3)}_{n,3}}=\frac{c_{h,q}(L)^4}{n^{4}g^4}\Order{nh^{4q}g^2+n^2h^{4q}g^3}=\Order{n^{-2}g^{-1}}.\label{gofdens:VarTn3_3}
	\end{align}
	
	\textit{Term $T^{(1)}_{n,3}$.} This is the hardest part, as it presents more combinations. As with the previous terms,
	\begin{align*}
	T^{(1)}_{n,3}=&\,\frac{c_{h,q}(L)^2}{n^{3}g^2}\sum_{i=1}^n\sum_{j=1}^n\sum_{k=1}^n\Iqr{LK_n((\bx,z),(\bX_i,Z_i)) L_n(\bx,\bX_j)K_n(z,Z_k)}{\bx}{z}.
	\end{align*}
	and now the triple summation can be split into five summations \begin{align*}
	\V{T^{(1)}_{n,3}}=&\,\frac{c_{h,q}(L)^4}{n^{6}g^4}\mathcal{O}\Big( n\V{T^{(1,1)}_{n,3}}+n^2\lrp{\V{T^{(1,2a)}_{n,3}}+\V{T^{(1,2b)}_{n,3}}+\V{T^{(1,2c)}_{n,3}}}\\
	&+n^3\V{T^{(1,3)}_{n,3}}\Big),
	\end{align*}
	where:
	\begin{align*}
	T^{(1,1)}_{n,3}=&\,\Iqr{LK_n((\bx,z),(\bX_1,Z_1))L_n(\bx,\bX_1)K_n(z,Z_1)}{\bx}{z},\\
	T^{(1,2a)}_{n,3}=&\,\Iqr{LK_n((\bx,z),(\bX_1,Z_1))L_n(\bx,\bX_2)K_n(z,Z_2)}{\bx}{z},\\
	T^{(1,2b)}_{n,3}=&\,\Iqr{LK_n((\bx,z),(\bX_1,Z_1))L_n(\bx,\bX_1)K_n(z,Z_2)}{\bx}{z},\\
	T^{(1,2c)}_{n,3}=&\,\Iqr{LK_n((\bx,z),(\bX_1,Z_1))L_n(\bx,\bX_2)K_n(z,Z_1)}{\bx}{z},\\
	T^{(1,3)}_{n,3}=&\,\Iqr{LK_n((\bx,z),(\bX_1,Z_1))L_n(\bx,\bX_2)K_n(z,Z_3)}{\bx}{z}.
	\end{align*}
	
	The first term is computed by 
	\begin{align*}
	\mathbb{V}\mathrm{ar}\Big[T^{(1,1)}_{n,3}\Big]\leq&\,\E{\bigg(\Iqr{LK_n((\bx,z),(\bX,Z))L_n(\bx,\bX)K_n(z,Z)}{\bx}{z}\bigg)^2}\\
	=&\,\mathbb{E}\Bigg[\bigg(\Iqr{\lrc{LK\lrp{\frac{1-\bx^T\bX}{h^2},\frac{z-Z}{g}}-\Order{h^qg}}\\
		&\times\lrc{L\lrp{\frac{1-\bx^T\bX}{h^2}}-\Order{h^q}}\lrc{K\lrp{\frac{z-Z}{g}}-\Order{g}} }{\bx}{z}\bigg)^2\Bigg]\\
	=&\,\bigg(\int_{\Om{q-1}}\!\int_{0}^{2h^{-2}}\!\!\!\!\Ir{\lrc{LK\lrp{r,t}-\Order{h^qg}}\lrc{L\lrp{r}-\Order{h^q}}\\
		&\times\lrc{K\lrp{t}-\Order{g}} h^q(2-h^2r)^{\frac{q}{2}-1}r^{\frac{q}{2}-1}g\,dr\,\om{q-1}(d\bxi)}{z}\bigg)^2 \\
	=&\,\Order{h^{2q}g^2},
	\end{align*}
	by the same arguments as for $T^{(2,1)}_{n,3}$. The fifth addend is
	\begin{align*}
	\mathbb{V}\mathrm{ar}\Big[T^{(1,3)}_{n,3}\Big]\leq&\,\mathbb{E}\Bigg[\Iqr{\Iqr{LK_n((\bx_1,z_1),(\bX_1,Z_1))L_n(\bx_1,\bX_2)K_n(z_1,Z_3)\\
			&\times LK_n((\bx_2,z_2),(\bX_1,Z_1))L_n(\bx_2,\bX_2)K_n(z_2,Z_3) }{\bx_1}{z_1}}{\bx_2}{z_2}\Bigg]\\
	=&\,\Iqr{\Iqr{CLK_n((\bx_1,z_1),(\bX_1,Z_1);(\bx_2,z_2),(\bX_1,Z_1))\\
			&\times CL_n(\bx_1,\bX_2;\bx_2,\bX_2) CK_n(z_1,Z_3;z_2,Z_3) }{\bx_1}{z_1}}{\bx_2}{z_2}\\
	\leq &\, \Order{h^qg}\Iqr{\Iqr{CLK_n((\bx_1,z_1),(\bX_1,Z_1);(\bx_2,z_2),(\bX_1,Z_1))\\
			&\times}{\bx_1}{z_1}}{\bx_2}{z_2}\\
	\leq &\, \Order{h^{3q}g^3},
	\end{align*}
	again by the same arguments used for $T^{(2,2)}_{n,3}$. It only remains to obtain the variance of $T^{(1,2a)}_{n,3}$, $T^{(1,2b)}_{n,3}$ and $T^{(1,2c)}_{n,3}$. The first one arises from
	\begin{align*}
	\mathbb{V}\mathrm{ar}\Big[T^{(1,2a)}_{n,3}\Big]\leq&\,\mathbb{E}\Bigg[\Iqr{\!\Iqr{\!LK_n((\bx_1,z_1),(\bX_1,Z_1))L_n(\bx_1,\bX_2)K_n(z_1,Z_2)\\
			&\times LK_n((\bx_2,z_2),(\bX_1,Z_1))L_n(\bx_2,\bX_2)K_n(z_2,Z_2)}{\bx_1}{z_1}}{\bx_2}{z_2}\Bigg]\\
	=&\,\Iqr{\!\Iqr{\!CLK_n((\bx_1,z_1),(\bX_1,Z_1);(\bx_2,z_2),(\bX_1,Z_1))\\
			&\times CL_n(\bx_1,\bX_2;\bx_2,\bX_2) CK_n(z_1,Z_2;z_2,Z_2) }{\bx_1}{z_1}}{\bx_2}{z_2}\\
	=&\,\Order{h^{3q}g^3},
	\end{align*}
	in virtue of the assumption of independence and the computation of $\mathbb{V}\mathrm{ar}\big[T^{(1,1)}_{n,3}\big]$. The second one is
	\begin{align*}
	\mathbb{V}\mathrm{ar}\Big[T^{(1,2b)}_{n,3}\Big]\leq&\,\mathbb{E}\Bigg[\Iqr{\!\Iqr{\!LK_n((\bx_1,z_1),(\bX_1,Z_1))L_n(\bx_1,\bX_1)K_n(z_1,Z_2)\\
			&\times LK_n((\bx_2,z_2),(\bX_1,Z_1))L_n(\bx_2,\bX_1)K_n(z_2,Z_2) }{\bx_1}{z_1}}{\bx_2}{z_2}\Bigg]\\
	=&\,\Iqr{\!\Iqr{\!\mathbb{E}\big[LK_n((\bx_1,z_1),(\bX_1,Z_1))LK_n((\bx_2,z_2),(\bX_1,Z_1))\\
			&\times L_n(\bx_1,\bX_1)L_n(\bx_2,\bX_1)\big] CK_n(z_1,Z_2;z_2,Z_2)}{\bx_1}{z_1}}{\bx_2}{z_2}\\
	=&\,\Order{g} \E{\bigg(\!\Iqr{\!\!LK_n((\bx,z),(\bX_1,Z_1))L_n(\bx,\bX_1)}{\bx}{z}\bigg)^2}\\
	=&\,\Order{h^{2q}g^3},
	\end{align*}
	where the order of the expectation is obtained again using the change of variables described in the proof of Lemma \ref{gofdens:lemma:convunif},
	\begin{align*}
	\mathbb{E}\Bigg[\bigg(\Iqr{&LK_n((\bx,z),(\bX_1,Z_1))L_n(\bx,\bX_1)}{\bx}{z}\bigg)^2\Bigg]\\
	=&\,\mathbb{E}\Bigg[\bigg(\Iqr{\lrc{LK\lrp{\frac{1-\bx^T\bX_1}{h^2},\frac{z-Z_1}{g}}-\Order{h^qg}}\\
		&\times\lrc{L\lrp{\frac{1-\bx^T\bX_1}{h^2}}-\Order{h^q}}}{\bx}{z}\bigg)^2\Bigg]\\
	=&\,\mathbb{E}\Bigg[\bigg(\int_{\Om{q-1}}\int_{0}^{2h^{-2}}\!\!\!\Ir{\lrc{LK\lrp{r,u}-\Order{h^qg}}\lrc{L\lrp{r}-\Order{h^q}}\\
		&\times h^{q}(2-h^2r)^{\frac{q}{2}-1}r^{\frac{q}{2}-1}g}{u}\,dr\,\om{q-1}(d\bxi)\bigg)^2\Bigg]\\
	=&\,\Order{h^{2q}g^2}.
	\end{align*}
	The variance of $T^{(1,2c)}_{n,3}$ is obtained analogously:
	\begin{align*}
	\mathbb{V}\mathrm{ar}\Big[T^{(1,2c)}_{n,3}\Big]\leq&\,\mathbb{E}\Bigg[\Iqr{\Iqr{LK_n((\bx_1,z_1),(\bX_1,Z_1))L_n(\bx_1,\bX_2)K_n(z_1,Z_1)\\
			&\times LK_n((\bx_2,z_2),(\bX_1,Z_1))L_n(\bx_2,\bX_2)K_n(z_2,Z_1) }{\bx_1}{z_1}}{\bx_2}{z_2}\Bigg]\\
	=&\,\Iqr{\Iqr{\mathbb{E}\big[LK_n((\bx_1,z_1),(\bX_1,Z_1))LK_n((\bx_2,z_2),(\bX_1,Z_1))\\
			&\times K_n(z_1,Z_1)K_n(z_2,Z_1)\big] CL_n(\bx_1,\bX_2;\bx_2,\bX_2)}{\bx_1}{z_1}}{\bx_2}{z_2}\\
	=&\,\Order{h^q} \E{\bigg(\Iqr{LK_n((\bx,z),(\bX_1,Z_1))K_n(z,Z_1)}{\bx}{z}\bigg)^2}\\
	=&\,\Order{h^{3q}g^2}.
	\end{align*}
	Then, putting together the variances of $T^{(1,1)}_{n,3}$, $T^{(1,2a)}_{n,3}$, $T^{(1,2b)}_{n,3}$, $T^{(1,2c)}_{n,3}$ and $T^{(1,3)}_{n,3}$, it follows
	\begin{align}
	\V{T^{(1)}_{n,3}}=&\,\frac{c_{h,q}(L)^4}{n^{6}g^4}\Order{ nh^{2q}g^2+n^2(h^{3q}g^3+h^{2q}g^3+h^{3q}g^2)+n^3 h^{3q}g^3}\nonumber\\
	=&\,\frac{c_{h,q}(L)^4}{n^{6}g^4}\Order{n^3h^{3q}g^3}\nonumber\\
	=&\,\Order{n^{-3}h^{-q}g^{-1}}.\label{gofdens:VarTn3_1}
	\end{align}
	Finally, joining (\ref{gofdens:VarTn3_2}), (\ref{gofdens:VarTn3_3}) and (\ref{gofdens:VarTn3_1}),
	\begin{align*}
	\V{T_{n,3}}=\Order{n^{-3}h^{-q}g^{-1}}+\Order{n^{-2}h^{-q}}+\Order{n^{-2}g^{-1}}=\Order{n^{-2}(h^{-q}+g^{-1})},
	\end{align*}
	which proves the lemma.
\end{proof}

%-------------------------------------------------%
\subsection{Goodness-of-fit test for models with directional data}
%-------------------------------------------------%

\begin{lem}
	\label{gofdens:lem:gof:1}
	Under $H_0: f=f_{\btheta_0}$, with $\btheta_0\in\Theta$ unknown and \ref{gofdens:assump:a1}--\ref{gofdens:assump:a3} and \ref{gofdens:assump:a5}--\ref{gofdens:assump:a6},
	$n(h^qg)^\frac{1}{2} R_{n,1}\stackrel{p}{\longrightarrow}0$ and $n(h^qg)^\frac{1}{2} R_{n,4}\stackrel{p}{\longrightarrow}0$.
\end{lem}

\begin{proof}[Proof of Lemma \ref{gofdens:lem:gof:1}]
	Under the null $f=f_{\btheta_0}$, for a known $\btheta_0\in\Theta$.\\
	
	\textit{Term $R_{n,4}$}. Using a first order Taylor expansion of $f_{\hat\btheta}$ in $\btheta_0$,
	\begin{align*}
	R_{n,4}=&\,\Iqr{\lrp{LK_{h,g}\lrp{f_{\btheta_0}(\bx,z)-f_{ \hat \btheta}(\bx,z)}}^2}{\bx}{z}\\
	=&\,\Iqr{\bigg(LK_{h,g}\bigg(\big(\hat\btheta-\btheta_0\big)^T\frac{\partial f_{\btheta}(\bx,z)}{\partial \btheta}\Big|_{\btheta=\btheta_n}\bigg)\bigg)^2}{\bx}{z}\\
	\leq&\,\big|\big|\hat\btheta-\btheta_0\big|\big|^2\Iqr{\bigg(LK_{h,g}\bigg(\bigg|\bigg|\frac{\partial f_{\btheta}(\bx,z)}{\partial \btheta}\Big|_{\btheta=\btheta_n}\bigg|\bigg|\bigg)\bigg)^2}{\bx}{z}\\
	=&\,\Orderp{n^{-1}}\Orderp{1}\\
	=&\,\Orderp{n^{-1}},
	\end{align*}
	where $\btheta_n\in\Theta$ is a certain parameter depending on the sample. The order holds because, on the one hand, $\big|\big|\hat\btheta-\btheta_0\big|\big|^2=\Orderp{n^{-1}}$ by \ref{gofdens:assump:a6} and on the other, by \ref{gofdens:assump:a5} and Lemma \ref{gofdens:lemma:convunif},
	\begin{align*}
	\Iqr{\bigg(LK_{h,g}&\bigg(\bigg|\bigg|\frac{\partial f_{\btheta}(\bx,z)}{\partial \btheta}\Big|_{\btheta=\btheta_n}\bigg|\bigg|\bigg)\bigg)^2}{\bx}{z}\\
	=&\,\bigg(\Iqr{\bigg|\bigg|\frac{\partial f_{\btheta}(\bx,z)}{\partial \btheta}\Big|_{\btheta=\btheta_n}\bigg|\bigg|^2}{\bx}{z}\bigg)(1+\order{1})\\
	=&\,\Orderp{1}.
	\end{align*}
	Therefore, $R_{n,4}=\Orderp{n^{-1}}$ and, by \ref{gofdens:assump:a3}, $n(h^qg)^\frac{1}{2} R_{n,4}\stackrel{p}{\longrightarrow}0$.\\
	
	\textit{Term $R_{n,1}$}. It follows also by a Taylor expansion of second order centred at $\btheta_0$:
	\begin{align*}
	R_{n,1}=&\,2\frac{c_{h,q}(L)}{ng}\sum_{i=1}^n\Iqr{\!\!LK_n\lrp{(\bx,z),(\bX_i,Z_i)}LK_{h,g}\lrp{f_{\btheta_0}(\bx,z)-f_{ \hat \btheta}(\bx,z)}}{\bx}{z}\\
	=&\,2\frac{c_{h,q}(L)}{ng}\sum_{i=1}^n\Iqr{\!\!LK_n\lrp{(\bx,z),(\bX_i,Z_i)} LK_{h,g}\bigg(\big(\hat\btheta-\btheta_0\big)^T\frac{\partial f(\bx,z)}{\partial \btheta}\Big|_{\btheta=\btheta_0}\\
		&+\big(\hat\btheta-\btheta_0\big)^T\frac{\partial^2 f(\bx,z)}{\partial \btheta\partial \btheta^T}\Big|_{\btheta=\btheta_n}\big(\hat\btheta-\btheta_0\big)\bigg)}{\bx}{z}\\
	\leq&\,2\frac{c_{h,q}(L)}{ng}\sum_{i=1}^n\Iqr{\!\!LK_n\lrp{(\bx,z),(\bX_i,Z_i)} \bigg[\big|\big|\hat\btheta-\btheta_0\big|\big|LK_{h,g}\bigg(\bigg|\bigg|\frac{\partial f(\bx,z)}{\partial \btheta}\Big|_{\btheta=\btheta_0}\bigg|\bigg|\bigg)\\
		&+\big|\big|\hat\btheta-\btheta_0\big|\big|^2 LK_{h,g}\bigg( \bigg|\bigg|\frac{\partial^2 f(\bx,z)}{\partial \btheta\partial \btheta^T}\Big|_{\btheta=\btheta_n}\bigg|\bigg|_F \bigg)\bigg]}{\bx}{z}\\
	=&\,\big|\big|\hat\btheta-\btheta_0\big|\big| R_{n,1}^{(1)}+\big|\big|\hat\btheta-\btheta_0\big|\big|^2 R_{n,1}^{(2)},
	\end{align*}
	where $\norm{A}_F$ stands for the Frobenious norm of the matrix $A$. By Lemma \ref{gofdens:lemma:convunif} and \ref{gofdens:assump:a5}, 
	\[
	R_{n,1}^{(i)}=\Orderp{\frac{c_{h,q}(L)}{ng}\sum_{i=1}^n\Iqr{LK_n\lrp{(\bx,z),(\bX_i,Z_i)}}{\bx}{z}},
	\]
	for $i=1,2$. As a consequence of this and \ref{gofdens:assump:a6}, the first addend of $R_{n,1}$ dominates the second. The proof now is based on proving that $R_{n,1}^{(1)}=\mathcal{O}_\mathbb{P}\big(n^{-\frac{1}{2}}\big)$ using the Chebychev inequality and the fact that the integrand of $R_{n,1}^{(1)}$ is deterministic. Now recall that $\mathbb{E}\big[R_{n,1}^{(i)}\big]=0$ and by the proof of (\ref{gofdens:lemma:Hn2:1}) in Lemma \ref{gofdens:lemma:Hn2},
	\begin{align*}
	\mathbb{V}\mathrm{ar}\Big[R_{n,1}^{(1)}\Big]=&\,\frac{c_{h,q}(L)^2}{ng^2}\E{\bigg(\Iqr{LK_n\lrp{(\bx,z),(\bX,Z)} }{\bx}{z}\bigg)^2}\\
	=&\,\frac{c_{h,q}(L)^2}{ng^2}\Iqr{\Iqr{\E{LK_n\lrp{(\bx,z),(\bX,Z)}LK_n\lrp{(\by,t),(\bX,Z)}}\\
			&\times }{\bx}{z}}{\by}{t}\\
	=&\,\frac{c_{h,q}(L)^2}{ng^2}\Iqr{\Iqr{\lrp{E_1\lrp{(\bx,z),(\by,t)}-E_2\lrp{(\bx,z),(\by,t)}} \\
			&\times }{\bx}{z}}{\by}{t}\\
	=&\,\frac{c_{h,q}(L)^2}{ng^2} \Order{h^{2q}g^2}\\
	=&\,\Order{n^{-1}},
	\end{align*}
	so by the Chebychev inequality, $R_{n,1}^{(1)}=\mathcal{O}_\mathbb{P}\big(n^{-\frac{1}{2}}\big)$ and as a consequence of \ref{gofdens:assump:a5}, $R_{n,1}=\Orderp{n^{-1}}$ and $n(h^qg)^\frac{1}{2} R_{n,1}\stackrel{p}{\longrightarrow}0$ follows.
\end{proof}

\begin{lem}
	\label{gofdens:lem:gof:2}
	Under the alternative hypothesis (\ref{gofdens:pit}) and \ref{gofdens:assump:a1}--\ref{gofdens:assump:a3}, \ref{gofdens:assump:a5} and \ref{gofdens:assump:a7}, $n(h^qg)^\frac{1}{2} \widetilde R_{n,1}\allowbreak\stackrel{p}{\longrightarrow}0$ and $n(h^qg)^\frac{1}{2} \widetilde R_{n,4}\stackrel{p}{\longrightarrow}R(\Delta)$.
\end{lem}

\begin{proof}[Proof of Lemma \ref{gofdens:lem:gof:2}]
	The convergence in probability is obtained using the decompositions $\widetilde R_{n,1}=R_{n,1}+\widetilde R_{n,1}^{(1)}$ and $\widetilde R_{n,4}=R_{n,4}+\widetilde R_{n,4}^{(1)}+\widetilde R_{n,4}^{(2)}$. \\
	
	\textit{Terms $R_{n,1}$ and $R_{n,4}$.} The proofs of $n(h^qg)^\frac{1}{2} R_{n,1}\stackrel{p}{\longrightarrow}0$ and $n(h^qg)^\frac{1}{2} R_{n,4}\stackrel{p}{\longrightarrow}0$ are analogous to the ones of Lemma \ref{gofdens:lem:gof:1} and follow just replacing \ref{gofdens:assump:a6} by \ref{gofdens:assump:a7} and $H_0$ by $H_{1P}$. \\
	
	\textit{Term $R_{n,1}^{(1)}$}. Recall that $\mathbb{E}\big[\widetilde R_{n,1}^{(1)}\big]=0$ and its variance, using the same steps as in the proof of $R_{n,1}^{(1)}$ in Lemma \ref{gofdens:lem:gof:1}, is
	\begin{align*}
	\V{\widetilde R_{n,1}^{(1)}}=&\,4\frac{c_{h,q}(L)^2}{n^2h^\frac{q}{2}g^\frac{3}{2}}\Iqr{\Iqr{\lrp{E_1\lrp{(\bx,z),(\by,t)}-E_2\lrp{(\bx,z),(\by,t)}}\\
			&\times LK_{h,g}\Delta(\bx,z)LK_{h,g}\Delta(\by,t)}{\bx}{z}}{\by}{t}\\
	=&\,4\frac{c_{h,q}(L)^2}{n^2h^\frac{q}{2}g^\frac{3}{2}}\Order{h^{2q}g^2}\\
	=&\,\Order{\big(n^2h^{\frac{q}{2}}g^{\frac{1}{2}}\big)^{-1}}.
	\end{align*}
	Then, $\widetilde R_{n,1}^{(1)}=\mathcal{O}_\mathbb{P}\big((nh^\frac{q}{4}g^\frac{1}{4})^{-1}\big)$ and $n(h^qg)^\frac{1}{2} \widetilde R_{n,1}^{(1)}\stackrel{p}{\longrightarrow}0$. \\
	
	\textit{Term $R_{n,4}^{(1)}$.} Applying Lemma \ref{gofdens:lemma:convunif},
	\begin{align*}
	\widetilde R_{n,4}^{(1)}=&\,\frac{1}{n(h^qg)^\frac{1}{2}}\Iqr{\lrp{LK_{h,g}\Delta(\bx,z)}^2}{\bx}{z}=\frac{1}{n(h^qg)^\frac{1}{2}}R(\Delta)(1+\order{1})
	\end{align*}
	and as a consequence $n(h^qg)^\frac{1}{2}\widetilde R_{n,4}^{(1)}\stackrel{p}{\longrightarrow}R(\Delta)$. \\
	
	\textit{Term $R_{n,4}^{(2)}$.} Applying the Cauchy--Schwartz inequality:
	\begin{align*}
	\frac{\sqrt{nh^\frac{q}{2}g^\frac{1}{2}}}{2}\widetilde R_{n,4}^{(2)}\leq \big(R_{n,4}\big)^\frac{1}{2} \Big(nh^\frac{q}{2}g^\frac{1}{2} \widetilde R_{n,4}^{(1)}\Big)^\frac{1}{2}=\mathcal{O}_\mathbb{P}\Big(n^{-\frac{1}{2}}\Big)\Orderp{1}=\mathcal{O}_\mathbb{P}\Big(n^{-\frac{1}{2}}\Big),
	\end{align*}
	Therefore, $\widetilde R_{n,4}^{(2)}=\mathcal{O}_\mathbb{P}\big((nh^\frac{q}{4}g^\frac{1}{4})^{-1}\big)$ and $n(h^qg)^\frac{1}{2}\widetilde R_{n,4}^{(2)}=\mathcal{O}_\mathbb{P}\big( (h^qg)^\frac{1}{4} \big)\stackrel{p}{\longrightarrow}0$.
\end{proof}

%-------------------------------------------------%
\subsection{General purpose lemmas}
%-------------------------------------------------%

For the proofs of some lemmas, three auxiliary lemmas have been used.

\begin{lem}
	\label{gofdens:lemma:convunif}
	Under \ref{gofdens:assump:a1}--\ref{gofdens:assump:a3}, for any function $\varphi:\Om{q}\times \R\rightarrow\R$ that is uniformly continuous and bounded, the smoothing operator (\ref{gofdens:smoothing}) satisfies
	\begin{align}
	\sup_{(\bx,z)\in\Om{q}\times\R}\abs{LK_{h,g}\varphi(\bx,z)-\varphi(\bx,z)}\xrightarrow[n\to\infty]{} 0.\label{gofdens:lemma:convunif:1}
	\end{align} 
	Thus, $LK_{h,g}\varphi(\bx,z)$ converges to $\varphi(\bx,z)$ uniformly in $\Om{q}\times\R$.
\end{lem}

\begin{proof}[Proof of Lemma \ref{gofdens:lemma:convunif}]
	Let denote $D_{n}=\abs{LK_{h,g}\varphi(\bx,z)-\varphi(\bx,z)}$. Since $\varphi(\bx,z)$ can be written as $\frac{c_{h,q}(L)}{g}\Iqr{LK\big(\frac{1-\bx^T\by}{h^2},\frac{z-t}{g}\big)\varphi(\by,t)}{\bx}{z}$, then
	\begin{align*}
	D_{n}=&\,\bigg|\frac{c_{h,q}(L)}{g}\Iqr{LK\lrp{\frac{1-\bx^T\by}{h^2},\frac{z-t}{g}}\lrp{\varphi(\by,t)-\varphi(\bx,z)}}{\by}{t}\bigg|\\
	\leq&\,\frac{c_{h,q}(L)}{g}\Iqr{LK\lrp{\frac{1-\bx^T\by}{h^2},\frac{z-t}{g}}\abs{\varphi(\by,t)-\varphi(\bx,z)}}{\by}{t}\\
	\leq&\,D_{n,1}+D_{n,2}, %
	\end{align*}
	where:
	\begin{align*}
	D_{n,1}=&\,\frac{c_{h,q}(L)}{g}\int_{A_{\delta}}LK\lrp{\frac{1-\bx^T\by}{h^2},\frac{z-t}{g}}\abs{\varphi(\by,t)-\varphi(\bx,z)}\,dt\,\om{q}(d\by),\\
	D_{n,2}=&\,\frac{c_{h,q}(L)}{g}\int_{\bar{A_\delta}}LK\lrp{\frac{1-\bx^T\by}{h^2},\frac{z-t}{g}}\abs{\varphi(\by,t)-\varphi(\bx,z)}\,dt\,\om{q}(d\by),\\
	A_{\delta}=&\,\lrb{(\by,t)\in\Om{q}\times\R:\max\Big(\sqrt{2(1-\bx^T\by)},\abs{z-t}\Big)<\delta},\\
	A_{1,\delta}=&\,\bigg\{(\by,t)\in\Om{q}\times\R:1-\bx^T\by<\frac{\delta^2}{2}\bigg\},\\
	A_{2,\delta}=&\,\lrb{(\by,t)\in\Om{q}\times\R:\abs{z-t}<\delta}
	\end{align*}
	and $\bar A_\delta$ denotes the complementary set to $A_\delta$ for a $\delta>0$. Recall that $A_\delta=A_{1,\delta}\cap A_{2,\delta}$ and as a consequence $\bar A_\delta=\bar A_{1,\delta}\cup\bar A_{2,\delta}$. \\
	
	As stated in \ref{gofdens:assump:a1}, the uniform continuity of the functions defined in $\Om{q}\times\R$ is understood with respect to the product Euclidean norm, that is
	\[
	\norm{(\bx,z)}_2=\sqrt{\norm{\bx}^2_{\Om{q}}+\norm{z}^2_\R},\text{ where }\norm{\cdot}_{\Om{q}}=\norm{\cdot}_2\text{ and }\norm{\cdot}_\R=\abs{\cdot}.
	\]
	Nevertheless, given the equivalence between the product $2$-norm and the product $\infty$-norm, defined as $\norm{(\bx,z)}_\infty=\max\big(\norm{\bx}_{\Om{q}},\norm{z}_\R\big)$, and for the sake of simplicity, the second norm will be used in the proof. Then, by the uniform continuity of $\varphi$, it holds that for any $\varepsilon>0$, there exists a $\delta>0$ such that 
	\[
	\forall (\bx,z),(\by,t)\in\Om{q}\times\R,\,\norm{(\bx,z)-(\by,t)}_\infty<\delta \implies \abs{\varphi(\bx,z)-\varphi(\by,t)}<\varepsilon.
	\]
	Therefore the first term is dominated by
	\[
	D_{n,1}<\varepsilon\frac{c_{h,q}(L)}{g}\int_{A_{\delta}}LK\lrp{\frac{1-\bx^T\by}{h^2},\frac{z-t}{g}}\,dt\,\om{q}(d\by)\leq\varepsilon,
	\]
	for any $\varepsilon>0$, so as a consequence $D_{n,1}=\order{1}$ uniformly in $(\bx,z)\in\Om{q}\times\R$.\\
	
	For the second term, let consider the change of variables introduced in the proof of Lemma \ref{gofdens:lemma:Hn2} (see Lemma 2 of \cite{Garcia-Portugues:dirlin} for a detailed derivation):
	\begin{align*}
	\lb\begin{array}{l}
	\by=u\bx+(1-u^2)^\frac{1}{2}\bB_{\bx}\bxi,\\
	\om{q}(d\by)=(1-u^2)^{\frac{q}{2}-1}\,du\,\om{q-1}(d\bxi),
	\end{array}\ri
	\end{align*}
	where $u\in(-1,1)$, $\bxi\in\Om{q-1}$ and $\bB_{\bx} = (\bb_1,\ldots, \bb_{q})_{(q+1)\times q}$ is the semi-orthonormal matrix resulting from the completion of $\bx$ to the orthonormal basis $\lrb{\bx,\bb_1,\ldots,\bb_{q}}$. Applying this change of variables and then using the standard changes of variables $r=\frac{1-u}{h^2}$ (for the first addend) and $s=\frac{z-t}{g}$ (second addend), it follows:
	\begin{align*}
	D_{n,2}=&\,\frac{c_{h,q}(L)}{g}\int_{\bar{A}_\delta}LK\lrp{\frac{1-\bx^T\by}{h^2},\frac{z-t}{g}}\abs{\varphi(\by,t)-\varphi(\bx,z)}\,dt\,\om{q}(d\by)\\
	\leq &\, \frac{c_{h,q}(L)}{g}\int_{\bar{A}_{1,\delta}}LK\lrp{\frac{1-\bx^T\by}{h^2},\frac{z-t}{g}}\abs{\varphi(\by,t)-\varphi(\bx,z)}\,dt\,\om{q}(d\by)\\
	&+\frac{c_{h,q}(L)}{g}\int_{\bar{A}_{2,\delta}}LK\lrp{\frac{1-\bx^T\by}{h^2},\frac{z-t}{g}}\abs{\varphi(\by,t)-\varphi(\bx,z)}\,dt\,\om{q}(d\by)\\
	\leq &\, 2\frac{c_{h,q}(L)}{g}\sup_{(\by,t)\in\Om{q}\times\R}\abs{\varphi(\by,t)}\Bigg\{ \int_{\bar{A}_{1,\delta}}LK\lrp{\frac{1-\bx^T\by}{h^2},\frac{z-t}{g}}\,dt\,\om{q}(d\by)\\
	&+\int_{\bar{A}_{2,\delta}}LK\lrp{\frac{1-\bx^T\by}{h^2},\frac{z-t}{g}}\,dt\,\om{q}(d\by)\Bigg\}\\
	\leq &\, 2\sup_{(\by,t)\in\Om{q}\times\R}\abs{\varphi(\by,t)}\Bigg\{ c_{h,q}(L)\om{q-1}\int_{-1}^{1-\frac{\delta^2}{2}}L\lrp{\frac{1-u}{h^2}}(1-u^2)^{\frac{q}{2}-1}\,du\\
	&+2\int_{\delta g^{-1}}^{\infty}K\lrp{s}\,ds\Bigg\}\\
	\leq&\, 2\sup_{(\by,t)\in\Om{q}\times\R}\abs{\varphi(\by,t)}\Bigg\{ c_{h,q}(L)\om{q-1}\int_{-1}^{1}(1-u^2)^{\frac{q}{2}-1}\,du\times\sup_{r\geq \delta^2/(2h^2)} L(r)r^\frac{q}{2}r^{-\frac{q}{2}}  \\
	&+2\int_{\delta g^{-1}}^{\infty}K\lrp{s}\,ds\Bigg\}\\
	\leq&\,\Order{1}\Bigg\{ \lambda_{h,q}(L)^{-1}\om{q-1}2^{-\frac{q}{2}}\delta^{-q}\int_{-1}^{1}(1-u^2)^{\frac{q}{2}-1}\,du\times\sup_{r\geq \delta^2/(2h^2)} L(r)r^\frac{q}{2}   +\order{1}\Bigg\}\\
	=&\,\Order{1}\lrp{\Order{1}\order{1}+\order{1}}\\
	=&\,\order{1},
	\end{align*}
	by relation (\ref{gofdens:normalizing}), the fact $\int_{-1}^{1}(1-u^2)^{\frac{q}{2}-1}\,du<\infty$ for all $q\geq1$ and because by \ref{gofdens:assump:a2}, $\lambda_{q+2}(L)<\infty$, which implies that $\displaystyle\lim_{r\to\infty} L(r)r^{\frac{q}{2}}=0$. \\
	
	Then, $D_n\to0$ as $n\to\infty$ and this holds regardless the point $(\bx,z)$, since $\varphi$ is uniformly continuous, so (\ref{gofdens:lemma:convunif:1}) is satisfied and $LK_{h,g}\varphi(\bx,z)$ converges to $\varphi(\bx,z)$ uniformly in $\Om{q}\times\R$.
\end{proof}

\begin{lem}
	\label{gofdens:lemma:biasvar}
	Under \ref{gofdens:assump:a1}--\ref{gofdens:assump:a3}, the bias and the variance for the directional-linear estimator in a point $(\bx,z)\in\Omega_{q}\times\R$ is given by 
	\begin{align*}
	\E{\hat f_{h,g}(\bx,z)}=&\,f(\bx,z)+\frac{b_q(L)}{q}\tr{\bHcal_\bx f(\bx,z)}h^2+\frac{1}{2}\mu_2(K)\Hcal_z f(\bx,z)g^2+\order{h^2+g^2},\\
	\V{\hat f_{h,g}(\bx,z)}=&\,\frac{\lambda_q(L^2)\lambda_q(L)^{-2}R(K)}{nh^qg}f(\bx,z)+\order{(nh^qg)^{-1}},
	\end{align*}
	where the remainder orders are uniform.
\end{lem}

\begin{proof}[Proof of Lemma \ref{gofdens:lemma:biasvar}]
	The asymptotic expressions of the bias and the variance are given in \cite{Garcia-Portugues:dirlin}. Recalling the extension of $f$ in \ref{gofdens:assump:a1}, the partial derivative of $f$ for the direction $\bx$ and evaluated at $(\bx,z)$, that is $\bx^T\bnab_\bx f(\bx,z)$, is null:
	\[
	\bx^T\bnab_\bx f(\bx,z)=\lim_{h\to0}\frac{f\lrp{(1+h)\bx,z}-f\lrp{\bx,z}}{h}=\lim_{h\to0}\frac{f\lrp{\bx,z}-f\lrp{\bx,z}}{h}=0.
	\]
	Using this fact, it also follows that $\bx^T\bHcal_\bx f(\bx,z)\bx=0$, since
	\[
	\bx^T\lrp{\frac{\partial}{\partial \bx}\bx^T \bnab_\bx f(\bx,z)}=\bx^T\lrp{\bnab_\bx f(\bx,z)+\bHcal_\bx f(\bx,z)\bx}=0.
	\]
	Therefore, the operator $\boldsymbol\Psi_\bx(f,\bx,z)$ appearing in the bias expansion given in \cite{Garcia-Portugues:dirlin} can be written in the simplified form
	\begin{align*}
	\boldsymbol\Psi_\bx(f,\bx,z)=-\bx^T\boldsymbol \bnab_\bx f(\bx,z)+\frac{1}{q}\lrp{\nabla^2f(\bx,z)-\bx^T\bHcal_\bx f(\bx,z)\bx}=\frac{1}{q}\tr{\bHcal_\bx f(\bx,z)},
	\end{align*}
	because $\nabla^2 f(\bx,z)$ represents the directional Laplacian of $f$ (the trace of $\bHcal_\bx f(\bx,z)$).\\
	
	The uniformity of the orders, not considered in the above paper, can be obtained by using the extra-smoothness assumption \ref{gofdens:assump:a1} and the integral form of the remainder in the Taylor's theorem\nolinebreak[4] on\nolinebreak[4] $f$:
	\begin{align*}
	f(\by+\ba)-f(\by)=\ba^T\bnab f(\by)+\frac{1}{2}\ba^T \bHcal f(\by)\ba+R,
	\end{align*}
	with $\by\equiv(\bx,z)$, $\ba\in\Om{q}\times\R$ and where the remainder has the exact form
	\begin{align*}
	R=&\,\int_0^1\frac{(1-t)^2}{2} \sum_{i,j,k=1}^{q+1}\frac{\partial^3}{\partial x_i\partial x_j\partial x_k}f(\bx+t\ba)\alpha_i\alpha_j\alpha_k \,dt\leq \frac{1}{6} M \sum_{i,j,k=1}^{q+1}{\alpha_i\alpha_j\alpha_k}=\order{\ba^T\ba},
	\end{align*}
	where $M$ is the bound of the third derivatives of $f$ and in the last equality it is used the second point of Lemma \ref{gofdens:lemma:orders}. Then the remainder does not depend on the point $\by\equiv(\bx,z)$ and following the proofs of \cite{Garcia-Portugues:dirlin} the convergence of the bias and variance is uniform on\nopagebreak[4] $\Om{q}\times\R$.
\end{proof}

\begin{lem}
	\label{gofdens:lemma:orders}
	Let $a_n$, $b_n$ and $c_n$ sequences of positive real numbers. Then:
	\begin{enumerate}[label=\textit{\roman{*}}.]
		\item If $a_n,b_n\to0$, then $a_nb_n=\order{a_n+b_n}$.
		\item If $a_n,b_n,c_n\to0$, then $a_n b_n c_n=\order{a_n^2+b_n^2+c_n^2}$.
		\item $a_n^ib_n^j=\mathcal{O}\big(a_n^k+b_n^k\big)$, for any integers $i,j\geq0$ such that $i+j=k$.
		\item $(a_n+b_n)^k=\mathcal{O}\big(a_n^k+b_n^k\big)$, for any integer $k\geq1$.
	\end{enumerate} 
\end{lem}

\begin{proof}[Proof of Lemma \ref{gofdens:lemma:orders}]
	The first statement follows immediately from the definition of $\order{\cdot}$,
	\begin{align*}
	a_nb_n=\order{a_n+b_n} :\iff \lim_{n\to\infty}\frac{a_nb_n}{a_n+b_n}=\lim_{n\to\infty}\frac{1}{\frac{1}{b_n}+\frac{1}{a_n}}=\frac{1}{\infty}=0.
	\end{align*}
	For the second, suppose that, when $n\to\infty$, $a_n=\max(a_n,b_n,c_n)$ to fix notation. Then
	\begin{align*}
	\lim_{n\to\infty}\frac{a_nb_nc_n}{a_n^2+b_n^2+c_n^2}\leq \lim_{n\to\infty}\frac{a_n^3}{a_n^2+b_n^2+c_n^2}=\lim_{n\to\infty}\frac{1}{\frac{1}{a_n}+\frac{b_n^2}{a_n^3}+\frac{c_n^2}{a_n^3}}=\frac{1}{\infty}=0.
	\end{align*}
	Let $C$ be a positive constant. The third statement follows from the definition of $\Order{\cdot}$, 
	\begin{align*}
	\lim_{n\to\infty}\frac{a_n^ib_n^j}{a_n^k+b_n^k}=\lim_{n\to\infty}\frac{1}{\lrp{\frac{a_n}{b_n}}^j+\lrp{\frac{b_n}{a_n}}^i}=\lb\begin{array}{ll}
	\frac{1}{0+\infty},&a_n=\order{b_n},\\
	\frac{1}{\infty+0},&b_n=\order{a_n},\\
	\frac{1}{C^j+C^{-i}},&a_n\sim C b_n.
	\end{array}\ri
	\end{align*}
	Then the limit is bounded and $a_n^i b_n^j=\mathcal{O}\big(a_n^k+b_n^k\big)$. The last statement arises as a consequence of this result and the Newton binomial:
	\begin{align*}
	\lrp{a_n+b_n}^k=\sum_{i=0}^k\binom{k}{i} a_n^{k-i}b_n^{i}=\sum_{i=0}^k\binom{k}{i}\Order{a_n^k+b_n^k}=\Order{a_n^k+b_n^k}.
	\end{align*}
\end{proof}

%-------------------------------------------------%
\section{Further results for the independence test}
\label{gofdens:su:num}
%-------------------------------------------------%

%-------------------------------------------------%
\subsection{Closed expressions}
%-------------------------------------------------%

Consider $K$ and $L$ a normal and a von Mises kernel, respectively. In this case $R(K)=\big(2\pi^\frac{1}{2}\big)^{-1}$, $\lambda_q(L)=(2\pi)^\frac{q}{2}$ and $\lambda_q(L^2)\lambda_q(L)^{-2}=\big(2\pi^\frac{1}{2}\big)^{-q}$. Furthermore, it is possible to compute exactly the form of the contributions of these two kernels to the asymptotic variance, resulting:
\begin{align*}
\gamma_q \lambda_q(L)^{-4}\int_0^{\infty} r^{\frac{q}{2}-1}\lrb{\int_0^{\infty} \rho^{\frac{q}{2}-1} L(\rho) \varphi_q(r,\rho) \,d\rho}^2\,dr&=(8\pi)^{-\frac{q}{2}},\\
\int_\R \lrb{\int_\R  K(u)K(u+v)\,du}^2\,dv&=(8\pi)^{-\frac{1}{2}}.
\end{align*}
\begin{coro}
	\label{gofdens:coro:indep}
	If $L(r)=e^{-r}$ and $K$ is a normal density, then the asymptotic bias and variance in Theorem \ref{gofdens:theo:indep} are
	\begin{align*}
	A_n=&\,\frac{1}{2^{q+1}\pi^\frac{q+1}{2}nh^qg}-\frac{R(f_Z)}{2^q\pi^\frac{q}{2}nh^q}-\frac{R(f_\bX)}{2\pi^\frac{1}{2}ng},\quad\sigma_I^2=(8\pi)^{-\frac{q+1}{2}} R(f_\bX) R(f_Z).
	\end{align*}
	In addition, if $f_\bX=f_{\mathrm{vM}}(\cdot;\bmu,\kappa)$ and $f_Z$ is the density of a $\mathcal{N}(m,\sigma^2)$, then $R(f_\bX)=\big(2\pi^\frac{q+1}{2}\big)^{-1}\kappa^\frac{q-1}{2}\allowbreak\ \mathcal{I}_{\frac{q-1}{2}}(2\kappa) \mathcal{I}_{\frac{q-1}{2}}(\kappa)^{-2}$ and $R(f_Z)=\big(2\pi^\frac{1}{2}\sigma\big)^{-1}$.
\end{coro}

\begin{proof}[Proof of Corollary \ref{gofdens:coro:indep}]
	The expressions for $R(K)$, $R(f_Z)$ and $\int_\R \lrb{\int_\R  K(u)K(u+v)\,du}^2\,dv=(8\pi)^{-\frac{1}{2}}$ follow easily from the convolution properties of normal densities. The expressions for $\lambda_q(L)$ and $\lambda_q(L^2)$ can be derived from the definition of the Gamma function. Similarly,
	\begin{align}
	R(f_\bX)=&\,C_q(\kappa)^2\Iq{e^{2\kappa\bx^T\bmu}}{\bx}=\frac{C_q(\kappa)^2}{C_q(2\kappa)}=\frac{\kappa^\frac{q-1}{2}\mathcal{I}_\frac{q-1}{2}(2\kappa)}{2\pi^\frac{q+1}{2}\mathcal{I}_\frac{q-1}{2}(\kappa)^2},\nonumber\\
	\gamma_q^{-1} \lambda_q(L)^{4}=&\,\left\{\begin{array}{ll}
	2^{-\frac{5}{4}}\pi^2,& q=1,\\
	2^{\frac{q}{2}}\pi^{\frac{q}{2}+1}\Gamma\lrp{\frac{q}{2}}\Gamma\big(\frac{q-1}{2}\big)^2,& q>1.
	\end{array}\right.\label{gofdens:gamma}
	\end{align}
	
	For $q=1$ the contribution of the directional kernel to the asymptotic variance can be computed using (\ref{gofdens:gamma}) and 
	\[
	\int_0^\infty \rho^{-\frac{1}{2}} e^{-2\lrp{\rho\pm\sqrt{r\rho}}}\,d\rho %
	=\sqrt{2\pi}e^{\frac{r}{2}}\lrp{1-\Phi\lrp{\mp\sqrt{r}}},
	\]
	where $\Phi$ is the cumulative distribution function of a $\mathcal{N}(0,1)$. Then: 
	\begin{align*}
	\gamma_1 \lambda_1(L)^{-4}&\int_0^{\infty} r^{-\frac{1}{2}}\lrb{\int_0^{\infty} \rho^{-\frac{1}{2}} L(\rho) \varphi_1(r,\rho) \,d\rho}^2\,dr\\
	=&\,\gamma_1 \lambda_1(L)^{-4}\int_0^{\infty} r^{-\frac{1}{2}}e^{-2r}\bigg\{\int_0^{\infty} \rho^{-\frac{1}{2}} e^{-2\rho-2(r\rho)^{\frac{1}{2}}}\,d\rho+\int_0^{\infty} \rho^{-\frac{1}{2}}e^{-2\rho+2(r\rho)^{\frac{1}{2}}}\,d\rho \bigg\}^2\,dr\\
	=&\,2^{-\frac{1}{2}}\lrp{2\pi}^{-1}\int_0^{\infty} r^{-\frac{1}{2}}e^{-r} \,dr\\
	=&\,\lrp{8\pi}^{-\frac{1}{2}}.
	\end{align*}
	
	For $q>1$, the integral with respect to $\theta$ is computed from the definition of the modified Bessel function and the integral with respect to $\rho$ is
	\begin{align*}
	\int_0^{\infty} \rho^{\frac{q}{4}-\frac{1}{2}} e^{-2\rho} \mathcal{I}_{\frac{q}{2}-1}\lrp{2\sqrt{r\rho}}\,d\rho %
	=&\,2^{-\frac{q}{2}}r^{\frac{q}{4}-\frac{1}{2}}e^{\frac{r}{2}}.
	\end{align*}
	Using these two facts, it results:
	\begin{align*}
	\gamma_q \lambda_q(L)^{-4} &\int_0^{\infty} r^{\frac{q}{2}-1}\lrb{\int_0^{\infty} \rho^{\frac{q}{2}-1} L(\rho) \varphi_q(r,\rho) \,d\rho}^2\,dr\\
	=&\,2^{-\frac{q}{2}}\pi^{-\lrp{\frac{q}{2}+1}}\Gamma\lrp{\frac{q}{2}}^{-1}\Gamma\lrp{\frac{q-1}{2}}^{-2} \\
	&\times\int_0^{\infty} r^{\frac{q}{2}-1} \Bigg\{\int_0^{\infty} \rho^{\frac{q}{2}-1} e^{-\lrp{r+2\rho}}\lrc{\pi^{\frac{1}{2}}\Gamma\lrp{\frac{q-1}{2}}(r\rho)^{-\frac{q-2}{4}}\mathcal{I}_{\frac{q}{2}-1}\lrp{2\sqrt{r\rho}}  }\,d\rho\Bigg\}^2\,dr\\
	=&\,2^{-\frac{q}{2}}\pi^{-\frac{q}{2}}\Gamma\lrp{\frac{q}{2}}^{-1}\int_0^{\infty} r^{\frac{q}{2}-1}e^{-2r}\lrb{2^{-\frac{q}{2}}e^{\frac{r}{2}}
	}^2\,dr\\
	=&\,\lrp{8\pi}^{-\frac{q}{2}}.
	\end{align*}
\end{proof}

%-------------------------------------------------%
\subsection{Extension to the directional-directional case}
%-------------------------------------------------%

Under the directional-directional analogue of \ref{gofdens:assump:a4}, that is, $h_{1,n}^{q_1}h_{2,n}^{-q_2}\to c$, with $0<c<\infty$, the directional-linear independence test can be directly adapted to this setting, considering the following test statistic:
\[
T_n=\int_{\Om{q_1}\times\Om{q_2}}\lrp{\hat f_{(\bX,\bY);h_1,h_2}(\bx,\by)-\hat f_{\bX;h_1}(\bx)\hat f_{\bY;h_2}(\by)}^2\,\om{q_2}(d\by)\,\om{q_1}(d\bx).
\]
\begin{coro}[Directional-directional independence test]
	\label{gofdens:coro:indep:dirdir}
	Under the directional-directional analogues of \ref{gofdens:assump:a1}--\ref{gofdens:assump:a4} and the null hypothesis of independence,
	\[
	n(h_1^{q_1}h_2^{q_2})^\frac{1}{2}\lrp{T_n-A_n}\stackrel{d}{\longrightarrow}\mathcal{N}(0,2\sigma_I^2),
	\]
	where  
	\begin{align*}
	A_n=&\,\frac{\lambda_{q_1}(L_1^2)\lambda_{q_1}(L_1)^{-2}\lambda_{q_2}(L_2^2)\lambda_{q_2}(L_2)^{-2}}{nh_1^{q_1}h_{2}^{q_2}}\\
	&-\frac{\lambda_{q_1}(L_1^2)\lambda_{q_1}(L_1)^{-2}R(f_\bY)}{nh_1^{q_1}}-\frac{\lambda_{q_2}(L_2^2)\lambda_{q_2}(L_2)^{-2}R(f_\bX)}{nh_2^{q_2}},
	\end{align*}
	and $\sigma_I^2$ is defined as $\sigma^2$ in Corollary \ref{gofdens:coro:clt:dirdir} but with $R(f)=R(f_\bX)R(f_\bY)$. Further, if $L_1$ and $L_2$ are the von Mises kernel,
	\[
	A_n=\frac{1}{2^{q_1+q_2}\pi^\frac{q_1+q_2}{2}nh_1^{q_1}h_2^{q_2}}-\frac{R(f_\bY)}{2^{q_1}\pi^\frac{q_1}{2}nh_1^{q_1}}-\frac{R(f_\bX)}{2^{q_2}\pi^\frac{q_2}{2}nh_2^{q_2}}
	\]
	and $\sigma_I^2=(8\pi)^{-\frac{q_1+q_2}{2}} R(f_\bX) R(f_\bY)$. If $f_\bX$ and $f_\bY$ are von Mises densities, $R(f_\bX)$  and $R(f_\bY)$ are given as in Corollary \ref{gofdens:coro:indep}.
\end{coro}
\begin{proof}[Proof of Corollary \ref{gofdens:coro:indep:dirdir}]
	The proof follows from adapting the proofs of Theorem \ref{gofdens:theo:indep} and Corollary \ref{gofdens:coro:indep} to the directional-directional situation.
\end{proof}

%-------------------------------------------------%
\subsection{Some numerical experiments}
%-------------------------------------------------%

The purpose of this subsection is to provide some numerical experiments to illustrate the degree of misfit between the true distribution of the standardized statistic (approximated by Monte Carlo) and its asymptotic distribution, for increasing sample sizes. \\

For simplicity, independence will be assessed in a circular-linear framework ($q=1$), with a $\mathrm{vM}((0,1),1)$ for the circular variable and a $\mathcal{N}(0,1)$ for the linear one. Kernel density estimation is done using von Mises and normal kernels, as in Corollary \ref{gofdens:coro:indep}. Sample sizes considered are $n=5^j\times 10^k$, $j=0,1$, $k=3,5$ (see supplementary material for $k=1,2,4$). The sequence of bandwidths is taken as $h_n=g_n=2n^{-\frac{1}{3}}$, as a compromise between fast convergence and numerical problems avoidance. Figure \ref{gofdens:fig1:ext} presents the histogram of $1000$ values from $(nh_n^qg_n)^\frac{1}{2}\lrp{T_n-A_n}$ for different sample sizes, jointly with the $p$-values of the Kolmogorov--Smirnov test for the distribution $\mathcal{N}(0,2\sigma_I^2)$ and of the Shapiro--Wilk test for normality. Both tests are significant, until a very large sample size (close to $500,000$ data) is reached.\\

It should be noted that, in practical problems, the use of the asymptotic distribution does not seem feasible, and a resampling mechanism for the calibration of the test is required. This issue is addressed in \cite{Garcia-Portugues:testindep}, considering a permutation approach. The reader is referred to the aforementioned paper for the details concerning the practical application.

\begin{figure}[htpb]
	\centering
	\vspace*{-0.75cm}
	\includegraphics[height=0.25\textheight]{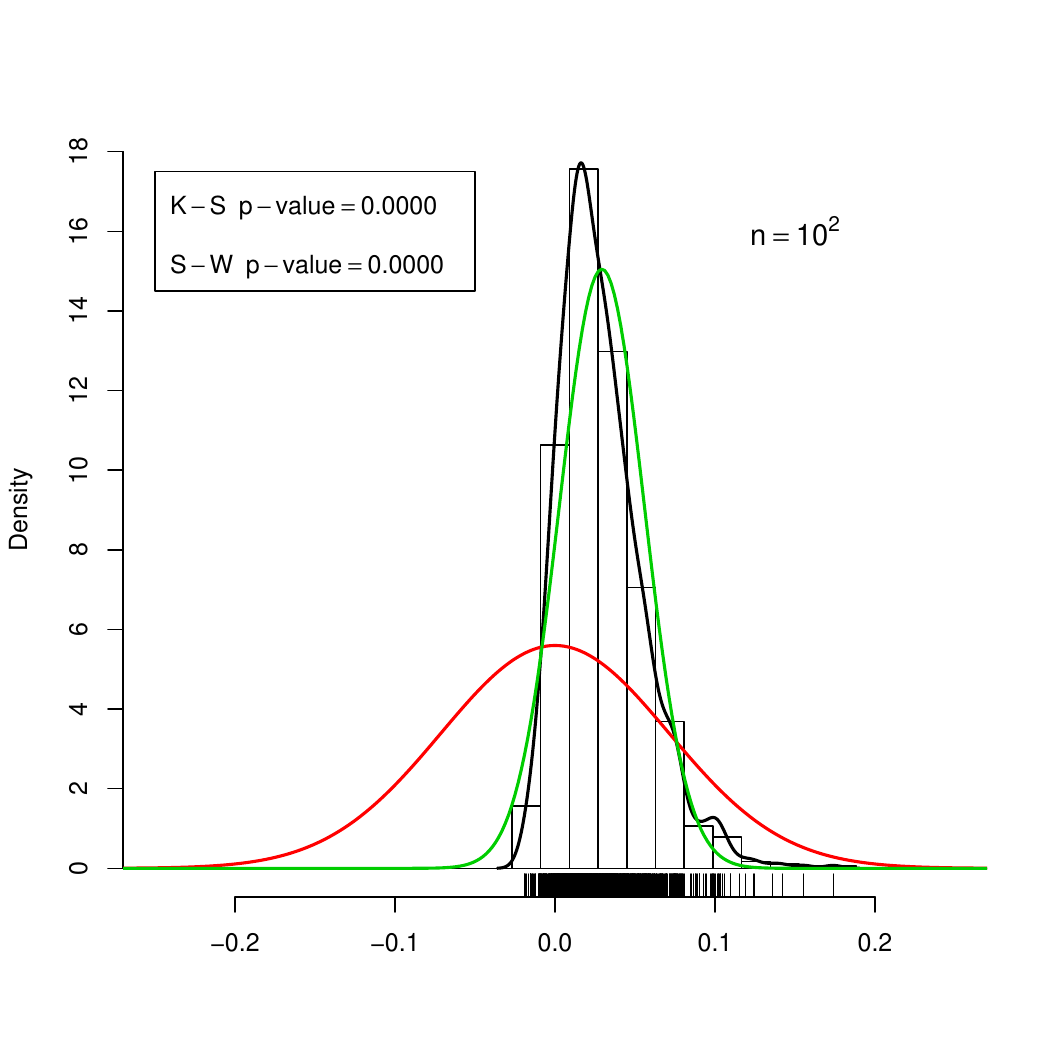}\includegraphics[height=0.25\textheight]{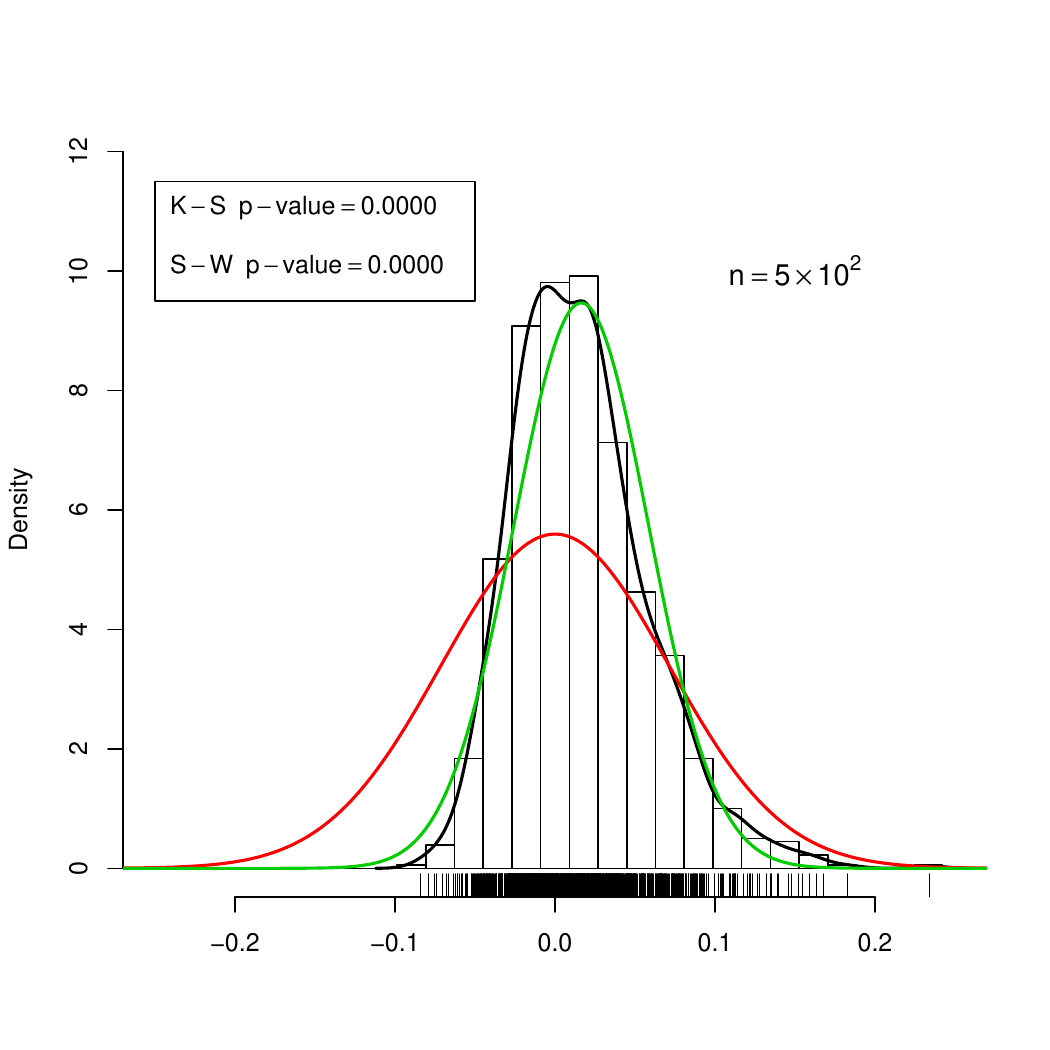}\\[-0.5cm]
	\includegraphics[height=0.25\textheight]{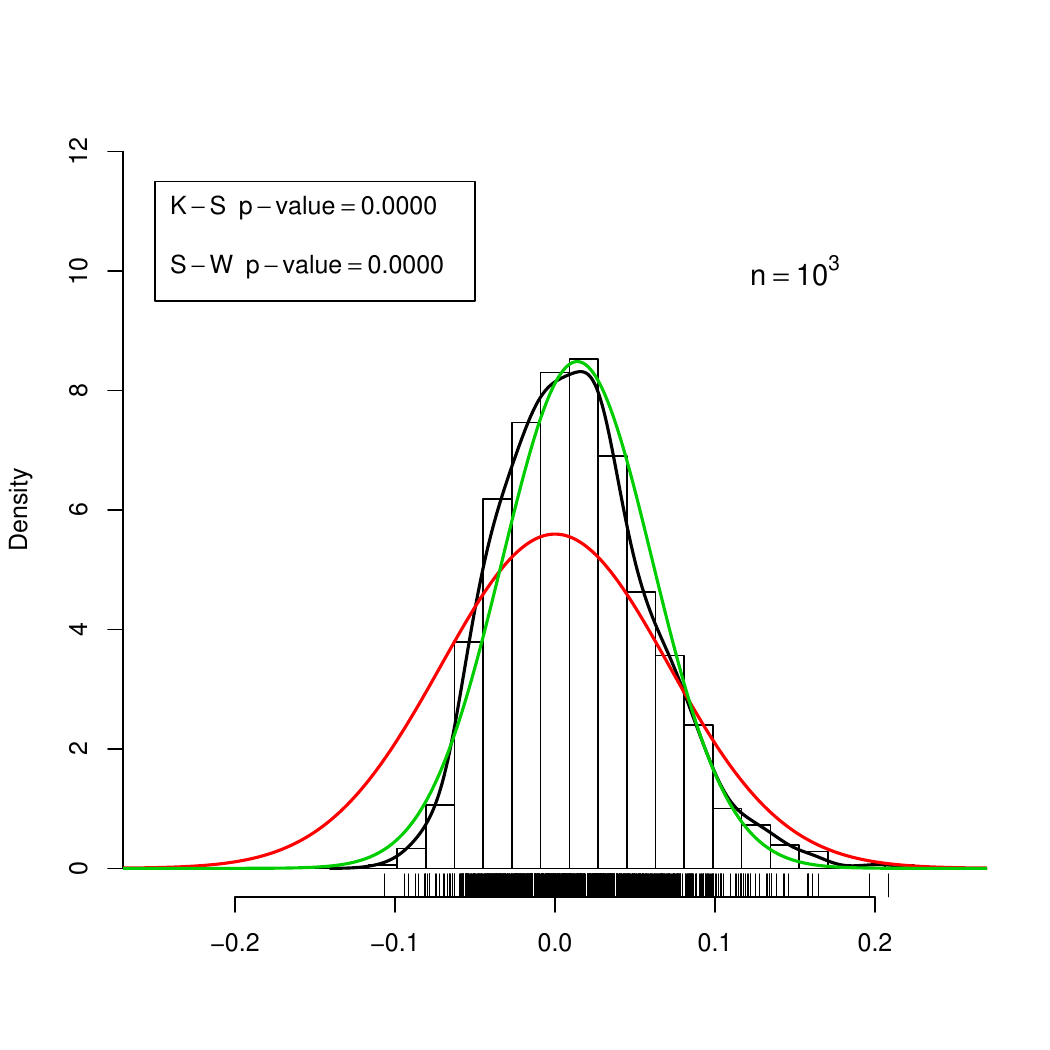}\includegraphics[height=0.25\textheight]{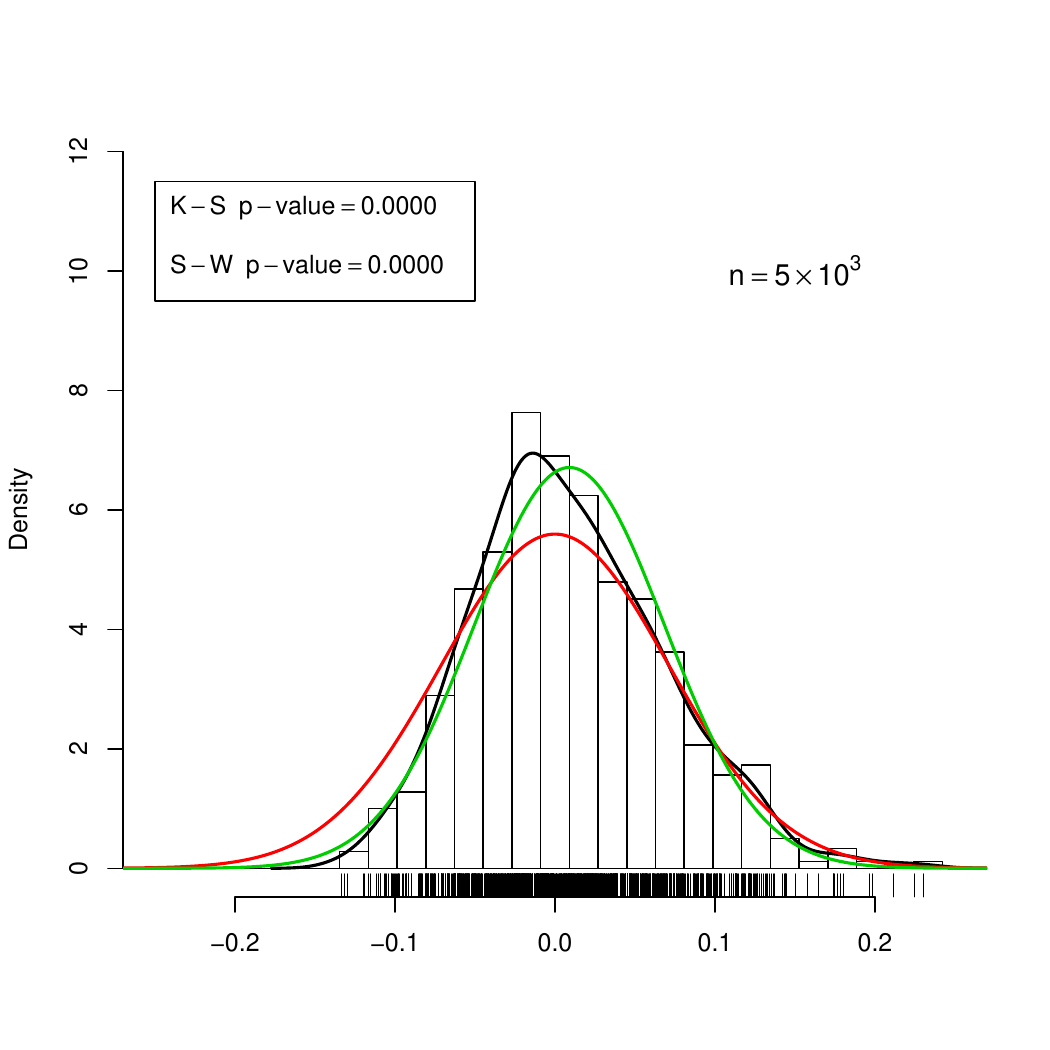}\\[-0.5cm]
	\includegraphics[height=0.25\textheight]{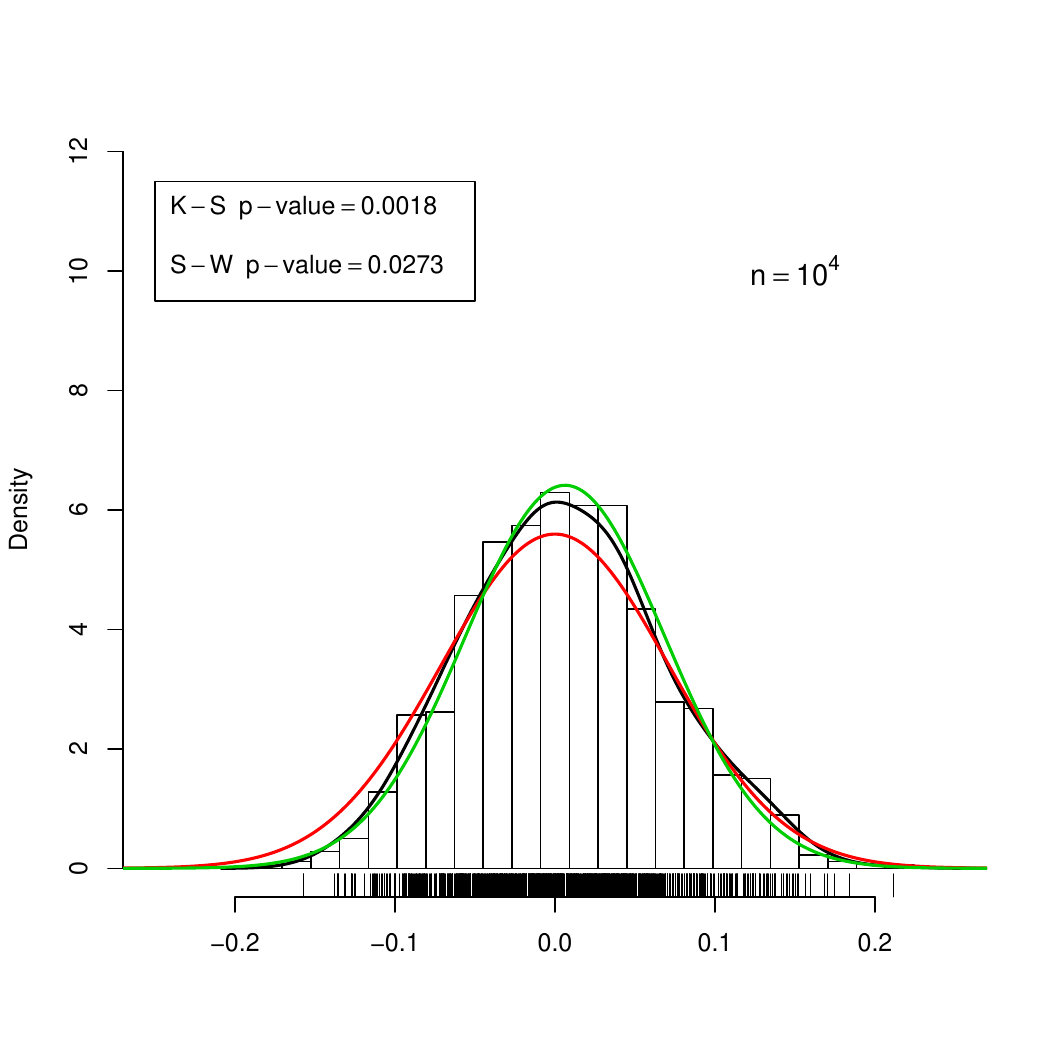}\includegraphics[height=0.25\textheight]{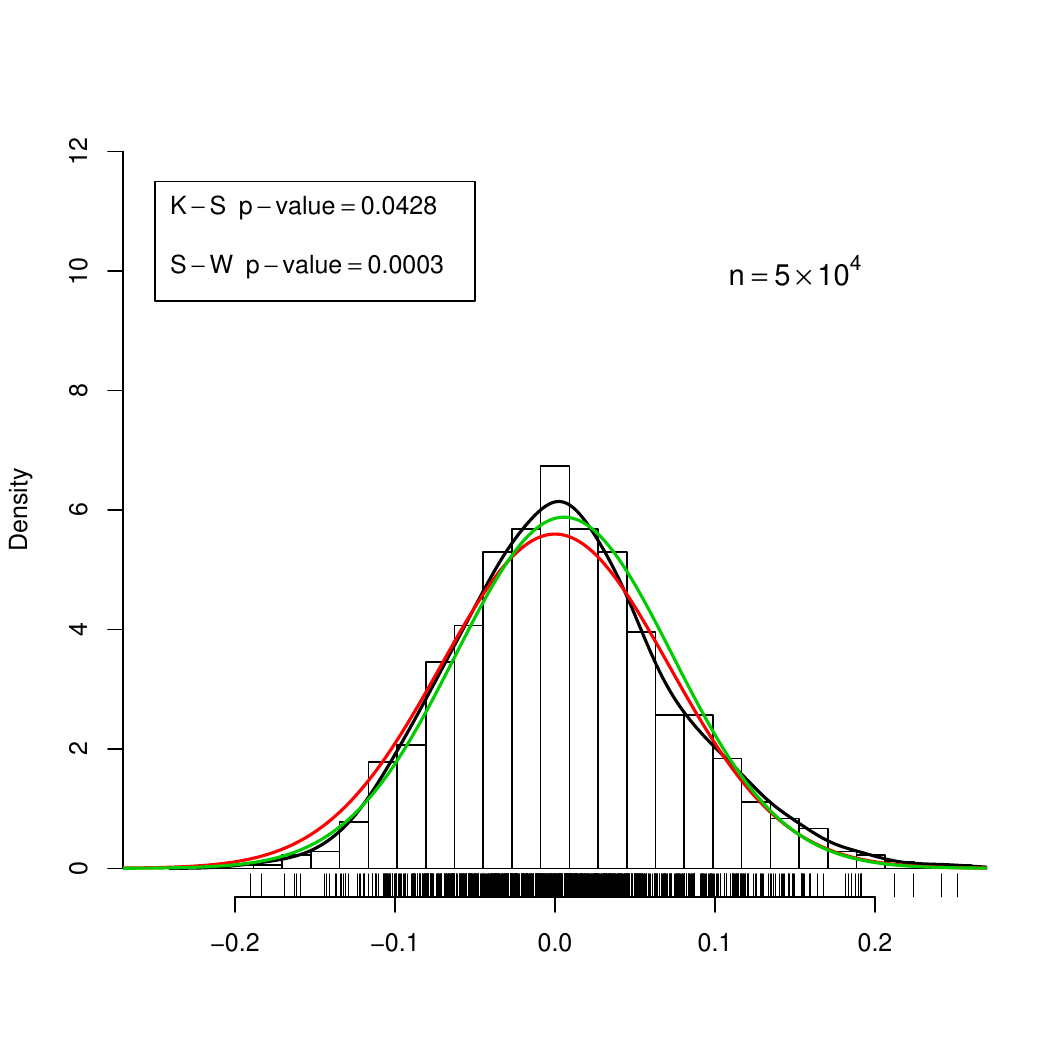}\\[-0.5cm]
	\includegraphics[height=0.25\textheight]{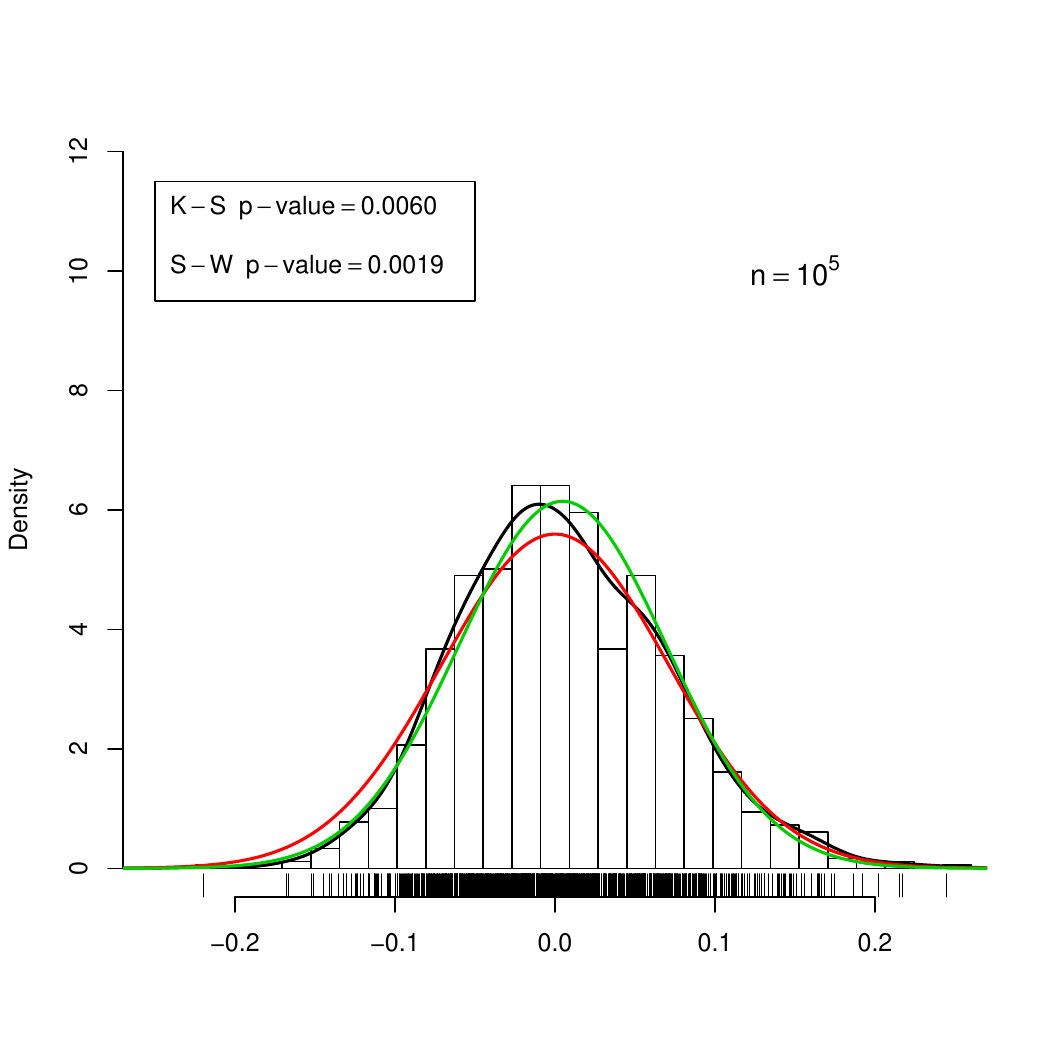}\includegraphics[height=0.25\textheight]{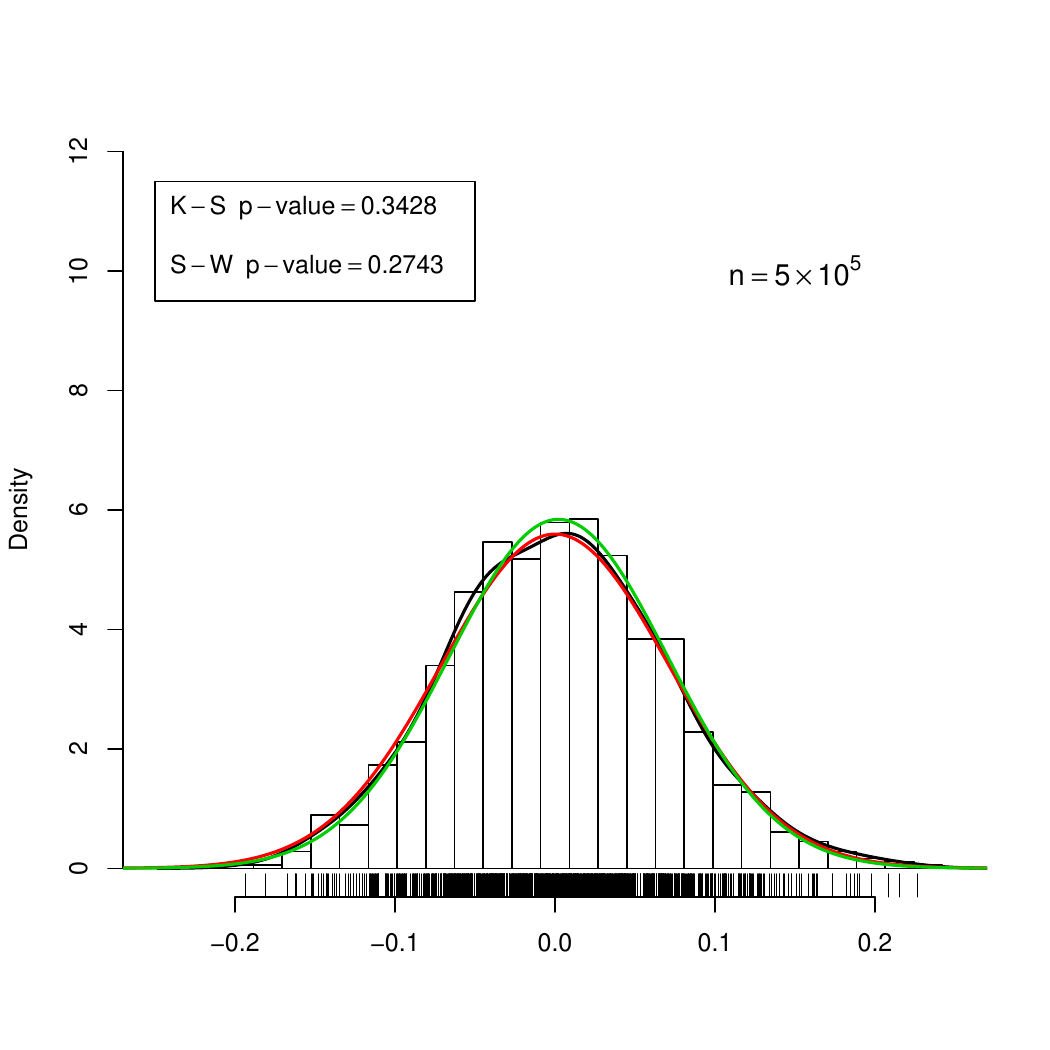}\\[-0.5cm]
	\caption{\small Comparison of the asymptotic and empirical distributions of $(nh_n^qg_n)^\frac{1}{2}\lrp{T_n-A_n}$ for sample sizes $n=5^j\times10^k$, $j=0,1$, $k=2,3,4,5$. Black curves represent a kernel estimation from $1000$ simulations, green curves represent a normal fit to the unknown density and red curves represent the theoretical asymptotic distribution. \label{gofdens:fig1:ext}}
\end{figure}

%-------------------------------------------------%
\section{Extended simulation study}
\label{gofdens:su:simus}
%-------------------------------------------------%

Some technical details concerning the simulation study and further results are provided in this section. First, the simulated models considered will be described. For constructing the test statistic, parametric estimators as well as simulation methods are required. Different Maximum Likelihood Estimators (MLE) and simulation approaches have been considered, playing copulas a remarkable role in both problems (see \cite{Nelsen2006} for a comprehensive review). Some details on the construction of alternative models and bandwidth choice will be also given, jointly with extended results showing the performance of the tests (for circular-linear and circular-circular cases) for different significance levels. 

%-------------------------------------------------%
\subsection{Parametric models}
%-------------------------------------------------%

Two collections of Circular-Linear (CL) and Circular-Circular (CC) parametric scenarios have been considered. The corresponding density contours can be seen in Figures \ref{gofdens:fig:cl} and \ref{gofdens:fig:cc} in the paper. For the circular-linear case, the first five models (CL1--CL5) contain parametric densities with independent components and different kinds of marginals, for which estimation and simulation are easily accomplished. The models are based on von Mises, wrapped Cauchy, wrapped normal, normal, log-normal, gamma and mixtures of these densities. Models CL6--CL7 represent two parametric choices of the model in \cite{Mardia1978} for cylindrical variables, which is constructed conditioning a normal density on a von Mises one. Models CL8--CL9 include two parametric densities of the semiparametric circular-linear model given in Theorem 5 of \cite{Johnson1978}. This family is indexed by a circular density $g$ that defines the underlying circular-linear copula density, allowing for flexibility both in the specification of the link density and the marginals. CL10 is the model given in Theorem 1 of \cite{Johnson1978}, which considers an exponential density conditioned on a von Mises. CL11 is constructed considering the QS copula density of \cite{Garcia-Portugues:so2} and cardioid and log-normal marginals. Finally, CL12 is an adaptation of the circular-circular copula density of \cite{Kato2009} to the circular-linear scenario, using an identity matrix in the joint structure and von Mises and log-normal marginals. \\

The first models (CC1--CC5) of the circular-circular case include also parametric densities with independent components and different kinds of marginals (von Mises, wrapped Cauchy, cardioid and mixtures of them). Models CC6--CC7 represent two parametric choices of the sine model given by \cite{Singh2002}. This model introduces elliptical contours for bivariate circular densities and also allows for certain multimodality. Models CC8--CC9 are two densities of the semiparametric models of \cite{Wehrly1980}, which are based on the previous work of \cite{Johnson1978} and comprise as a particular case the bivariate von Mises model of \cite{Shieh2005}. Models CC10--CC11 are two parametric choices of the wrapped normal torus density given in \cite{Johnson1977}, a natural extension of the circular wrapped normal to the circular-circular setting. Finally, CC12 employs the copula density of \cite{Kato2009} with von Mises marginals.

\begin{table}[h]
	\centering
	\small
	\renewcommand{\arraystretch}{1.4} %
	\begin{tabular}{l|l}
		\toprule\toprule
		Density name &  Expression \\
		\midrule
		Normal & $f_{\mathcal{N}}(z;m,\sigma)=\frac{1}{\sqrt{2\pi}\sigma}\exp\lrb{-\frac{(z-m)^2}{2\sigma^2}}$\\
		Log-normal & $f_{\mathcal{LN}}(z;m,\sigma)=\frac{1}{z\sqrt{2\pi}\sigma}\exp\lrb{-\frac{(\log z-m)^2}{2\sigma^2}}\mathbbm{1}_{(0,\infty)}(z)$\\
		Gamma & $f_\Gamma(z;a,p)=\frac{a^p}{\Gamma(p)}z^{p-1}e^{-az}\mathbbm{1}_{(0,\infty)}(z)$\\
		Bivariate normal & $f_\mathcal{N}(z_1,z_2;m_1,m_2,\sigma_1,\sigma_2,\rho)=\frac{1}{2\pi\sigma_1\sigma_2\sqrt{1-\rho^2}}$\\ & $\times\exp\Big\{-\frac{1}{2(1-\rho^2)}\Big(\frac{(z_1-m_1)^2}{\sigma_1^2}+\frac{(z_2-m_2)^2}{\sigma_2^2}-\frac{2\rho(z_1-m_1)(z_2-m_2)}{\sigma_1\sigma_2}\Big)\Big\}$\\
		\midrule
		Von Mises & $f_{\mathrm{vM}}(\theta;\mu,\kappa)=\frac{1}{2\pi\mathcal{I}_0(\kappa)}\exp\lrb{\kappa\cos(\theta-\mu)}$\\
		Cardioid & $f_{\mathrm{Ca}}(\theta;\mu,\rho)=\frac{1}{2\pi}\lrp{1+2\rho\cos(\theta-\mu)}$\\
		Wrapped Cauchy & $f_{\mathrm{WC}}(\theta;m,\sigma)=\frac{1-\rho^2}{2\pi\lrp{1+\rho^2-2\rho\cos(\theta-\mu)}}$\\
		Wrapped Normal & $f_{\mathrm{WN}}(\theta;\mu,\rho)=\sum_{p=-\infty}^\infty f_\mathcal{N}(\theta+2\pi p;m,\sigma)$\\
		\bottomrule\bottomrule
	\end{tabular}
	\caption{\small Notation for the densities described in Tables \ref{gofdens:tab:cirlin} and \ref{gofdens:tab:circir}.\label{gofdens:tab:basics}}
\end{table}

The notation and density expressions used for the construction of the parametric models are collected in Table \ref{gofdens:tab:basics}, whereas Tables \ref{gofdens:tab:cirlin} and \ref{gofdens:tab:circir} show the explicit expressions and parameters for the circular-linear and circular-circular models displayed in Figure \ref{gofdens:fig:cl}. Most of the circular densities considered in the simulation study are purely circular (and hence not directional) and their circular formulation has been used in order to simplify expressions. The directional notation can be obtained taking into account that $\bx=(\cos\theta,\sin\theta)$, $\by=(\cos\psi,\sin\psi)$ and $\bmu=(\cos\mu,\sin\mu)$. The distribution function of a circular variable with density $f$, with $\theta\in[0,2\pi)$ will be denoted by $F(\theta)=\int_0^{\theta} f(\varphi)\,d\varphi$.

%-------------------------------------------------%
\subsection{Estimation}
%-------------------------------------------------%

In the scenarios considered, for most of the marginal densities, MLE are available through specific libraries of \texttt{R}. For the normal and log-normal densities closed expressions are used and for the gamma density the \texttt{fitdistr} function of the \texttt{MASS} \citep{MASS} library is employed. The estimation of the von Mises parameters is done exactly for the mean and numerically for the concentration parameter, whereas for the wrapped Cauchy and wrapped normal densities the numerical routines of the \texttt{circular} \citep{circular} package are used. The MLE for the cardioid density are obtained by numerical optimization. Finally, the fitting of mixtures of normals and von Mises was carried out using the Expectation-Maximization algorithms given in packages \texttt{nor1mix} \citep{nor1mix} and \texttt{movMF} \citep{movMF}, respectively.\\

The fitting of the independent models CL1--CL5 and CC1--CC5 is easily accomplished by marginal fitting of each component. For models CL6--CL7, the closed expressions for the MLE given in \cite{Mardia1978} are used. For models CL8--CL9, CL11--CL12, CC8--CC9 and CC12 a two-step Maximum Likelihood (ML) estimation procedure based on the copula density decomposition is used: first, the marginals are fitted by ML and then the copula is estimated by ML using the pseudo-observations computed from the fitted marginals. This procedure is described in more detail in Section 3 of \cite{Garcia-Portugues:so2}. In models CL8--CL9 and CC8--CC9 the MLE for the copula are obtained by estimating univariate von Mises or mixtures of von Mises, whereas numerical optimization is required for the copula estimation. For models CC6--CC7 and CC10--CC11, MLE can be also carried out by numerical optimization. Finally, MLE for model CL10 in \cite{Johnson1978} were obtained analytically: given the circular-linear sample $\lrb{(\Theta_i,Z_i)}_{i=1}^n$,
\[
\hat{\lambda}=\frac{\bar Z}{(\bar Z)^2-(\bar{Z_c})^2},\quad\hat{\kappa}=\sqrt{\hat{\lambda}^2-\hat\lambda \bar Z^{-1}}\quad\text{and}\quad\sum_{i=1}^n Z_i\sin(\Theta_i-\hat{\mu})=0,
\]
with $\bar Z=\frac{1}{n}\sum_{i=1}^nZ_i$ and $\bar Z_c=\frac{1}{n}\sum_{i=1}^nZ_i\cos(\Theta_i-\hat{\mu})$.

%-------------------------------------------------%
\subsection{Simulation}
\label{gofdens:subsec:sim}
%-------------------------------------------------%

Simulating from the linear marginals is easily accomplished by the built-in functions in \texttt{R}. The simulation of the wrapped Cauchy and wrapped normal is done with the \texttt{circular} library, the von Mises is sampled implementing the algorithm described in \cite{Wood1994} and the cardioid by the inversion method, whose equation is solved numerically. Sampling from the independence models is straightforward. Conditioning on the circular variable, it is easy to sample from models CL6--CL7 (sample the circular observation from a von Mises and then the linear from a normal with mean depending on the circular), CL10 (von Mises marginal and exponential with varying rate) and CC6--CC7 (using the properties detailed in \cite{Singh2002} and the inversion method). Simulation in CC10--CC11 is straightforward: sample from a bivariate normal and then wrap around $[0,2\pi)$ by applying a modulus of $2\pi$. Finally, simulation in two steps using copulas was required for models CL8--CL9, CL12, CC8--CC9 and CC12, where first a pair of uniform random variables $(U,V)$ is sampled from the copula of the density and then the inversion method is applied marginally. See Section 3.1 of \cite{Garcia-Portugues:so2} for more details. The simulation of the pair $(U,V)$ was done by the conditional and inversion methods and, specifically, for the models based on the densities given by \cite{Johnson1978} and \cite{Wehrly1980}, a transformation method was obtained. It is summarized in the following algorithm.

\begin{sidewaystable}
	\centering
	\footnotesize
		\renewcommand{\arraystretch}{1.4} %
	\begin{tabular}{l|>{\centering\arraybackslash}p{7.5cm}|p{6cm}|p{5.25cm}}
		\toprule\toprule
		Model & \multicolumn{1}{c|}{Density} & \multicolumn{1}{c|}{Parameters} & \multicolumn{1}{c}{Description} \\
		\midrule
		CL1  & $f_{\mathrm{vM}}(\theta;\mu,\kappa)\times f_{\mathcal{N}}(z;m,\sigma)$ & $\mu=\frac{3\pi}{2}$, $\kappa=2$, $m=0$, $\sigma=1$ & Independent von Mises and normal\\\midrule
		CL2  & $f_{\mathrm{WC}}(\theta;\mu,\rho)\times f_{\mathcal{LN}}(z;m,\sigma)$ & $\mu=\frac{3\pi}{2}$, $\rho=\sigma=\frac{3}{4}$, $m=\frac{1}{2}$ &  Independent wrapped Cauchy and log-normal\\\midrule
		CL3  & $\big(p_1f_{\mathrm{vM}}(\theta;\mu_1,\kappa_1)+p_2f_{\mathrm{vM}}(\theta;\mu_2,\kappa_2)\big)\times f_{\Gamma}(z;a,p)$ & $\mu_1=\frac{\pi}{4}$, $\mu_2=\frac{5\pi}{4}$, $\kappa_1=\kappa_2=2$, $p_1=p_2=\frac{1}{2}$, $a=\frac{1}{3}$, $p=3$ &  Independent mixture of von Mises and gamma\\\midrule
		CL4  & $f_{\mathrm{WN}}(\theta;m_1,\sigma_1)$ $\times\lrp{p_1f_{\mathcal{N}}(z;m_2,\sigma_2)+p_2f_{\mathcal{N}}(z;m_3,\sigma_3)}$ & $m_1=\frac{3\pi}{2}$, $\sigma_1=\sigma_3=1$, $m_2=0$, $\sigma_2=\frac{1}{4}$, $m_3=2$, $p_1=p_2=\frac{1}{2}$ &  Independent wrapped normal and mixture of normals\\\midrule
		CL5  & $\big(p_1f_{\mathrm{vM}}(\theta;\mu_1,\kappa_1)+p_2f_{\mathrm{vM}}(\theta;\mu_2,\kappa_2)\big)$  $\times\big(p_3f_{\mathcal{N}}(z;m_1,\sigma_1)+p_4f_{\mathcal{N}}(z;m_2,\sigma_2)\big)$ & $\mu_1=\frac{5\pi}{4}$, $\mu_2=\frac{7\pi}{4}$, $\kappa_1=10$, $\kappa_2=3$, $m_1=-1$, $m_2=2$, $\sigma_1=1$, $\sigma_2=p_1=p_2=\frac{1}{2}$, $p_3=\frac{3}{4}$, $p_4=\frac{1}{4}$ &  Independent mixture of von Mises and of normals\\\midrule
		CL6  & \multirow{2}{7.5cm}{\centering\arraybackslash $f_{\mathrm{vM}}(\theta;\mu,\kappa)\times f_{\mathcal{N}}\big(z;m(\theta),\sigma(1-\rho_1-\rho_2)\big)$, with $m(\theta)=m+\sigma\kappa^\frac{1}{2}\{\rho_1(\cos(\theta)-\cos(\mu))$ $+\rho_2(\sin(\theta)-\sin(\mu))\}$} & $\mu=\frac{3\pi}{2}$, $\kappa=1$, $m=0$, $\rho_1=\rho_2=\sigma=\frac{1}{2}$ & \multirow{2}{5.25cm}{See equation (1.1) of \cite{Mardia1978}} \\\cmidrule{3-3}\cmidrule{1-1}
		CL7  &  & $\mu=\frac{3\pi}{2}$, $\kappa=5$, $m=0$, $\rho_1=\frac{1}{2}$, $\rho_2=-\frac{3}{4}$, $\sigma=\frac{3}{2}$ &  \\\midrule
		CL8  & $f_{\mathrm{vM}}\lrp{2\pi\big(\frac{\theta}{2\pi}+F_\mathcal{N}(z;m,\sigma)\big);\mu_g,\kappa_g}$ $\times f_\mathcal{N}(z;m,\sigma)$  & $m=0$, $\sigma=1$, $\mu_g=\frac{5\pi}{4}$, $\kappa_g=\frac{3}{2}$ & \multirow{2}{5.25cm}{See Theorem 5 of \cite{Johnson1978} considering a von Mises and a mixture of von Mises as the link functions} \\\cmidrule{1-3}
		CL9  & $g\big(
		2\pi\big(\frac{\theta}{2\pi}-F_\mathcal{N}(z;m,\sigma)\big)\big)\times f_\mathcal{N}(z;m,\sigma)$, with $g(\theta)=p_{g_1}f_{\mathrm{vM}}\lrp{\theta;\mu_{g_1},\kappa_{g_1}}$ $+p_{g_2}f_{\mathrm{vM}}\lrp{\theta;\mu_{g_2},\kappa_{g_2}}$ & $m=0$, $\sigma=p_{g_1}=p_{g_2}=\frac{1}{2}$, $\mu_{g_1}=\frac{\pi}{4}$, $\kappa_{g_1}=\kappa_{g_2}=3$,  $\mu_{g_2}=\frac{5\pi}{4}$ &   \\\midrule
		CL10 & $\frac{(\lambda^2-\kappa^2)^\frac{1}{2}}{2\pi}\exp\lrb{-\lambda z+\kappa z\cos(\theta-\mu)}$ & $\mu=\frac{3\pi}{2}$, $\kappa=2$, $\lambda=3$ & See Theorem 1 of \cite{Johnson1978} \\\midrule
		CL11 & $\big\{1+2\pi\alpha\cos(2\pi F_{\mathrm{Ca}}(\theta;\mu,\rho))(1-2F_\mathcal{N}(z;m,\sigma))\big\}$ $\times f_{\mathrm{Ca}}(\theta;\mu,\rho)f_\mathcal{N}(z;m,\sigma)$ &  $\mu=\frac{3\pi}{2}$, $\rho=\frac{9}{20}$, $m=1$, $\sigma=\frac{1}{2}$, $\alpha=\frac{1}{2\pi}$ & See equation (7) of \cite{Garcia-Portugues:so2} \\\midrule
		CL12 & $ \lrb{4\pi^2\lrp{1-2\rho F_{\mathrm{vM}}(\theta;\mu,\kappa)F_\mathcal{LN}(z;m,\sigma)+\rho^2}}^{-1}$ $\times(1-\rho^2) f_{\mathrm{vM}}(\theta;\mu,\kappa)f_\mathcal{LN}(z;m,\sigma)$ & $\mu=\frac{3\pi}{2}$, $\kappa=1$, $m=\frac{1}{2}$, $\sigma=\rho=\frac{3}{4}$ & See Section 4.1 in \cite{Kato2009}\\
		\bottomrule\bottomrule
	\end{tabular}
	\caption{\small Circular-linear models.\label{gofdens:tab:cirlin}}
\end{sidewaystable}

\begin{sidewaystable}
	\centering
	\footnotesize
		\renewcommand{\arraystretch}{1.4} %
		\begin{tabular}{l|>{\centering\arraybackslash}p{7.5cm}|p{6cm}|p{5.25cm}}
		\toprule\toprule
		Model & \multicolumn{1}{c|}{Density} & \multicolumn{1}{c|}{Parameters} & \multicolumn{1}{c}{Description} \\
		\midrule
		CC1  & $\frac{1}{2\pi}\times f_{\mathrm{vM}}(\psi;\mu,\kappa)$ & $\mu=0$, $\kappa=2$ & Independent uniform and von Mises\\\midrule
		CC2  & $f_{\mathrm{vM}}(\theta;\mu_1,\kappa_1)\times f_{\mathrm{vM}}(\psi;\mu_2,\kappa_2)$ & $\mu_1=\frac{3\pi}{2}$, $\kappa_1=1$, $\mu_2=\pi$, $\kappa_2=3$ & Independent von Mises and von Mises\\\midrule
		CC3  & $f_{\mathrm{vM}}(\theta;\mu_1,\kappa)\times f_{\mathrm{WC}}(\psi;\mu_2,\rho)$ & $\mu_1=\frac{3\pi}{2}$, $\kappa=2$, $\mu_2=\frac{\pi}{4}$, $\rho=\frac{7}{10}$ & Independent von Mises and wrapped Cauchy\\\midrule
		CC4  & $\big(p_1f_{\mathrm{vM}}(\theta;\mu_1,\kappa_1)+p_2f_{\mathrm{vM}}(\theta;\mu_2,\kappa_2)\big)\times f_{\mathrm{Ca}}(\psi;\mu_3,\rho)$ & $\mu_1=0$, $\kappa_1=\kappa_2=10$, $\mu_2=\frac{3\pi}{2}$, $\mu_3=0$, $\rho=\frac{1}{4}$, $p_1=p_2=\frac{1}{2}$ & Independent mixture von Mises and cardioid\\\midrule
		CC5  & $\big(p_1f_{\mathrm{vM}}(\theta;\mu_1,\kappa_1)+p_2f_{\mathrm{vM}}(\theta;\mu_2,\kappa_2)\big)$ $\times\big(p_3f_{\mathrm{vM}}(\psi;\mu_3,\kappa_3)+p_4f_{\mathrm{vM}}(\psi;\mu_4,\kappa_4))$ & $\mu_1=0$, $\kappa_1=\kappa_2=3$, $\mu_2=\frac{3\pi}{2}$, $\mu_3=\frac{\pi}{4}$, $\kappa_3=\kappa_4=5$, $\mu_4=\frac{7\pi}{4}$, $p_1=p_2=p_3=p_4=\frac{1}{2}$ & Independent mixture of von Mises and of von Mises \\\midrule
		CC6  & \multirow{2}{7.5cm}{\centering\arraybackslash$ C\exp\big\{\kappa_1\cos(\theta-\mu_1)+\kappa_2\cos(\psi-\mu_2)$ $+\lambda\sin(\theta-\mu_1)\sin(\psi-\mu_2)\big\}$ %
		} & $\mu_1=\frac{7\pi}{8}$, $\kappa_1=\frac{1}{2}$, $\mu_2=0$, $\kappa_2=1$, $\lambda=-3$ & \multirow{2}{5.25cm}{See equation (1.1) of \cite{Singh2002}} \\\cmidrule{1-1}\cmidrule{3-3}
		CC7  &  & $\mu_1=\mu_2=0$, $\kappa_1=5$, $\kappa_2=1$, $\lambda=-5$ &  \\\midrule
		CC8  &  $f_{\mathrm{vM}}\lrp{2\pi\big(F_{\mathrm{Ca}}(\theta;\mu,\rho)-\frac{\psi}{2\pi}\big);\mu_g,\kappa_g}\times f_{\mathrm{Ca}}(\theta;\mu,\rho)$  & $\mu=0$, $\rho=\frac{1}{2}$, $\mu_g=\pi$, $\kappa_g=7$ & \multirow{2}{5.25cm}{See equations (1) and (2) in \cite{Wehrly1980} with a von Mises and a mixture of von Mises as links}  \\\cmidrule{1-3}
		CC9  &  $\frac{1}{2\pi}\big( p_{g_1}f_{\mathrm{vM}}\lrp{\theta+\psi;\mu_{g_1},\kappa_{g_1}}+p_{g_2}f_{\mathrm{vM}}\lrp{\theta+\psi;\mu_{g_2},\kappa_{g_2}}\big)$ & $\mu_{g_1}=\frac{\pi}{4}$, $\kappa_{g_1}=\kappa_{g_2}=10$, $\mu_{g_2}=\frac{7\pi}{4}$, $p_{g_1}=p_{g_2}=\frac{1}{2}$ & \\\midrule
		CC10 & \multirow{2}{7.5cm}{\centering\arraybackslash$\sum_{p_1=-\infty}^\infty\sum_{p_1=-\infty}^\infty $ $f_\mathcal{N}(\theta+2\pi p_1,\psi+2\pi p_2;m_1,m_2,\sigma_1,\sigma_2,\rho)$} & $m_1=0$, $m_2=\frac{\pi}{6}$, $\sigma_1=\frac{3}{2}$, $\sigma_2=\frac{1}{4}$, $\rho=0$ & \multirow{2}{5.25cm}{See Example 7.3 in \cite{Johnson1977}} \\\cmidrule{3-3}\cmidrule{1-1}
		CC11 &  & $m_1=m_2=0$, $\sigma_1=\sigma_2=1$, $\rho=-\frac{9}{10}$ &  \\\midrule
		CC12 &  $ \lrb{4\pi^2\lrp{1-2\rho F_{\mathrm{vM}}(\theta;\mu_1,\kappa_1)F_{\mathrm{vM}}(\psi,\mu_2,\kappa_2)+\rho^2}}^{-1}$ $\times(1-\rho^2) f_{\mathrm{vM}}(\theta;\mu_1,\kappa_1)f_{\mathrm{vM}}(\psi;\mu_2,\kappa_2)$ & $\mu_1=\frac{3\pi}{4}$, $\kappa_1=5$, $\mu_2=0$, $\kappa_2=1$, $\rho=\frac{1}{2}$ & See Section 4.1 of \cite{Kato2009}\\
		\bottomrule\bottomrule
	\end{tabular}
	\caption{\small Circular-circular models. \label{gofdens:tab:circir}}
\end{sidewaystable}
\begin{algo}
	Let $g$ be a circular density. A pair $(U,V)$ of uniform variables with joint density $c_g(u,v)=2\pi g(2\pi(u\pm v))$ is obtained as follows:
	\begin{enumerate}[label=\textit{\roman{*}}.]
		\item Sample $\Psi$, a random variable with circular density $g$.
		\item Sample $V$, a uniform variable in $[0,1]$.
		\item Set $U=\frac{(\Psi\mp 2\pi V)\mod 2\pi}{2\pi}$.
	\end{enumerate}
\end{algo}

%-------------------------------------------------%
\subsection{Alternative models}
%-------------------------------------------------%

The alternative hypothesis for the goodness-of-fit test, both in the circular-linear and circular-circular cases, is stated as:
\[
H_{k,\delta}: f=(1-\delta)f^k_{\btheta_0}+\delta\Delta,\quad 0\leq\delta\leq1.
\]
Three mixing densities $\Delta$ are considered, two for the circular-linear situation and one for the circular-circular:
\begin{align*}
\Delta_1(\theta,z)&=f_{\mathrm{vM}}(\theta;\mu_1,\kappa)\times f_\mathcal{N}(z;m_1,\sigma_1),\\
\Delta_2(\theta,z)&=f_{\mathrm{vM}}(\theta;\mu_1,\kappa)\times f_\mathcal{LN}(z;m_2,\sigma_2),\\
\Delta_3(\theta,\psi)&=f_{\mathrm{vM}}(\theta;\mu_2,\kappa)\times f_{\mathrm{vM}}(\psi;\mu_1,\kappa),
\end{align*}
where $\mu_1=\pi$, $\mu_2=0$, $\kappa=3$, $m_1=2$, $\sigma_1=1$ and $m_2=\sigma_2=\frac{1}{2}$. To account for similar ranges in the linear data obtained under $H_{k,0}$ and under $H_{k,\delta}$, $\Delta_1$ is used in models CL1, CL4--CL11 and CL13, whereas $\Delta_2$ in the other models. In the circular-circular case, the deviation for all models is $\Delta_3$.

\begin{figure}[htpb]
	\centering
	\vspace*{-0.25cm}
	\includegraphics[trim=2cm 0cm 7cm 7cm,clip=true,scale=0.175]{grid_CL1.png}\includegraphics[trim=2cm 0cm 7cm 7cm,clip=true,scale=0.175]{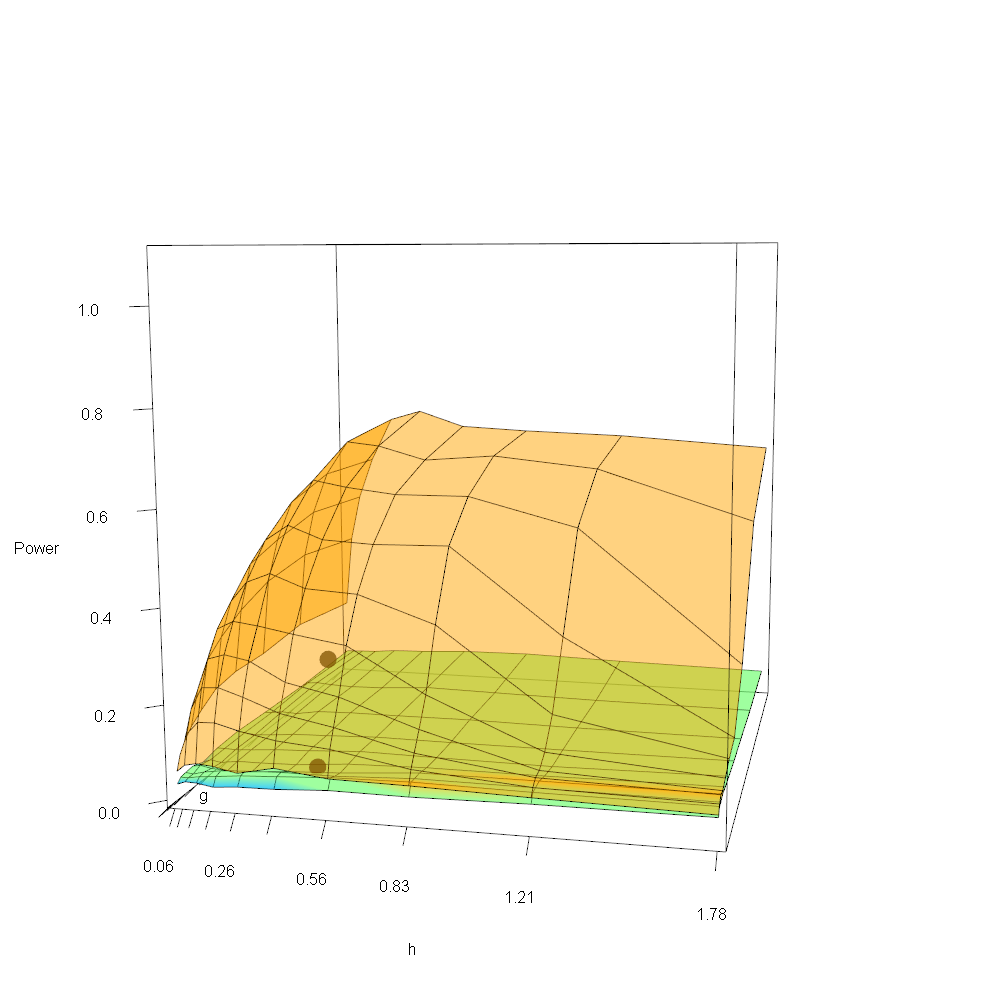}\includegraphics[trim=2cm 0cm 7cm 7cm,clip=true,scale=0.175]{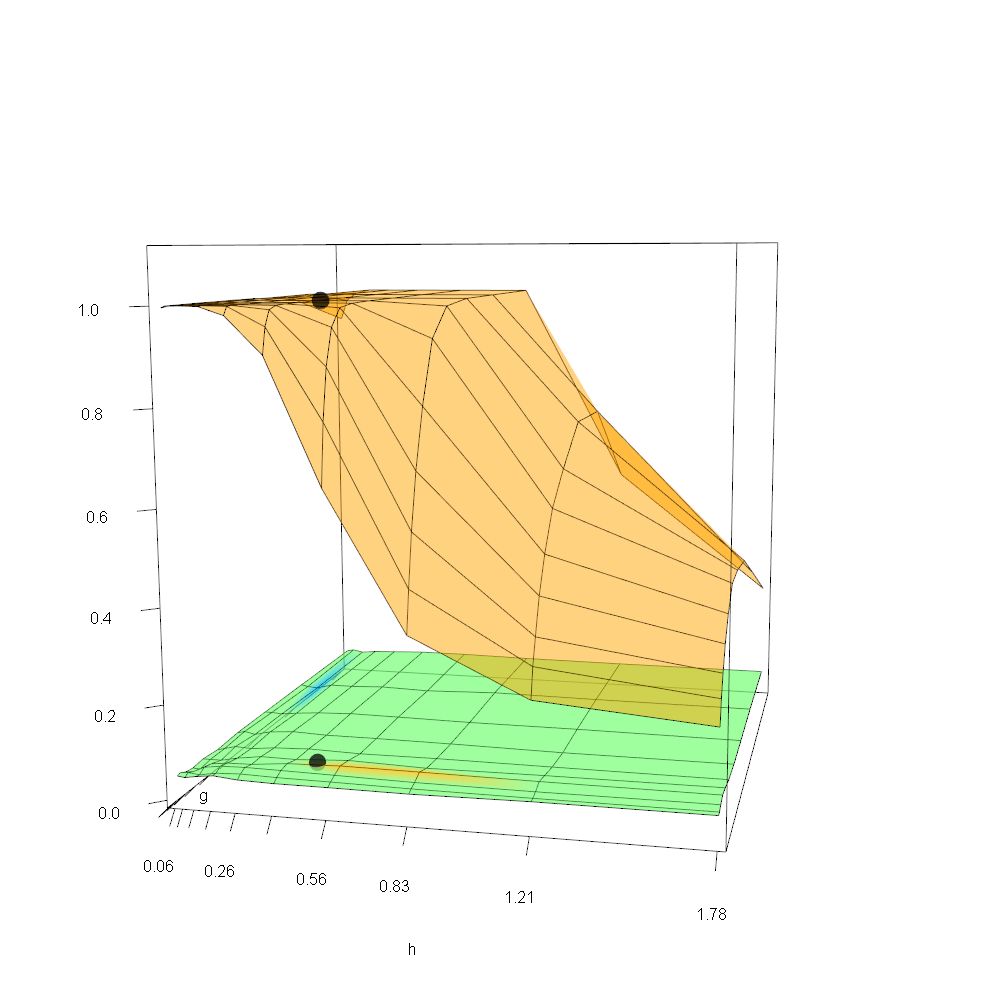}\\
	\includegraphics[trim=2cm 0cm 7cm 7cm,clip=true,scale=0.175]{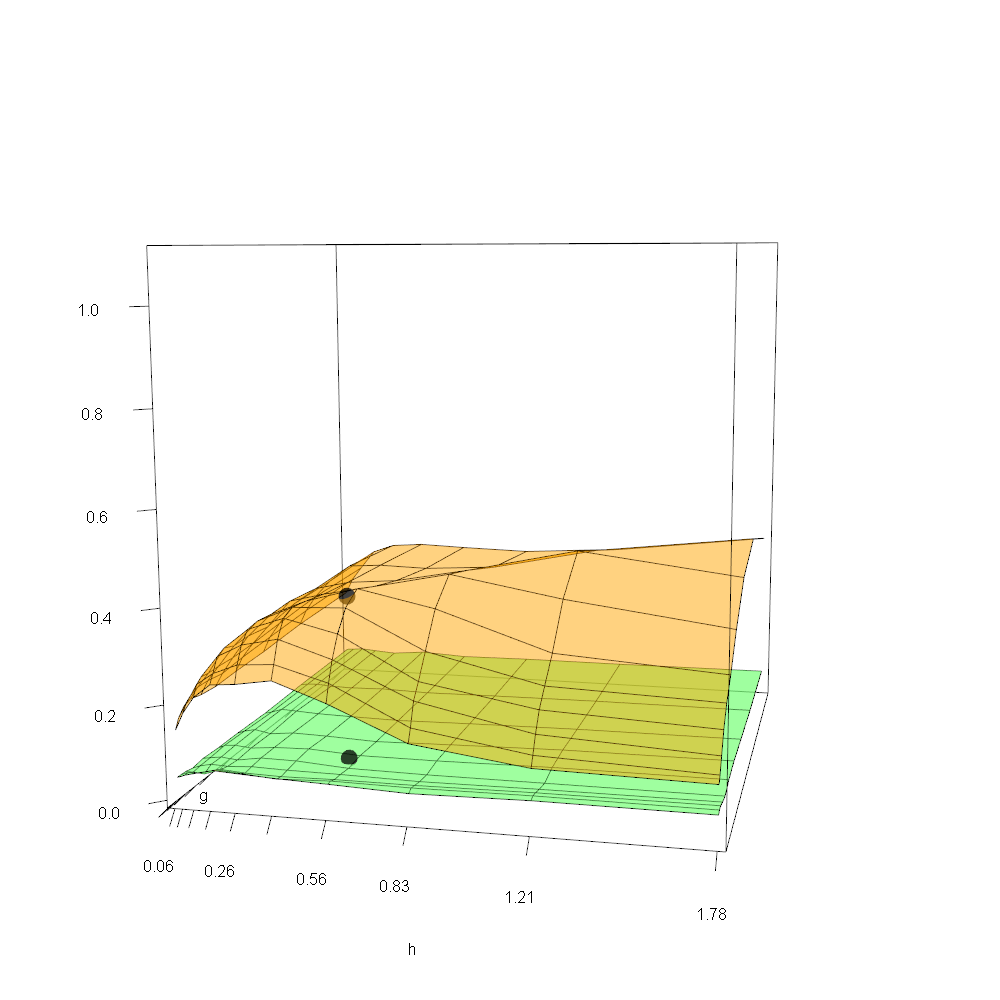}\includegraphics[trim=2cm 0cm 7cm 7cm,clip=true,scale=0.175]{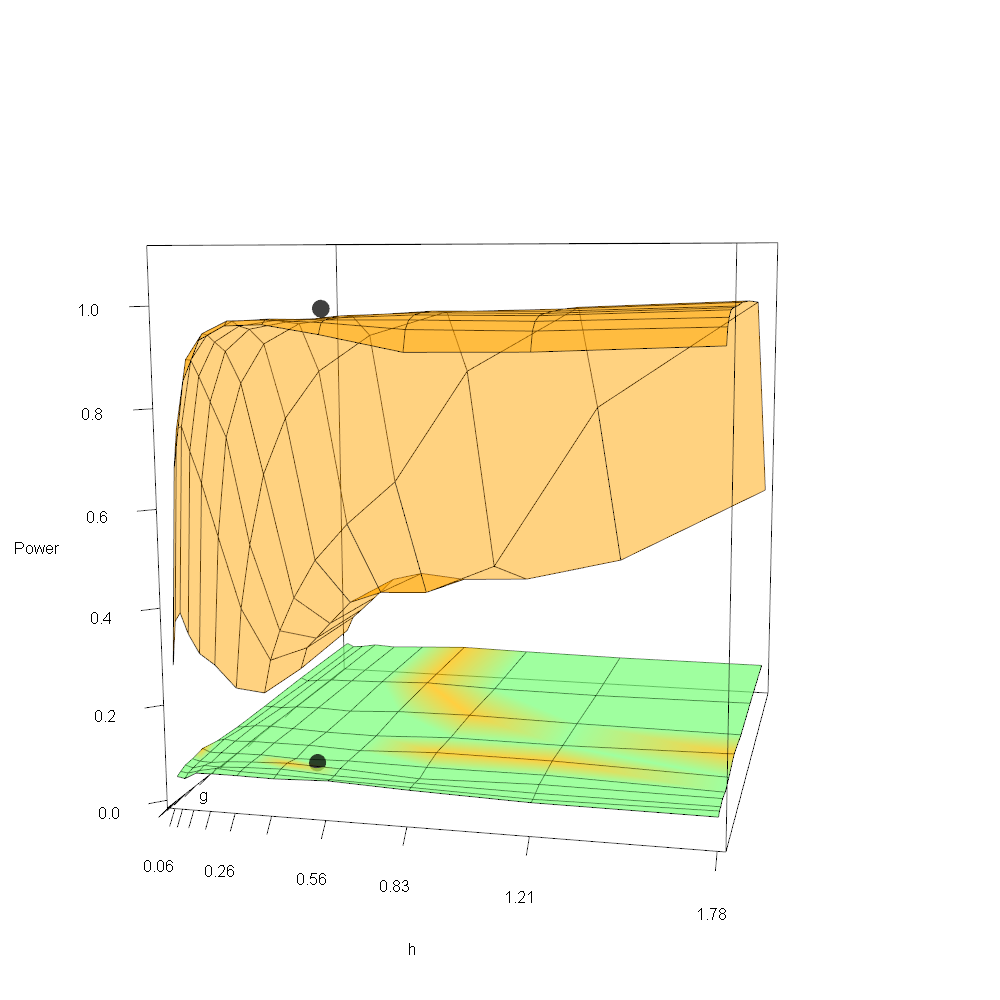}\includegraphics[trim=2cm 0cm 7cm 7cm,clip=true,scale=0.175]{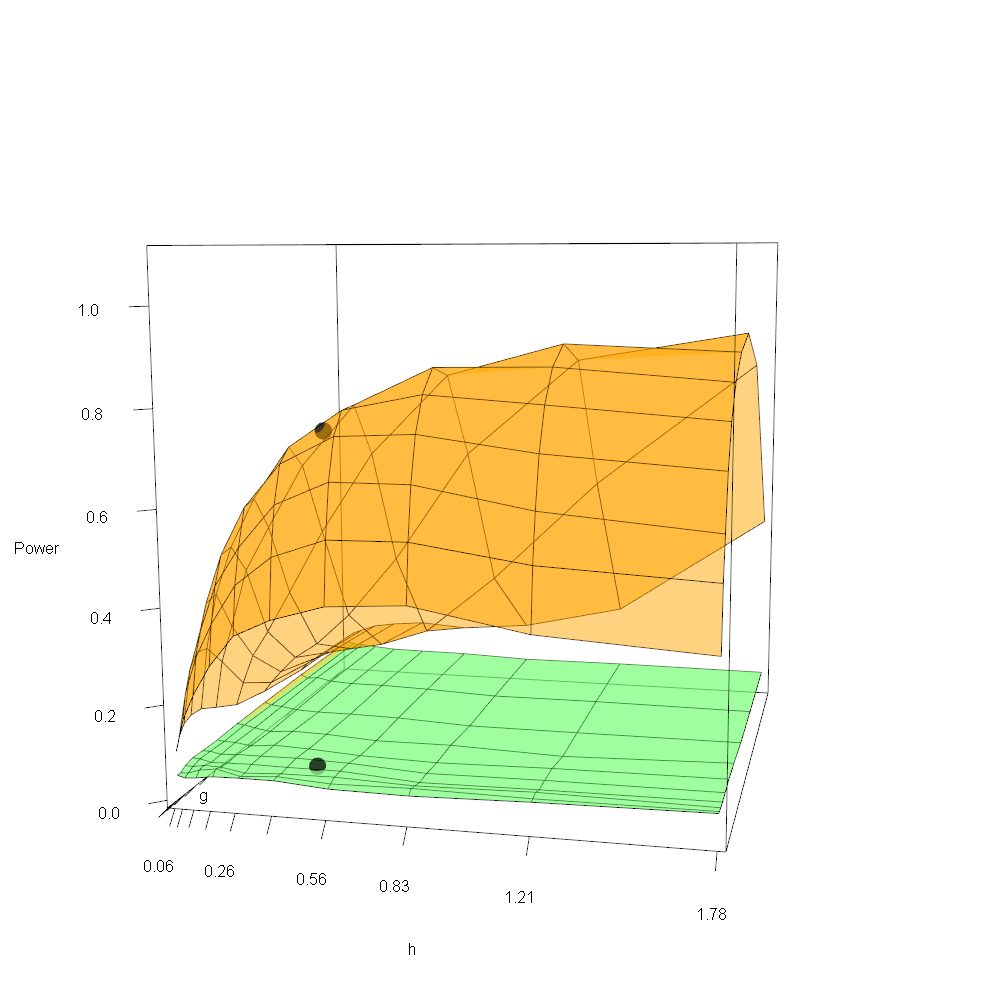}\\
	\includegraphics[trim=2cm 0cm 7cm 7cm,clip=true,scale=0.175]{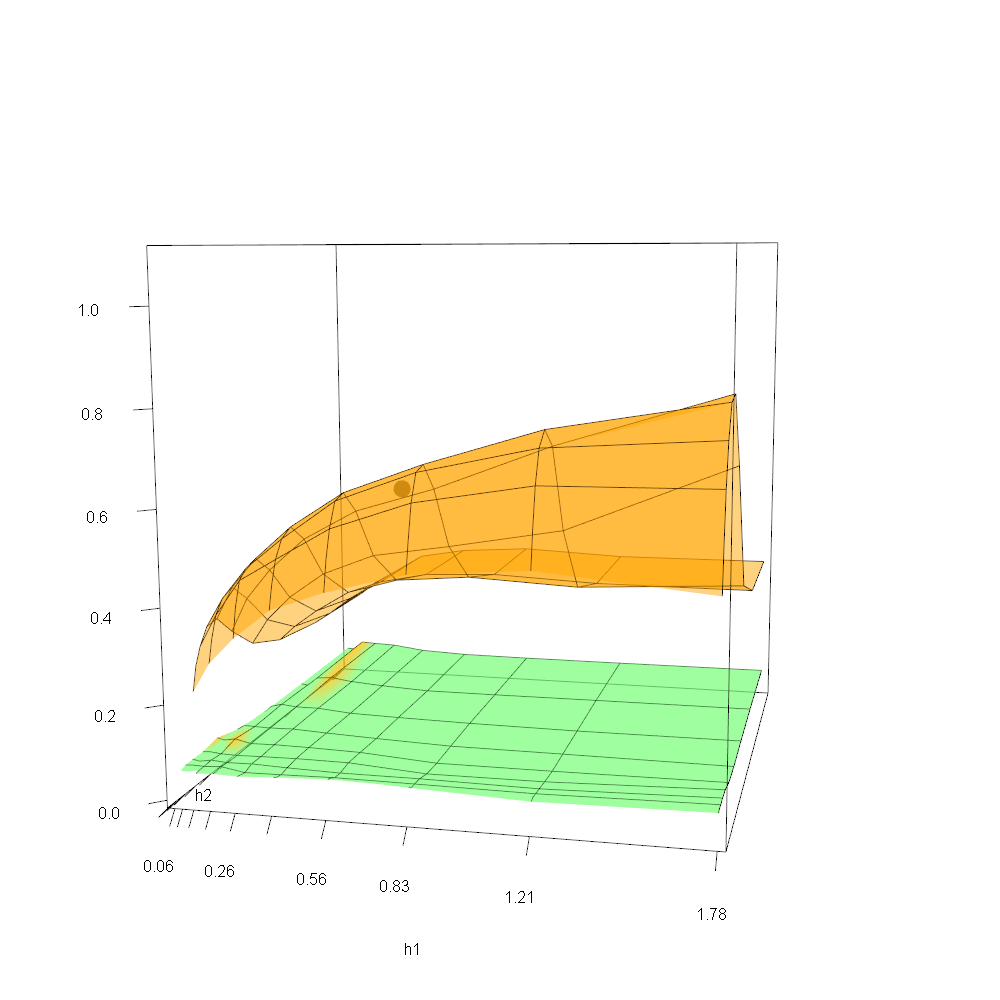}\includegraphics[trim=2cm 0cm 7cm 7cm,clip=true,scale=0.175]{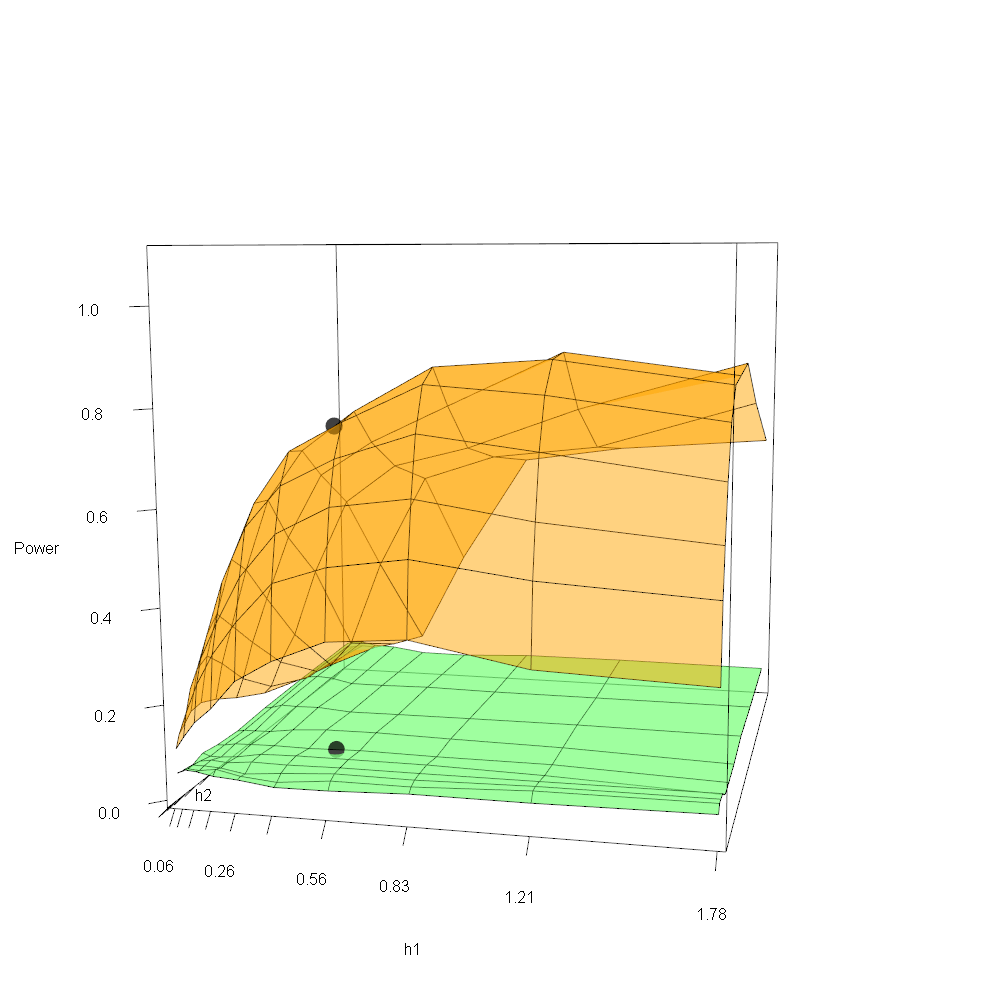}\includegraphics[trim=2cm 0cm 7cm 7cm,clip=true,scale=0.175]{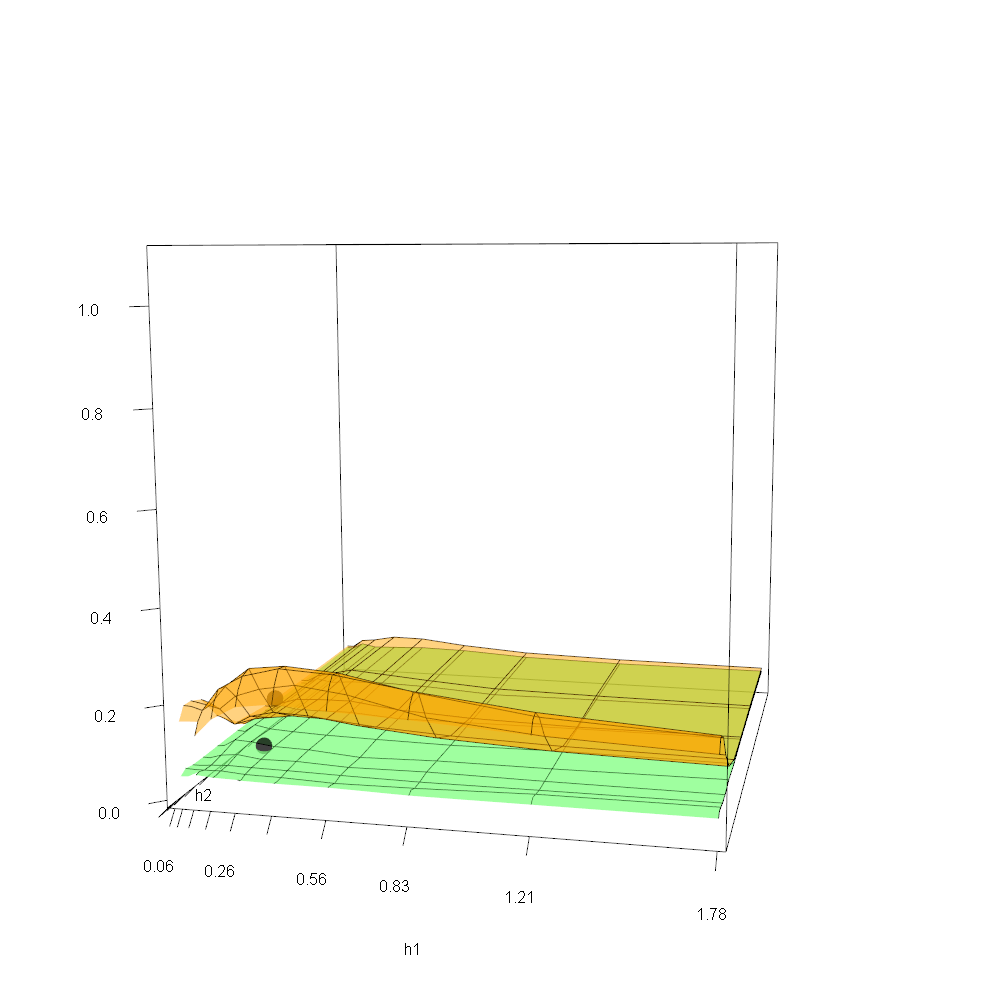}\\
	\includegraphics[trim=2cm 0cm 7cm 7cm,clip=true,scale=0.175]{grid_CC8.png}\includegraphics[trim=2cm 0cm 7cm 7cm,clip=true,scale=0.175]{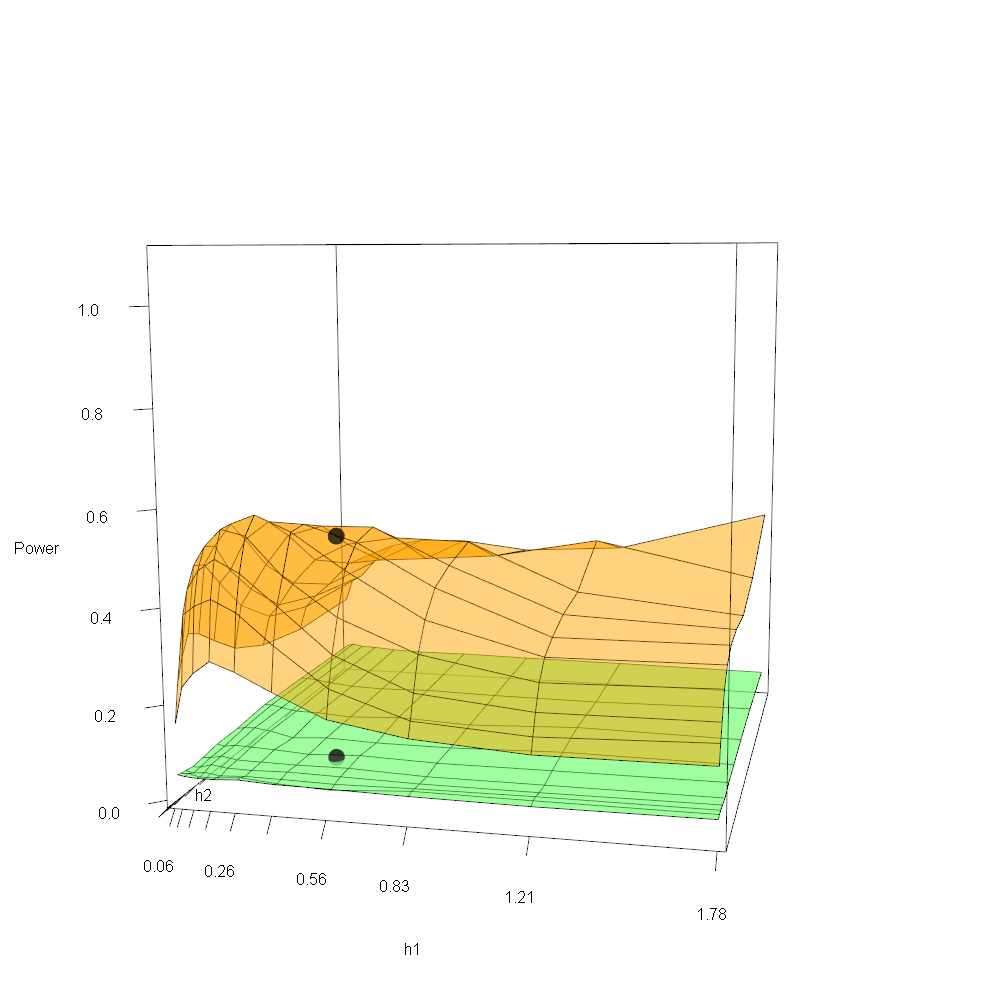}\includegraphics[trim=2cm 0cm 7cm 7cm,clip=true,scale=0.175]{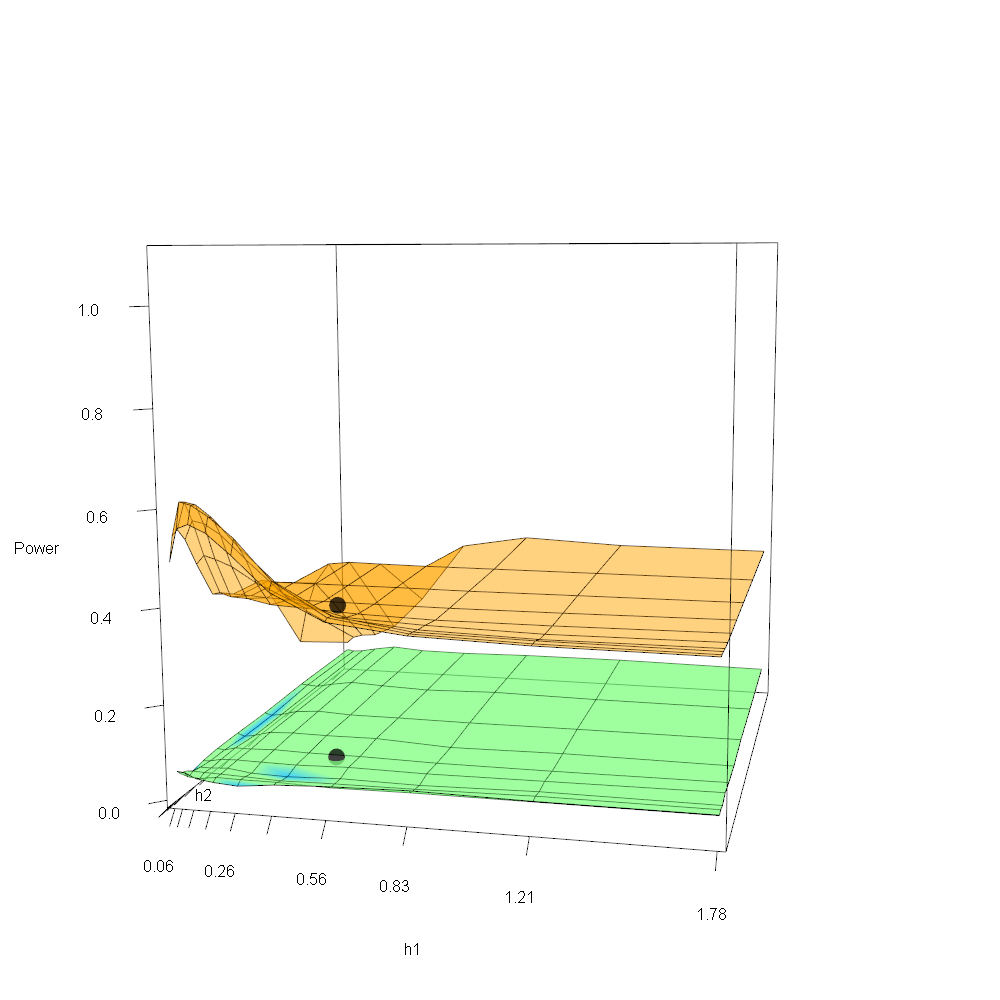}\\[-0.25cm]
	\caption{\small Empirical size and power of the goodness-of-fit tests for a $10\times10$ grid of bandwidths. First two rows, from left to right and up to down: models CL1, CL5, CL7, CL8, CL9 and CL11. Last two rows: CC1, CC5, CC7, CC8, CC9 and CC11. Lower surface represents the empirical rejection rate under $H_{0.00}$ and upper surface under $H_{0.15}$. Green colour represent that the empirical rejection is in the $95\%$ confidence interval of $\alpha=0.05$, blue that is lower and orange that is larger. Black points represent the sized and powers obtained with the median of the LCV bandwidths (for model CC1 under $H_0$ is outside the grid). \label{gofdens:fig:band:ext}}
\end{figure}

\begin{table}[htpb!]
	\centering
	\footnotesize
	\begin{tabular}{l|ccc|ccc|ccc}
		\toprule\toprule
		\multirow{3}{*}{Model} & \multicolumn{9}{c}{Sample size $n$ and significance level $\alpha$} \\\cmidrule(lr){2-10}
		& \multicolumn{3}{c|}{$n=100$} & \multicolumn{3}{c|}{$n=500$} & \multicolumn{3}{c}{$n=1000$}  \\\cmidrule(lr){2-4} \cmidrule(lr){5-7} \cmidrule(lr){8-10}
		& $\alpha$=0.10 & $\alpha$=0.05 & $\alpha$=0.01 &  $\alpha$=0.10 & $\alpha$=0.05 & $\alpha$=0.01  &  $\alpha$=0.10 & $\alpha$=0.05 & $\alpha$=0.01  \\
		\midrule
		$H_{1,0.00}$  & $0.111$ & $0.051$ & $0.010$ & $0.107$ & $0.052$ & $0.013$ & $0.102$ & $0.048$ & $0.013$ \\
		$H_{2,0.00}$  & $0.094$ & $0.051$ & $0.013$ & $0.096$ & $0.049$ & $0.010$ & $0.107$ & $0.050$ & $0.009$ \\
		$H_{3,0.00}$  & $0.095$ & $0.048$ & $0.014$ & $0.101$ & $0.046$ & $0.014$ & $0.090$ & $0.050$ & $0.009$ \\
		$H_{4,0.00}$  & $0.102$ & $0.045$ & $0.009$ & $0.096$ & $0.039$ & $0.011$ & $0.102$ & $0.045$ & $0.008$ \\
		$H_{5,0.00}$  & $0.094$ & $0.049$ & $0.009$ & $0.102$ & $0.049$ & $0.009$ & $0.101$ & $0.041$ & $0.009$ \\
		$H_{6,0.00}$  & $0.095$ & $0.039$ & $0.010$ & $0.104$ & $0.043$ & $0.010$ & $0.110$ & $0.050$ & $0.015$ \\
		$H_{7,0.00}$  & $0.086$ & $0.042$ & $0.013$ & $0.093$ & $0.043$ & $0.008$ & $0.091$ & $0.049$ & $0.016$ \\
		$H_{8,0.00}$  & $0.095$ & $0.049$ & $0.011$ & $0.108$ & $0.050$ & $0.003$ & $0.108$ & $0.044$ & $0.006$ \\
		$H_{9,0.00}$  & $0.106$ & $0.062$ & $0.016$ & $0.086$ & $0.043$ & $0.010$ & $0.104$ & $0.064$ & $0.015$ \\
		$H_{10,0.00}$ & $0.094$ & $0.045$ & $0.007$ & $0.103$ & $0.056$ & $0.018$ & $0.097$ & $0.045$ & $0.005$ \\
		$H_{11,0.00}$ & $0.102$ & $0.059$ & $0.009$ & $0.104$ & $0.056$ & $0.010$ & $0.113$ & $0.056$ & $0.013$ \\
		$H_{12,0.00}$ & $0.120$ & $0.073$ & $0.020$ & $0.113$ & $0.054$ & $0.013$ & $0.109$ & $0.051$ & $0.010$ \\\midrule
		$H_{1,0.10}$  & $0.665$ & $0.552$ & $0.355$ & $1.000$ & $0.997$ & $0.981$ & $1.000$ & $1.000$ & $1.000$ \\
		$H_{2,0.10}$  & $0.361$ & $0.244$ & $0.107$ & $0.885$ & $0.805$ & $0.579$ & $0.995$ & $0.982$ & $0.898$ \\
		$H_{3,0.10}$  & $0.185$ & $0.107$ & $0.032$ & $0.502$ & $0.362$ & $0.166$ & $0.775$ & $0.659$ & $0.421$ \\
		$H_{4,0.10}$  & $0.255$ & $0.172$ & $0.060$ & $0.687$ & $0.568$ & $0.322$ & $0.927$ & $0.868$ & $0.697$ \\
		$H_{5,0.10}$  & $0.416$ & $0.272$ & $0.087$ & $0.987$ & $0.972$ & $0.894$ & $1.000$ & $1.000$ & $0.999$ \\
		$H_{6,0.10}$  & $0.997$ & $0.996$ & $0.988$ & $1.000$ & $1.000$ & $1.000$ & $1.000$ & $1.000$ & $1.000$ \\
		$H_{7,0.10}$  & $1.000$ & $1.000$ & $0.999$ & $1.000$ & $1.000$ & $1.000$ & $1.000$ & $1.000$ & $1.000$ \\
		$H_{8,0.10}$  & $0.325$ & $0.204$ & $0.069$ & $0.940$ & $0.893$ & $0.723$ & $1.000$ & $1.000$ & $0.983$ \\
		$H_{9,0.10}$  & $0.947$ & $0.914$ & $0.796$ & $1.000$ & $1.000$ & $1.000$ & $1.000$ & $1.000$ & $1.000$ \\
		$H_{10,0.10}$ & $0.340$ & $0.218$ & $0.089$ & $0.829$ & $0.723$ & $0.481$ & $0.962$ & $0.944$ & $0.838$ \\
		$H_{11,0.10}$ & $0.618$ & $0.510$ & $0.296$ & $0.996$ & $0.993$ & $0.963$ & $1.000$ & $1.000$ & $1.000$ \\
		$H_{12,0.10}$ & $0.230$ & $0.152$ & $0.057$ & $0.788$ & $0.655$ & $0.442$ & $0.991$ & $0.969$ & $0.895$ \\\midrule
		$H_{1,0.15}$  & $0.883$ & $0.822$ & $0.621$ & $1.000$ & $1.000$ & $1.000$ & $1.000$ & $1.000$ & $1.000$ \\
		$H_{2,0.15}$  & $0.650$ & $0.525$ & $0.311$ & $1.000$ & $0.997$ & $0.977$ & $1.000$ & $1.000$ & $1.000$ \\
		$H_{3,0.15}$  & $0.281$ & $0.163$ & $0.055$ & $0.776$ & $0.682$ & $0.420$ & $0.970$ & $0.940$ & $0.860$ \\
		$H_{4,0.15}$  & $0.399$ & $0.297$ & $0.127$ & $0.910$ & $0.869$ & $0.724$ & $0.998$ & $0.993$ & $0.981$ \\
		$H_{5,0.15}$  & $0.663$ & $0.514$ & $0.235$ & $0.999$ & $0.999$ & $0.999$ & $1.000$ & $1.000$ & $1.000$ \\
		$H_{6,0.15}$  & $1.000$ & $1.000$ & $1.000$ & $1.000$ & $1.000$ & $1.000$ & $1.000$ & $1.000$ & $1.000$ \\
		$H_{7,0.15}$  & $1.000$ & $1.000$ & $1.000$ & $1.000$ & $1.000$ & $1.000$ & $1.000$ & $1.000$ & $1.000$ \\
		$H_{8,0.15}$  & $0.522$ & $0.379$ & $0.168$ & $0.999$ & $0.997$ & $0.976$ & $1.000$ & $1.000$ & $1.000$ \\
		$H_{9,0.15}$  & $0.996$ & $0.989$ & $0.962$ & $1.000$ & $1.000$ & $1.000$ & $1.000$ & $1.000$ & $1.000$ \\
		$H_{10,0.15}$ & $0.505$ & $0.378$ & $0.154$ & $0.988$ & $0.975$ & $0.893$ & $1.000$ & $1.000$ & $0.996$ \\
		$H_{11,0.15}$ & $0.838$ & $0.763$ & $0.567$ & $1.000$ & $1.000$ & $1.000$ & $1.000$ & $1.000$ & $1.000$ \\
		$H_{12,0.15}$ & $0.373$ & $0.254$ & $0.114$ & $0.989$ & $0.967$ & $0.872$ & $1.000$ & $1.000$ & $1.000$ \\  
		\bottomrule\bottomrule
	\end{tabular}
	\caption{\small Empirical size and power of the circular-linear goodness-of-fit test for models CL1--CL12 with different sample sizes, deviations and significance levels. \label{gofdens:tab:results:cl:ext}}
\end{table}

\begin{table}[htpb!]
	\centering
	\footnotesize
	\begin{tabular}{l|ccc|ccc|ccc}
		\toprule\toprule
		\multirow{3}{*}{Model} & \multicolumn{9}{c}{Sample size $n$ and significance level $\alpha$} \\\cmidrule(lr){2-10}
		& \multicolumn{3}{c|}{$n=100$} & \multicolumn{3}{c|}{$n=500$} & \multicolumn{3}{c}{$n=1000$}  \\\cmidrule(lr){2-4} \cmidrule(lr){5-7} \cmidrule(lr){8-10}
		& $\alpha$=0.10 & $\alpha$=0.05 & $\alpha$=0.01 &  $\alpha$=0.10 & $\alpha$=0.05 & $\alpha$=0.01  &  $\alpha$=0.10 & $\alpha$=0.05 & $\alpha$=0.01  \\
		\midrule
		$H_{1,0.00}$  & $0.102$ & $0.061$ & $0.016$ & $0.094$ & $0.047$ & $0.004$ & $0.103$ & $0.048$ & $0.008$ \\       
		$H_{2,0.00}$  & $0.094$ & $0.054$ & $0.007$ & $0.100$ & $0.043$ & $0.011$ & $0.096$ & $0.056$ & $0.012$ \\       
		$H_{3,0.00}$  & $0.103$ & $0.061$ & $0.009$ & $0.096$ & $0.042$ & $0.011$ & $0.113$ & $0.058$ & $0.011$ \\       
		$H_{4,0.00}$  & $0.094$ & $0.049$ & $0.010$ & $0.089$ & $0.048$ & $0.008$ & $0.108$ & $0.052$ & $0.016$ \\       
		$H_{5,0.00}$  & $0.117$ & $0.059$ & $0.011$ & $0.091$ & $0.050$ & $0.003$ & $0.090$ & $0.051$ & $0.009$ \\       
		$H_{6,0.00}$  & $0.101$ & $0.069$ & $0.055$ & $0.082$ & $0.045$ & $0.009$ & $0.074$ & $0.034$ & $0.009$ \\        
		$H_{7,0.00}$  & $0.095$ & $0.048$ & $0.010$ & $0.100$ & $0.059$ & $0.014$ & $0.105$ & $0.044$ & $0.005$ \\        
		$H_{8,0.00}$  & $0.094$ & $0.043$ & $0.014$ & $0.100$ & $0.054$ & $0.013$ & $0.097$ & $0.050$ & $0.011$ \\        
		$H_{9,0.00}$  & $0.094$ & $0.043$ & $0.009$ & $0.104$ & $0.057$ & $0.017$ & $0.098$ & $0.042$ & $0.012$ \\        
		$H_{10,0.00}$ & $0.095$ & $0.047$ & $0.005$ & $0.096$ & $0.041$ & $0.006$ & $0.088$ & $0.042$ & $0.010$ \\        
		$H_{11,0.00}$ & $0.088$ & $0.041$ & $0.008$ & $0.096$ & $0.047$ & $0.010$ & $0.108$ & $0.053$ & $0.013$ \\        
		$H_{12,0.00}$ & $0.117$ & $0.062$ & $0.023$ & $0.116$ & $0.058$ & $0.013$ & $0.092$ & $0.048$ & $0.016$ \\\midrule
		$H_{1,0.10}$  & $0.587$ & $0.456$ & $0.240$ & $0.996$ & $0.995$ & $0.961$ & $1.000$ & $1.000$ & $1.000$ \\ 
		$H_{2,0.10}$  & $0.634$ & $0.506$ & $0.300$ & $0.998$ & $0.994$ & $0.976$ & $1.000$ & $1.000$ & $1.000$ \\ 
		$H_{3,0.10}$  & $0.786$ & $0.706$ & $0.466$ & $1.000$ & $1.000$ & $1.000$ & $1.000$ & $1.000$ & $1.000$ \\ 
		$H_{4,0.10}$  & $0.890$ & $0.837$ & $0.665$ & $1.000$ & $1.000$ & $1.000$ & $1.000$ & $1.000$ & $1.000$ \\ 
		$H_{5,0.10}$  & $0.601$ & $0.431$ & $0.176$ & $1.000$ & $1.000$ & $0.999$ & $1.000$ & $1.000$ & $1.000$ \\ 
		$H_{6,0.10}$  & $0.237$ & $0.123$ & $0.059$ & $0.875$ & $0.759$ & $0.503$ & $0.982$ & $0.958$ & $0.859$ \\
		$H_{7,0.10}$  & $0.210$ & $0.112$ & $0.025$ & $0.838$ & $0.724$ & $0.429$ & $0.996$ & $0.989$ & $0.916$ \\
		$H_{8,0.10}$  & $0.794$ & $0.693$ & $0.480$ & $1.000$ & $1.000$ & $1.000$ & $1.000$ & $1.000$ & $1.000$ \\
		$H_{9,0.10}$ & $0.471$ & $0.325$ & $0.112$ & $1.000$ & $1.000$ & $1.000$ & $1.000$ & $1.000$ & $1.000$ \\  
		$H_{10,0.10}$ & $1.000$ & $1.000$ & $1.000$ & $1.000$ & $1.000$ & $1.000$ & $1.000$ & $1.000$ & $1.000$ \\
		$H_{11,0.10}$ & $0.985$ & $0.973$ & $0.910$ & $1.000$ & $1.000$ & $1.000$ & $1.000$ & $1.000$ & $1.000$ \\
		$H_{12,0.10}$ & $0.942$ & $0.899$ & $0.788$ & $1.000$ & $1.000$ & $1.000$ & $1.000$ & $1.000$ & $1.000$ \\\midrule
		$H_{1,0.15}$  & $0.847$ & $0.751$ & $0.521$ & $1.000$ & $1.000$ & $1.000$ & $1.000$ & $1.000$ & $1.000$ \\         
		$H_{2,0.15}$  & $0.862$ & $0.798$ & $0.627$ & $1.000$ & $1.000$ & $1.000$ & $1.000$ & $1.000$ & $1.000$ \\         
		$H_{3,0.15}$  & $0.958$ & $0.932$ & $0.830$ & $1.000$ & $1.000$ & $1.000$ & $1.000$ & $1.000$ & $1.000$ \\         
		$H_{4,0.15}$  & $0.981$ & $0.958$ & $0.885$ & $1.000$ & $1.000$ & $1.000$ & $1.000$ & $1.000$ & $1.000$ \\         
		$H_{5,0.15}$  & $0.847$ & $0.720$ & $0.445$ & $1.000$ & $1.000$ & $1.000$ & $1.000$ & $1.000$ & $1.000$ \\         
		$H_{6,0.15}$  & $0.443$ & $0.270$ & $0.097$ & $0.985$ & $0.960$ & $0.858$ & $0.997$ & $0.993$ & $0.982$ \\        
		$H_{7,0.15}$  & $0.357$ & $0.201$ & $0.043$ & $0.990$ & $0.976$ & $0.879$ & $1.000$ & $1.000$ & $1.000$ \\        
		$H_{8,0.15}$  & $0.969$ & $0.945$ & $0.842$ & $1.000$ & $1.000$ & $1.000$ & $1.000$ & $1.000$ & $1.000$ \\        
		$H_{9,0.15}$  & $0.719$ & $0.600$ & $0.345$ & $1.000$ & $1.000$ & $1.000$ & $1.000$ & $1.000$ & $1.000$ \\                 
		$H_{10,0.15}$ & $1.000$ & $1.000$ & $1.000$ & $1.000$ & $1.000$ & $1.000$ & $1.000$ & $1.000$ & $1.000$ \\      
		$H_{11,0.15}$ & $1.000$ & $1.000$ & $0.993$ & $1.000$ & $1.000$ & $1.000$ & $1.000$ & $1.000$ & $1.000$ \\       
		$H_{12,0.15}$  & $0.999$ & $0.993$ & $0.975$ & $1.000$ & $1.000$ & $1.000$ & $1.000$ & $1.000$ & $1.000$ \\
		\bottomrule\bottomrule
	\end{tabular}
	\caption{\small Empirical size and power of the circular-circular goodness-of-fit test for models CC1--CC12 with different sample sizes, deviations and significance levels. \label{gofdens:tab:results:cc:ext}}
\end{table}

%-------------------------------------------------%
\subsection{Bandwidth choice}
%-------------------------------------------------%

The delicate issue of the bandwidth choice for the testing procedure has been approached as follows. In the simulation results presented in Section \ref{gofdens:sec:sim}, a fixed pair of bandwidths has been chosen based on a Likelihood Cross Validation criterion. Ideally, one would like to run the test in a grid of several bandwidths to check how the test is affected by the bandwidth choice. This has been done for six circular-linear and circular-circular models, as shown in Figure \ref{gofdens:fig:band:ext}. Specifically, Figure \ref{gofdens:fig:band:ext} shows percentages of rejections under the null ($\delta=0.00$, green) and under the alternative ($\delta=0.15$, orange), computed from $M=1000$ Monte Carlo samples for each pair of bandwidths (the same collection of samples for each pair) on a logarithmic spaced $10\times10$ grid. The sample size considered is $n=100$ and the number of bootstrap replicates is $B=1000$.\\

As it can be seen, the test is correctly calibrated regardless the bandwidths value. In fact, for all the models explored, the rejection rates for each pair of bandwidths in the grid are inside the $95\%$ confidence interval of the proportion $\alpha=0.05$ (this happens for $95.75\%$ of the bandwidths in the grid). However, the power is notably affected by the choice of the bandwidths, with rather different behaviours depending on the model and on the alternative. Reasonable choices of the bandwidths based on an estimation criterion such as the one obtained by the median of the LCV bandwidths (\ref{gofdens:band:lcv}) lead in general to a competitive power.

%-------------------------------------------------%
\subsection{Further results}
%-------------------------------------------------%

Tables \ref{gofdens:tab:results:cl:ext} and \ref{gofdens:tab:results:cc:ext} collect the results of the simulation study for each combination of model (CL or CC), deviation ($\delta$), sample size ($n$) and significance level ($\alpha$). When the null hypothesis holds, the level of the test is correctly attained for all significance levels, sample sizes and models. Under the alternative, the tests perform satisfactorily, having both of them a quick detection of the alternative when only a $10\%$ and a $15\%$ of the data come from a density not belonging to the null parametric family. 

\begin{figure}[H]
	\centering
	\includegraphics[width=0.425\textwidth]{fires_scatterplot.pdf}\includegraphics[width=0.425\textwidth]{prot_scatterplot.pdf}\\[-0.75cm]
	\includegraphics[width=0.425\textwidth]{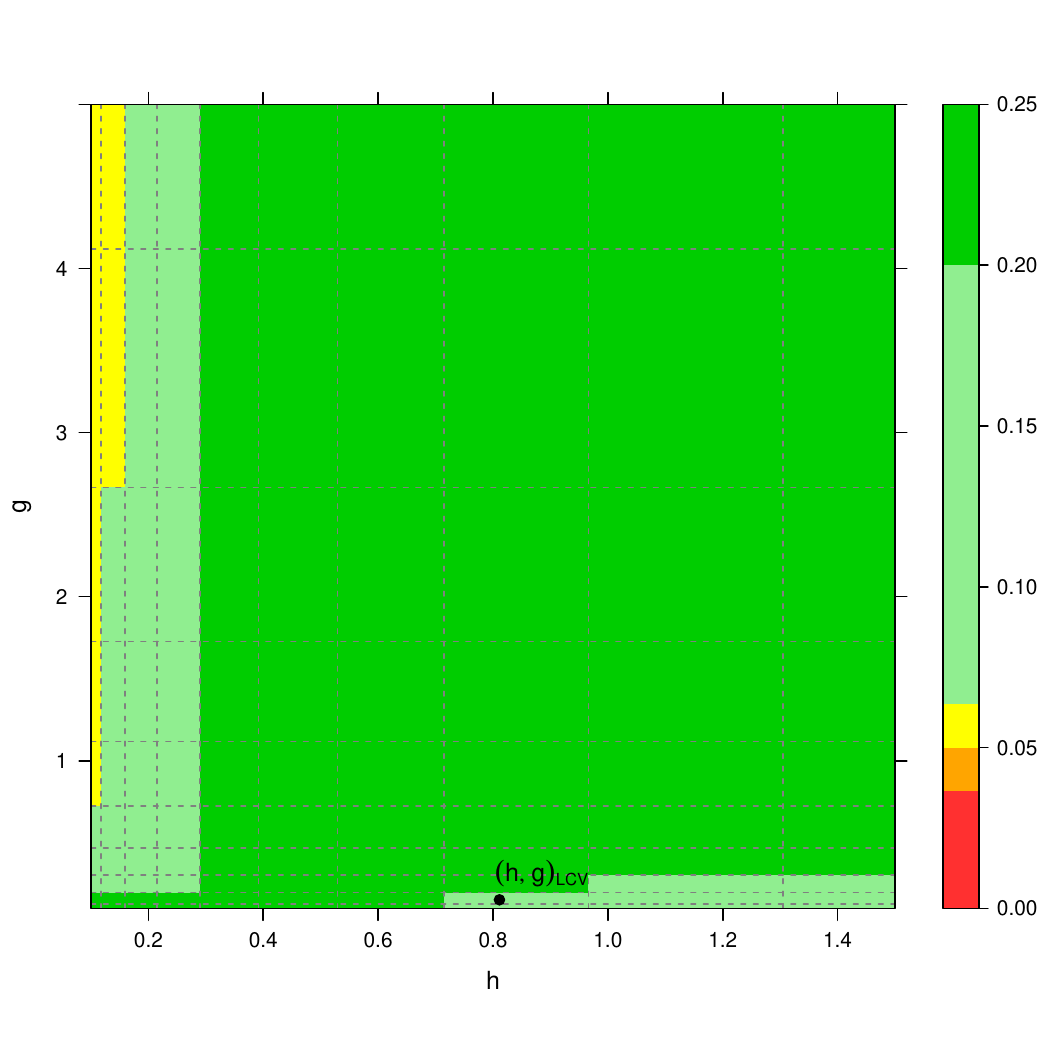}\includegraphics[width=0.425\textwidth]{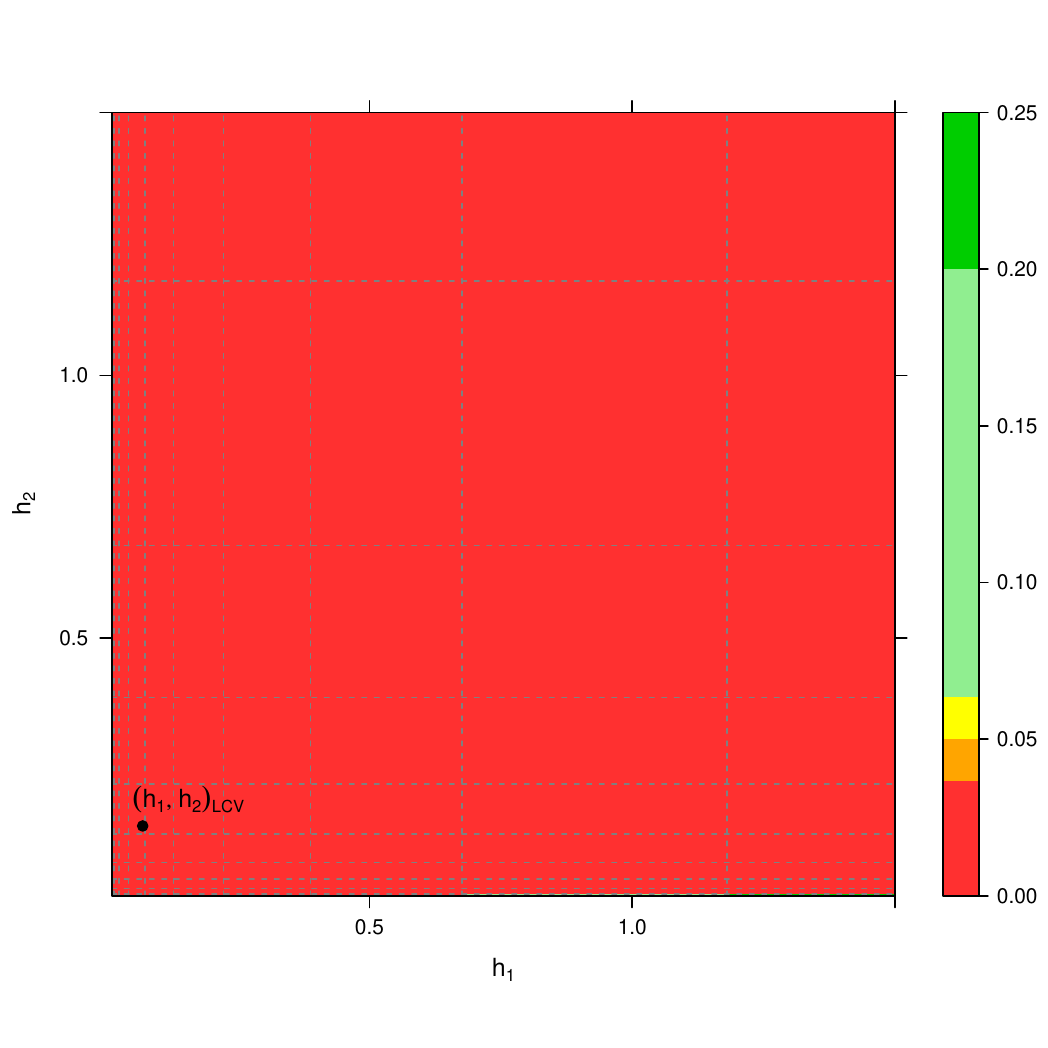}\\[-0.25cm]
	\caption{\small Upper row, from left to right: parametric fit (model from \cite{Mardia1978}) to the circular mean orientation and mean log-burnt area of the fires in each of the $102$ watersheds of Portugal; parametric fit (model from \cite{Fern'andez-Dur'an2007}) for the dihedral angles of the alanine-alanine-alanine segments. Lower row: $p$-values of the goodness-of-fit tests for a $10\times10$ grid, with the LCV bandwidth for the data. \label{gofdens:fig:data:mardia:ext}}
\end{figure}

%-------------------------------------------------%
\section{Extended data application}
\label{gofdens:su:data}
%-------------------------------------------------%

The analysis of the two real datasets presented in Section \ref{gofdens:sec:data} has been complemented by exploring the effect of different bandwidths in the test. To that aim, Figure \ref{gofdens:fig:data:mardia:ext} shows the $p$-values computed from $B=1000$ bootstrap replicates for a logarithmic spaced $10\times10$ grid, as well as bandwidths obtained by LCV for each dataset. The graphs shows that there are no evidences against the model of \cite{Mardia1978} for modelling the wildfires data and that the model used to describe the proteins dataset is not adequate. This model employs the copula structure of \cite{Wehrly1980} with marginals and link function given by circular densities based on NNTS, specifying \cite{Fern'andez-Dur'an2007} that the best fit in terms of BIC arises from considering three components for the NNTS's in the marginals and two for the link function. The fitting of the NNTS densities was performed using the \texttt{nntsmanifoldnewtonestimation} function of the package \texttt{CircNNTSR} \citep{CircNNTSR}, which computes the MLE of the NNTS parameters using a Newton algorithm on the hypersphere. The two-step ML procedure described in Section \ref{gofdens:su:simus} was employed to fit first the marginals and then the copula. The resulting contour levels of the parametric estimate are quite similar to the ones shown in Figure 5 of \cite{Fern'andez-Dur'an2007}. The dataset is available\nopagebreak[4] as \texttt{ProteinsAAA} in the \texttt{CircNNTSR} package.

\fi

\end{document}